%% file: manuscript.tex
\documentclass[11pt]{article}

\usepackage{amsmath, amssymb, amsthm, mathtools}

\usepackage[utf8]{inputenc}
\usepackage{newtxtext,newtxmath}

\usepackage[a4paper, margin=1in]{geometry}

\usepackage{graphicx}
\usepackage{booktabs}
\usepackage{caption}

\usepackage[authoryear]{natbib}
\usepackage{hyperref}
\usepackage[affil-it]{authblk}
\usepackage{enumitem}

\usepackage{algpseudocode}
\usepackage{algorithm}

\usepackage{tikz}
\usetikzlibrary{decorations.pathreplacing, patterns}
\usetikzlibrary{positioning, shapes}
\usetikzlibrary{matrix}
\usetikzlibrary{shapes.symbols}

\definecolor{CMUred}{RGB}{196, 18, 48}
\definecolor{CMUbeige}{RGB}{228, 218, 196}
\definecolor{CMUgray1}{RGB}{109, 110, 113}
\definecolor{CMUskyblue}{RGB}{0, 123, 192}
\definecolor{CMUgold}{RGB}{253, 181, 21}
\definecolor{CMUgreen}{RGB}{0, 150, 71}

\newcommand{\Exp}{\mathbb{E}}
\newcommand{\Var}{\mathbb{V}}
\newcommand{\Ind}{\mathbb{I}}

\newcommand{\Real}{\mathbb{R}}
\newcommand{\Mop}{\mathcal{M}}
\newcommand{\Lop}{\mathcal{L}}
\newcommand{\Cov}{\text{Cov}}
\newcommand{\Corr}{\text{Corr}}
\newcommand{\Data}{\mathcal{D}}
\newcommand{\Normal}{\mathcal{N}}

\newcommand{\indep}{\perp \!\!\!\! \perp}

\newcommand{\Mray}{\widetilde{\mathcal{M}}^{-1}}
\newcommand{\Ustr}{\mathcal{U}}
\newcommand{\Vcorr}{\mathcal{V}}
\newcommand{\Qcal}{\mathcal{Q}}
\newcommand{\Ycal}{\mathcal{Y}}
\newcommand{\rstar}{\rho^{\ast}}
\newcommand{\Func}{F}
\newcommand{\Rem}{\operatorname{Rem}}

\newcommand{\pto}{\overset{p}{\rightarrow}}
\newcommand{\dto}{\leadsto}

\theoremstyle{plain}
\newtheorem{theorem}{Theorem}
\newtheorem{lemma}{Lemma}
\newtheorem{corollary}{Corollary}
\newtheorem{proposition}{Proposition}
\newtheorem{assumption}{Assumption}

\theoremstyle{definition}
\newtheorem{definition}{Definition}
\newtheorem{example}{Example}

\theoremstyle{remark}
\newtheorem{remark}{Remark}

\title{Towards Efficient Inference under Nonmonotone Missingness with General Imputation}
\author[1]{Qi Xu}
\author[1,2]{Lorenzo Testa}
\author[1]{Jing Lei}
\author[1,3]{Kathryn Roeder\thanks{Corresponding author. Email: roeder@andrew.cmu.edu}}
\affil[1]{Department of Statistics \& Data Science, Carnegie Mellon University, Pittsburgh, PA 15213, USA}
\affil[2]{L'EMbeDS, Sant'Anna School of Advanced Studies, 56127 Pisa PI, Italy}
\affil[3]{Department of Computational Biology, Carnegie Mellon University, Pittsburgh, PA 15213, USA}
\date{\today}

\begin{document}
\maketitle

\begin{abstract}

Missing data are ubiquitous in classical survey and longitudinal studies as well as modern multi-modality data analysis. A longstanding challenge arises under nonmonotone missingness, where different units may observe arbitrary subsets of all variables. We study parameter estimation and inference problem under this setting. Semiparametric efficiency theory characterizes the efficient estimator through inversion of an operator constructed from pattern-specific conditional expectations. However, this estimator is generally not tractable due to compositions of conditional expectations across patterns.

We introduce the Restricted ANOVA hierarchY (RAY), a functional decomposition that reveals an almost-eigen structure of the operator under missing completely at random. This structure yields a closed-form, computable approximation to the efficient estimator. RAY estimator is applicable to general Z-estimation problems, and it remains unbiased for arbitrary independent imputation functions. In theory, we establish verifiable sufficient conditions where RAY attains the efficiency lower bound, and offer a general bound for the efficiency gap otherwise. We further develop adaptive RAY estimator, which attains the minimal asymptotic variance within a broader class containing RAY and other existing estimators. Finally, we investigate the extension of RAY under missing at random mechanism. Simulations and a single-cell multi-omics application demonstrate the efficiency gains of the proposed estimators.
\end{abstract}


\section{Introduction}
\label{sec: intro}

Missing data have been a significant obstacle since data were quantitatively collected. In modern scientific research, such challenge is becoming more pronounced. For example, in single-cell and spatial multi-omics, different assays measure different combinations of gene expression, chromatin accessibility, and protein abundance. Analogous structures arise in electronic health records and large surveys, where cost, study design, or collection logistics determine which variables are collected. When the observed patterns across units are arbitrary, or more formally, not totally ordered by inclusion, the missingness pattern is termed as \emph{nonmonotone}, also referred to as \emph{blockwise missingness}. Complete-case analysis excludes every unit not observing all variables and may consequently discard a substantial fraction of the available data. Direct substitution of imputed values does not resolve this limitation: inaccurate imputations may reduce efficiency, and treating imputed values as observed can invalidate uncertainty quantification. To this end, we target to combine information across all observed patterns while preserving valid inference given general imputation, which can be misspecified or biased.

\paragraph{Full data and target parameter.}
Let $Z = (Z_1, \dots, Z_p)$ denote the full-data vector of $p$ modalities. Each modality $Z_j$ takes values in $\mathcal{Z}_j \subseteq \Real^{d_j}$, so $Z$ takes values in $\mathcal{Z} = \prod_{j=1}^{p}\mathcal{Z}_j$. We write $Z \sim P$, where $P$ denotes the full-data law and places no restriction on dependence among the modalities. For any index set $r \subseteq [p]$, let $Z_r = (Z_j : j \in r)$ denote the sub-vector indexed by $r$.

We assume that $P$ belongs to a semiparametric model $\{P_{\theta,\eta}:\theta\in\Theta,\eta\in\mathrm H\}$, where $\Theta\subseteq\Real^d$ is open, $\theta$ is the parameter of interest, and $\eta$ is a possibly infinite-dimensional nuisance parameter. Let $\psi^F:\mathcal Z\times\Theta\to\Real^d$ be a measurable full-data estimating function. The target $\theta^\star=\theta(P)$ is the unique solution of
\begin{equation}
    \label{eq: full-data-ee}
    \Exp_P\!\left[\psi^F(Z, \theta^\star)\right] = \mathbf{0}_d.
\end{equation}
Standard Z-estimation conditions are assumed. Throughout, $\psi^F$ is treated as fixed, and all efficiency statements below are relative to this choice.

\begin{example}[Outcome mean and regression]
\label{ex: targets}
When $\theta^\star=\Exp_P[Z_1]$ is the mean of a scalar modality, one may take $\psi^F(Z,\theta)=Z_1-\theta$. For a (generalized) linear regression of a scalar response $Z_p$ on the remaining modalities, let $Z_{-p}:=(Z_1,\dots,Z_{p-1})$ denote the concatenated covariate vector, augmented by an intercept if desired. With link $g$, one may take $\psi^F(Z,\theta)=Z_{-p}\{Z_p-g^{-1}(\theta^\top Z_{-p})\}$. The resulting moment equation defines a population regression parameter and recovers the usual regression coefficient when the conditional mean model is correctly specified. Both examples fit into~\eqref{eq: full-data-ee}.
\end{example}

\paragraph{Missing data pattern.}
For each unit, the random \emph{pattern} $R\subseteq[p]$ indicates which modalities are observed, so the observed data are $(R,Z_R)$\footnote{Following \citet{tsiatis2006semiparametric}, the \emph{full data} are the vector $Z$ that would be recorded in the absence of missingness. The \emph{complete pattern} is $[p]$, and units with $R=[p]$ are \emph{complete cases}. Quantities defined on the full data carry the superscript $F$.}. Let $\mathbb P$ denote the joint law of $(Z,R)$, whose $Z$-marginal is $P$. We write $\Exp_P$ for expectation under the full-data law and use unsubscripted $\Exp$ for expectation under $\mathbb P$. Define
\[
\Qcal:=\{r\subseteq[p]:\mathbb P(R=r)>0\}
\]
as the collection of \emph{observed patterns}. Given $N$ independent units, let $n_r:=\#\{i:R_i=r\}$ and let $\Data_r:=(Z_{i,r})_{i:R_i=r}$ denote the subsample with pattern $r$. The observed data is $\Data:=((R_i,Z_{i,R_i}):i\in[N])$. We allow $\Qcal$ to be \emph{nonmonotone}, meaning that it need not be totally ordered by inclusion. The \emph{monotone} case, in which $\Qcal$ forms a chain $r_1\subsetneq\cdots\subsetneq r_K=[p]$, is treated as an important special case in Section~\ref{sec: exactness}.

\paragraph{Missingness mechanisms.}
We consider the two standard missing-data mechanisms stated in Assumptions~\ref{assumption: missing_positivity} and~\ref{assumption: mar} below. Both require that complete cases occur with positive probability, because without this overlap the target $\theta^\star$ is not identified in general.

\begin{assumption}[MCAR with positivity]\label{assumption: missing_positivity}
\leavevmode
\begin{enumerate}[label=(\alph*)]
    \item (MCAR) $R \indep Z$; equivalently, $\mathbb P(R=r\mid Z)=\mathbb P(R=r)=:\pi_r$ for all $r\in\Qcal$.
    \item (Positivity) There exists $c > 0$ such that $\pi_{[p]} \geq c$.
\end{enumerate}
\end{assumption}

\begin{remark}[A weaker missing probability condition]\label{remark: observed-block-independence}
The missing mechanism in Assumption~\ref{assumption: missing_positivity}(a) can be weakened. It is enough that
\begin{equation}
    \label{eq: observed-block-independence}
    \mathbb P(R=r\mid Z_r)=\mathbb P(R=r)=\pi_r,
    \qquad r\in\Qcal,
\end{equation}
including $r=[p]$. We call~\eqref{eq: observed-block-independence} as \emph{patternwise observed-block independence}. Under this condition, the conclusions in Sections~\ref{sec: estimation-mcar}--\ref{sec: exactness}, including those for RAY, IPI, and adaptive RAY, continue to hold. This condition is strictly weaker than MCAR because the probability of pattern $r$ may depend on modalities outside $r$. Such dependence is generally MNAR; indeed, imposing both~\eqref{eq: observed-block-independence} and MAR recovers MCAR. We therefore view~\eqref{eq: observed-block-independence} primarily as a mathematical relaxation. Its strict extension beyond MCAR is less directly interpretable as a data-collection mechanism.
\end{remark}

\begin{assumption}[MAR with strong positivity]\label{assumption: mar}
\leavevmode
\begin{enumerate}[label=(\alph*)]
    \item (MAR) $\mathbb P(R=r\mid Z)=\mathbb P(R=r\mid Z_r)=:\pi_r(Z_r)$ for all $r\in\Qcal$.
    \item (Positivity) There exists $c > 0$ such that $\pi_{[p]}(Z_{[p]}) \geq c$ almost surely.
\end{enumerate}
\end{assumption}

\begin{remark}[Complete-pattern versus target-specific positivity]\label{remark: target-specific-positivity}
Complete-pattern positivity is a convenient sufficient overlap condition for a general full-data estimand and underlies the operator theory developed below. Without structural restrictions, a target depending on the joint law of all modalities is not identified when complete cases are never observed. For a target depending only on partial modalities, say $Z_u$, however, weaker target-specific positivity can suffice. Under MCAR or~\eqref{eq: observed-block-independence}, if $\psi^F(Z,\theta)=\psi_u(Z_u,\theta)$, identification may require only $\mathbb P(R\supseteq u)>0$. For example, identifying the outcome mean $\Exp_P[Z_1]$ requires positive probability that $Z_1$ is observed, not positive probability of observing every modality jointly. The general RAY theory below retains complete-pattern positivity; allowing $\pi_{[p]}=0$ for such special targets would require a target-specific reformulation of the estimating function and its operator analysis.
\end{remark}

\paragraph{Semiparametric efficiency lower bound.} To unify notation across mechanisms, write $\pi_r(Z_r)$ for the pattern propensity, interpreting it under MCAR as the constant pattern probability $\pi_r$. Following \citet{Robins1994estimation} and \citet{tsiatis2006semiparametric}, define two linear operators on $L_0^2(P):=\{f\in L^2(P):\Exp_P f=0\}$:\footnote{The operators are defined on scalar functions. For vector-valued functions such as $\psi^F$, they act componentwise; equivalently, they act on the product space $L_0^2(P)^d$, with $\langle f,g\rangle_d:=\sum_{j=1}^d\Exp_P(f_jg_j)$ and $\|f\|_d^2:=\langle f,f\rangle_d$. When differentiating estimating equations, we use the canonical extension of $\Mop$ to $L^2(P)$. Indeed, $L^2(P)=\operatorname{span}\{1\}\oplus L_0^2(P)$, $\Mop$ preserves $L_0^2(P)$ and fixes constants because $\Mop(1)=\sum_{r\in\Qcal}\pi_r(Z_r)=1$. Thus $\Mop^{-1}$ acts as the identity on the constant component and as the inverse on $L_0^2(P)$.}
\begin{equation}
    \label{eq: operators}
    \Mop(f) \;=\; \sum_{r \in \Qcal} \pi_r(Z_r)\, \mathcal{A}_r(f),
    \qquad
    \Lop(f) \;=\; \sum_{r \in \Qcal} \Ind(R = r)\, \mathcal{A}_r(f),
\end{equation}
where $\mathcal A_r(f):=\Exp_P[f\mid Z_r]$. Under either Assumption~\ref{assumption: missing_positivity} or~\ref{assumption: mar}, $\Mop$ is bounded and self-adjoint.

\begin{theorem}[Semiparametric efficient estimating function]
\label{thm: oef}
Suppose the propensity functions are known and either Assumption~\ref{assumption: missing_positivity} or Assumption~\ref{assumption: mar} holds. Fix the full-data estimating function $\psi^F$, the optimal observed-data estimating function is given by
\begin{equation}
    \label{eqn: oef}
    \psi_{\mathrm{opt}}(R,Z_R,\theta^\star)
    :=\Lop\{\Mop^{-1}[\psi^F(Z,\theta^\star)]\}.
\end{equation}
\end{theorem}

Theorem~\ref{thm: oef} can be proved using Theorem 10.1, 10.6 and Lemma 10.5 in \citet{tsiatis2006semiparametric}. Note that the inverse of $\Mop$ admits the operator-norm-convergent Neumann expansion $\Mop^{-1}=\sum_{k=0}^\infty(\mathcal I-\Mop)^k$. Therefore, it can be approximated by letting $\psi_0(Z, \theta) = \psi^F(Z, \theta)$ and computing $\psi_{k+1} = (\mathcal I - \Mop)(\psi_k) + \psi^F$ successively. This approximation requires composition and interaction of conditional expectations, which can be rather sophisticated without special structural assumptions.

\subsection{Related work}

\paragraph{Semiparametric theory and nonmonotone missingness.} Estimation and inference with missing data have been studied extensively in the semiparametric literature \citep{Robins1994estimation, chen2004nonparametric, tsiatis2006semiparametric}, including the nonmonotone observed-pattern sets considered here \citep{robins1997nonb, robins1997nona}. Although the optimal influence function under nonmonotone missingness is well understood in theory, its empirical implementation is usually involved. Several recent works seek semiparametrically efficient estimators under nonmonotone patterns by introducing additional structural assumptions. In particular, \citet{chaudhuri2016gmm} derive a closed-form efficient influence function under a restricted missing-at-random mechanism in which the selection probability depends solely on the always-observed variables. \citet{yang2022double} consider a sequential MAR assumption and propose an efficient estimator under it, and \citet{huang2025efficient} attain the efficiency bound when each observed pattern contains either all modalities or only one modality, under MCAR. A separate line of work approximates optimal observed-data estimating functions through generalized ANOVA decompositions \citep{berrett2024efficient}, but solves the resulting problem by gradient descent rather than the closed-form expression we obtain.

\paragraph{Prediction-powered inference.} Prediction-powered inference (PPI; \citealt{angelopoulos2023prediction, angelopoulos2023ppi++, zrnic2024cross}) is a versatile framework for estimation and inference in a semi-supervised setting. In its vanilla version, PPI employs independent imputation model to predict outcomes on an unlabeled dataset, then corrects the resulting bias with a labeled dataset. Statistical efficiency improves when the imputation model is accurate and the unlabeled dataset is much larger than the labeled one, but the vanilla version does not guarantee an improvement over the complete-case estimator. Follow-up works address this through tuning parameters \citep{angelopoulos2023ppi++} or efficient influence functions \citep{miao2023assumption, gronsbell2024another, ji2025predictions}, and recent extensions cover broader estimator classes \citep{zou2026generalized}. During the final stage of preparing our work, \citet{duan2025imputation, zhao2025imputation} and \citet{chen2025unified} proposed imputation-powered methods for nonmonotone missing data under MCAR and MAR, respectively. These methods share a pattern-recalibration idea, termed \textit{pattern stratification} by \citet{chen2025unified}. Under MCAR, the construction of \citet{duan2025imputation, zhao2025imputation} is a special case of our estimator class.

\paragraph{Regression under nonmonotone missing data.} A parallel literature studies parametric or nonparametric regression and variable selection under nonmonotone missing data \citep{kundu2019generalized, yu2020optimal, xue2021integrating, xue2021statistical, jin2023modular, li2024adaptive, song2024semi, sell2024nonparametric, diakite2025adapdiscom}. Relying on specific structure of the normal equations or moment conditions, most of these methods are restricted to regression coefficients, whereas our framework targets the broader class of Z-estimation problems \citep{van2000asymptotic}. Some ideas are nonetheless related: the weighting strategy of \citet{li2024adaptive} matches its counterpart in the imputation-powered methods of \citet{chen2025unified, zhao2025imputation}, while our RAY and adaptive estimators are closer to the multiple block-imputation of \citet{xue2021integrating}, which applies several imputations to the same pattern to improve efficiency.

\subsection{Contributions}

\paragraph{A closed-form approximation to the optimal estimating function.} The RAY decomposition organizes pattern-specific conditional expectations into a hierarchy of components and assigns them closed-form weights. Under MCAR, these components act as almost-eigen-operators of the operator $\Mop$, enabling close-form approximation of $\Mop^{-1}$. As a result, this construction yields a valid, closed-form estimating function for arbitrary nonmonotone observed-pattern sets and accommodates arbitrary imputation functions.

\paragraph{Adaptive estimators.} A change of augmentation weights recovers the imputation-powered inference (IPI) estimator of \citet{duan2025imputation, zhao2025imputation}. Since RAY and IPI estimators do not dominates each other uniformly, we consider a large class of estimators that including RAY and IPI estimators as special cases. The adaptive RAY (aRAY) estimator is then proposed, which attains the minimal asymptotic variance within the estimator class.

\paragraph{Exactness and efficiency gap.} We give a necessary and sufficient residual condition for RAY to attain the fixed-$\psi^F$ efficiency lower bound and identify several verifiable sufficient conditions. These include intersection-closed observed-pattern sets under independent modalities, monotone observed-pattern sets under arbitrary dependence, and anchored observed-pattern sets under conditional independence. Beyond these sufficient settings, we bound any efficiency gap relative to the lower bound by two interpretable factors: dependence between the private modalities of overlapping patterns and a structural defect of the observed-pattern set.

\paragraph{The extension to missing at random.} We further investigate the RAY construction under general MAR. The direct and projected weights both satisfy valid population identities, but the direct weights generally depend on unobserved modalities. Projection yields an observed-data-measurable equation but introduces the projected tail propensity. We identify this quantity and show that its estimation can contribute to the target equation at first order. This calculation provides a route to root-$N$ inference through correctly specified finite-dimensional models and stacked estimation, and motivates special cancellation or orthogonalization for more general nuisance estimators.

\subsection{Organization and notation}
Section~\ref{sec: estimation-mcar} develops the RAY decomposition and estimators under MCAR. Section~\ref{sec: exactness} characterizes the conditions when RAY estimator attains semiparametric efficiency lower bound and derives an upper bound for the efficiency gap when those conditions fail. Section~\ref{sec: estimation-mar} investigates an extension to MAR and its obstacles. Section~\ref{sec: sim} reports MCAR simulations and a sensitivity analysis under MAR and MNAR departures, and Section~\ref{sec: real_data} applies the method to a single-cell multi-omics benchmark with independently assigned observed patterns. Proofs and additional material are collected in the Appendix. Throughout, $[p] := \{1, \dots, p\}$, the cardinality of a set $r$ is $|r|$, and $r \subseteq s$ denotes set inclusion. Expectation and variance are written $\Exp[\cdot]$ and $\Var[\cdot]$, and $\Ind(\cdot)$ is the indicator function.

\section{Estimation under MCAR}
\label{sec: estimation-mcar}

This section develops the RAY methodology under MCAR. Section~\ref{subsec: warmup} first revisits a three-modality problem to show how prediction-powered and imputation-powered corrections use incomplete patterns. Section~\ref{subsec: almost-eigen-decomposition} then introduces the RAY decomposition and uses its almost-eigen structure to construct a tractable surrogate for $\Mop^{-1}$. Section~\ref{subsec: ray-estimator} defines the estimator, develops safeguard tuning, and establishes imputation-model-robust validity under MCAR. Finally, because neither RAY nor imputation-powered inference is uniformly preferable, Section~\ref{subsec: adaptive-ray-estimator} embeds both in a larger adaptively weighted class, and produce the adaptive RAY (aRAY) estimator.

\subsection{A motivating example}
\label{subsec: warmup}

\begin{figure}[h]
    \centering
    \resizebox{\textwidth}{!}{\input{prototype}}
    \caption{\label{figure:prototype}A motivating example of three-modality data with four distinct observed patterns. Each column is an observed pattern; a colored cell marks an observed modality and a hatched cell a missing one.}
\end{figure}

Consider the three-modality dataset of Figure~\ref{figure:prototype}, with observed patterns $\Qcal = \{[3], \{1,2\}, \{1,3\}, \{1\}\}$, where $[3] = \{1,2,3\}$ and modality $Z_3$ is the outcome. Our target is the outcome mean $\theta^\star = \Exp[Z_3]$, so that $\psi^F(Z, \theta) = Z_3 - \theta$. Among the four patterns, only $[3]$ and $\{1,3\}$ observe the outcome $Z_3$. We refer to the units in these two patterns as the \emph{labeled} set and to the units in the remaining two patterns as the \emph{unlabeled} set, and write $\hat\pi_r=n_r/N$. The complete-case estimator\footnote{For this outcome-mean example, the baseline uses every pattern on which $\psi^F(Z,\theta)=Z_3-\theta$ is observed. It therefore includes pattern $\{1,3\}$, which is not the complete pattern $[3]$ under the terminology of Section~\ref{sec: intro}. We retain the notation $\hat\theta_{\mathrm{CC}}$ for convenience in this motivating example.} averages $Z_3$ over the labeled units,
\begin{equation}
    \label{eqn: warmup_cc}
    \hat\theta_{\mathrm{CC}} = \frac{1}{N}\sum_{i=1}^{N}\frac{\Ind\!\left(R_i \in \{[3], \{1,3\}\}\right)}{\hat\pi_{[3]} + \hat\pi_{\{1,3\}}}\,Z_{i,3},
\end{equation}
whose asymptotic variance is $\Var(Z_3)/(\pi_{[3]} + \pi_{\{1,3\}})$. It discards the unlabeled units entirely, even though they carry information about $Z_3$ through the covariate $Z_1$.

Prediction-powered inference \citep{angelopoulos2023prediction} recovers part of that information. Given a model $f(Z_1)$ predicting $Z_3$, it adds a mean-zero correction that contrasts the predictions on the unlabeled and labeled sets,
\begin{equation}
    \label{eqn: warmup_ppi}
    \hat\theta_{\mathrm{PPI}} = \hat\theta_{\mathrm{CC}} + \frac{1}{N}\sum_{i=1}^{N}\left( \frac{\Ind(R_i \in \{\{1,2\},\{1\}\})}{\hat\pi_{\{1,2\}} + \hat\pi_{\{1\}}} - \frac{\Ind(R_i \in \{[3],\{1,3\}\})}{\hat\pi_{[3]} + \hat\pi_{\{1,3\}}}\right) f(Z_{i,1}).
\end{equation}
Thus~\eqref{eqn: warmup_cc} is the labeled-sample mean, and~\eqref{eqn: warmup_ppi} adds the difference between the unlabeled- and labeled-sample means of $f(Z_1)$. Under MCAR this correction has population mean zero, so $\hat\theta_{\mathrm{PPI}}$ is consistent for any independent imputation model $f$. Its asymptotic variance is $\Var(f)/(\pi_{\{1,2\}}+\pi_{\{1\}}) + \Var(Z_3 - f)/(\pi_{[3]}+\pi_{\{1,3\}})$. An accurate predictor can therefore improve on $\hat\theta_{\mathrm{CC}}$, whereas an inaccurate predictor can increase variance. Prediction-powered inference with tuning (\texttt{PPI++}; \citealt{angelopoulos2023ppi++}) introduces a scalar coefficient $\alpha$ and minimizes a selected scalar variance criterion. At the population optimum, the criterion cannot exceed its value at $\alpha=0$, which recovers $\hat\theta_{\mathrm{CC}}$.

The estimators in \citet{angelopoulos2023prediction, angelopoulos2023ppi++} use only the covariate $Z_1$, even though the pattern $\{1,2\}$ also observes the second covariate $Z_2$, which a richer model $g(Z_1, Z_2)$ can exploit. To make use of it, the imputation-powered inference estimator (IPI; \citealt{duan2025imputation, zhao2025imputation}) adds to \eqref{eqn: warmup_ppi} a second correction that compares the pattern $\{1,2\}$ with the complete cases,
\begin{equation}
    \label{eqn: warmup_ipi}
    \frac{\alpha_g}{N}\sum_{i=1}^{N}\left( \frac{\Ind(R_i = \{1,2\})}{\hat\pi_{\{1,2\}}} - \frac{\Ind(R_i = [3])}{\hat\pi_{[3]}} \right) g(Z_{i,1}, Z_{i,2}),
\end{equation}
and tunes $\alpha_f, \alpha_g$ jointly. Both corrections have population mean zero under MCAR, so this estimator is consistent for any independent imputation models $f$ and $g$. Because zero tuning coefficients are feasible, population-optimal tuning cannot increase the selected scalar variance criterion relative to $\hat\theta_{\mathrm{CC}}$. Even when $f = \Exp[Z_3 \mid Z_1]$ and $g = \Exp[Z_3 \mid Z_1, Z_2]$ are the true conditional means, however, the estimator \eqref{eqn: warmup_ppi}--\eqref{eqn: warmup_ipi} does not generally attain the fixed-$\psi^F$ efficiency lower bound in Theorem \ref{thm: oef}.

\subsection{The almost-eigen decomposition}
\label{subsec: almost-eigen-decomposition}

Under MCAR the weights $\pi_r$ are scalars, yet $\Mop = \sum_{r \in \Qcal} \pi_r \mathcal{A}_r$ still mixes non-commuting conditional-expectation projections, so its inverse remains intractable by direct expansion. Instead, we decompose $\Mop$ through a Restricted ANOVA hierarchY (RAY) structure whose components $\mathcal{P}_s$ are approximate eigen-operators of $\Mop$: applying $\Mop$ to each $\mathcal{P}_s(f)$ returns a scalar multiple $\lambda_s\mathcal{P}_s(f)$ plus a remainder term, as we show below.

\begin{lemma}[RAY decomposition]
    \label{lemma: ray}
    For any $f \in L^2_0(P)$,
    \begin{equation}
        \label{eqn: ray_decomp}
        f = \sum_{s \subseteq[p]}\mathcal{P}_s(f), \qquad \mathcal{P}_s(f) = \sum_{r \subseteq s,\, r\in \Qcal} (-1)^{|s|-|r|} \mathcal{A}_r(f), \qquad \mathcal{P}_{\emptyset} = 0.
    \end{equation}
\end{lemma}

The component $\mathcal{P}_s(f)$ isolates the contribution indexed by $s$ after adjusting for lower-order observed-pattern projections, and the proof of the lemma is given in Appendix~\ref{app_sec: ray_property}. The RAY decomposition is conceptually related to the functional ANOVA and Hoeffding--Sobol decompositions \citep{hoeffding1948, hooker2004discovering, owen2013monte}, all of which follow from inclusion--exclusion principle. Two differences matter here. First, $\mathcal{P}_s(f)$ sums signed projections that are \textit{restricted} to $\Qcal$, not over all subsets of $[p]$. Second, the decomposition allows dependence among modalities, so the components need not be orthogonal and $\mathcal P_s$ need not itself be a projection. Despite this, the components interact with $\Mop$ in a structured way: for any $s \subseteq [p]$,
\begin{equation}
\begin{split}
    \label{eqn: Mop_eigen}
    \Mop[\mathcal{P}_s(f)]
    &= \Big( \sum_{r\in \Qcal,\, s\subseteq r}\pi_r \Big) \mathcal{P}_s(f) + \sum_{r\in \Qcal,\, s\not\subseteq r}\pi_r \mathcal{A}_{r}[\mathcal{P}_s(f)]
    \;\triangleq\; \lambda_{s}\mathcal{P}_s(f) + \Rem_{s}(f),
\end{split}
\end{equation}
where the first equality uses that $\mathcal{P}_s(f)$ is $\sigma(Z_r)$-measurable whenever $s \subseteq r$. We call $\mathcal{P}_s$ an \emph{almost-eigen-operator}\footnote{The term \emph{almost-eigen} mirrors its use, which has been used elsewhere, for example, the almost-eigenvectors of random regular graphs \citep{backhausz2019almost}.} of $\Mop$ with \emph{almost-eigenvalue} $\lambda_s = \sum_{r \in \Qcal:\, r \supseteq s}\pi_r = \mathbb P(R \supseteq s)$, the proportion of units whose patterns record every modality in $s$. By positivity, $\lambda_s \ge \pi_{[p]} > c > 0$ for all $s$. The \emph{remainder} $\Rem_s(f) = \sum_{r \in \Qcal:\, s \not\subseteq r}\pi_r \mathcal{A}_r[\mathcal{P}_s(f)]$ collects the compositions of conditional expectations, and it is the sole obstacle to an exact closed form.

Because $\Mop$ is linear, \eqref{eqn: Mop_eigen} gives
\begin{equation}
\label{eqn: Mop_inverse}
    \Mop^{-1}(f) = \sum_{s\subseteq [p]}\frac{1}{\lambda_s}\mathcal{P}_s(f) - \sum_{s\subseteq [p]}\frac{1}{\lambda_s}\Mop^{-1}[\Rem_s(f)],
\end{equation}
which splits $\Mop^{-1}$ into a tractable term, built from the almost-eigen-operators $\mathcal{P}_s$, and an intractable term that carries the remainders. Before addressing the intractable term, we record a useful global identity: the remainders cancel in aggregate.

\begin{remark}
\label{remark:mzero}
    The components satisfy $\Mop(f) = \sum_{s\subseteq [p]}\lambda_s \mathcal{P}_s(f)$, equivalently $\sum_{s\subseteq [p]}\Rem_s(f) = 0$, so the remainders can be ignored \emph{as a whole} when representing $\Mop$. The subtlety is that $\Mop^{-1}$ weights each $\Rem_s$ by $\lambda_s^{-1}$, so individual remainders need not be negligible. Section~\ref{sec: exactness} quantifies exactly when they are.
\end{remark}

\begin{remark}
\label{remark:remainder}
The three-modality design of Section~\ref{subsec: warmup} makes the remainders explicit. A short calculation from~\eqref{eqn: Mop_eigen} gives the only three potentially nonzero remainders,
\begin{align}
    &\Rem_{[3]}(f) = -\pi_{\{1,2\}}\Big\{\mathcal{A}_{\{1,2\}}\mathcal{A}_{\{1,3\}}(f)-\mathcal{A}_{\{1\}}(f)\Big\}
    -\pi_{\{1,3\}}\Big\{\mathcal{A}_{\{1,3\}}\mathcal{A}_{\{1,2\}}(f)-\mathcal{A}_{\{1\}}(f)\Big\}, \notag \\
    &\Rem_{\{1,2\}}(f) = \pi_{\{1,3\}}\Big\{ \mathcal{A}_{\{1,3\}}\mathcal{A}_{\{1,2\}}(f) - \mathcal{A}_{\{1\}}(f) \Big\}, \notag \\
    &\Rem_{\{1,3\}}(f) = \pi_{\{1,2\}}\Big\{ \mathcal{A}_{\{1,2\}}\mathcal{A}_{\{1,3\}}(f) - \mathcal{A}_{\{1\}}(f) \Big\}. \notag
\end{align}
Each remainder contains a composition of conditional expectations across two patterns. Such compositions are not directly evaluable pattern by pattern and require additional nested regression or iterative estimation. They also reveal when remainders vanish: if $Z_2 \indep Z_3 \mid Z_1$, then $\mathcal{A}_{\{1,2\}}\mathcal{A}_{\{1,3\}}(f) = \mathcal{A}_{\{1,3\}}\mathcal{A}_{\{1,2\}}(f) = \mathcal{A}_{\{1\}}(f)$, so all three remainders vanish and the surrogate is exact. Conditional independence means that $Z_2$ adds no predictive information about $Z_3$ beyond $Z_1$, reducing the problem to the two-modality semi-supervised case. Section~\ref{sec: exactness} generalizes this observation.

\end{remark}

Ignoring the remainder term in~\eqref{eqn: Mop_inverse} defines the \emph{RAY surrogate} $\Mray := \sum_{s\subseteq[p]}\lambda_s^{-1}\mathcal{P}_s$. This is the first of two approximations. The surrogate is still not directly usable, because forming the estimating function requires applying $\Lop$ to it, and
\begin{equation}
    \label{eqn: L_surrogate_split}
    \Lop\{\Mray[\psi^F]\} = \sum_{r\in\Qcal}\Ind(R=r)\Big\{ \underbrace{\sum_{s\subseteq r}\lambda_s^{-1}\mathcal{P}_s[\psi^F]}_{\text{a function of }Z_r} \;+\; \underbrace{\sum_{s\not\subseteq r}\lambda_s^{-1}\mathcal{A}_r[\mathcal{P}_s\psi^F]}_{\text{cross-pattern compositions}} \Big\},
\end{equation}
where the split uses $\mathcal{A}_r[\mathcal{P}_s\psi^F] = \mathcal{P}_s\psi^F$ for $s\subseteq r$. The first group is $\sigma(Z_r)$-measurable and is therefore directly computable. The second group contains cross-pattern compositions that are not directly evaluable for a unit with pattern $r$ and would require additional nested regression or iterative estimation. The RAY estimating function makes a \emph{second} approximation at this point by retaining only the directly computable first group. Reindexing that group by pattern, as detailed in Appendix~\ref{app_sec: valid_ee}, gives the \emph{RAY approximation} of the optimal estimating function,
\begin{equation}
    \label{eqn: aef}
    \psi(R,Z_R,\theta^\star)
    = \frac{\Ind(R=[p])}{\pi_{[p]}}\psi^F(Z,\theta^\star) + \sum_{r\in \Qcal,\, r\neq [p]}\omega_{r}\mathcal{A}_{r}[\psi^F(Z, \theta^\star)],
    \qquad
    \omega_r = \sum_{s:\,r\subseteq s\subseteq[p]}(-1)^{|s|-|r|}\frac{\Ind(R \supseteq s)}{\lambda_s}.
\end{equation}
We call $\omega_r$ the \emph{RAY weight}. Here and throughout the definitions and proofs of the RAY weights, the summation index ranges over all modality subsets between $r$ and $[p]$, not only over members of $\Qcal$. It is the almost-eigen weight that the surrogate attaches to the imputation function for pattern $r$. The two approximations, dropping $\sum_s\lambda_s^{-1}\Mop^{-1}\Rem_s$ from $\Mop^{-1}$ and dropping the composition group from $\Lop\{\Mray\psi^F\}$, are analyzed together in Section~\ref{sec: exactness}. The resulting estimating function~\eqref{eqn: aef} is directly computable from the observed modalities once its single-pattern conditional expectations are supplied. Under MCAR, $\omega_r$ depends only on $R$ and is independent of any function of $Z$; Appendix~\ref{app_sec: valid_ee} also shows $\Exp[\omega_r] = 0$ for every $r \neq [p]$. These two facts give the following validity result.

\begin{proposition}
    \label{prop: valid_ee}
    Under Assumption~\ref{assumption: missing_positivity}, the function in~\eqref{eqn: aef} satisfies $\Exp[\psi(R,Z_R,\theta^\star)] = \mathbf{0}_d$, so it is a valid estimating function for $\theta^\star$.
\end{proposition}

The same argument remains valid when $\mathcal{A}_r[\psi^F]$ is replaced by any independent imputation functions of $Z_r$: MCAR separates that function from the mean-zero RAY weight. Thus the imputation functions affect efficiency but not the population moment condition. Section~\ref{subsec: ray-estimator} makes this imputation-model-robust validity precise for the sample estimator.

\subsection{The RAY estimator}
\label{subsec: ray-estimator}

\paragraph{General imputation.} Imputation functions enter through the single-pattern quantities $\mathcal{A}_r[\psi^F(Z,\theta)] = \Exp[\psi^F(Z,\theta) \mid Z_r]$. In the \emph{imputation} mode, these quantities are approximated by $\psi^F(m_r(Z_r),\theta)$ for a fixed predictor $m_r(Z_r)$ of the missing modalities, matching the MCAR construction of \citet{zhao2025imputation}. In the \emph{expectation} mode, a fixed function directly approximates $\Exp[\psi^F(Z,\theta)\mid Z_r]$, for example through Monte Carlo evaluation under a conditional generative model. We write $\Func_r(Z_r;\theta)$ for either construction and treat the collection $\{\Func_r:r\in\Qcal, r\neq[p]\}$ as fixed and given. Correct specification is not required for validity under MCAR.

\paragraph{Safeguard tuning and sample calibration.} A misspecified imputation function does not change the population moment condition under MCAR, but it can inflate variance. We therefore attach a tuning parameter $\alpha_r$ to each $\Func_r$ and first define the population estimating function
\begin{equation}
    \label{eqn: mmpp_aef}
    \psi_{\alpha}(R,Z_R,\theta) = \frac{\Ind(R=[p])}{\pi_{[p]}}\psi^F(Z,\theta) + \sum_{r\in \Qcal,\, r\neq [p]}\alpha_r\omega_r\Func_{r}(Z_{r};\theta).
\end{equation}
For implementation, let $\hat\pi_r=n_r/N$, $\hat\lambda_s=\sum_{t\in\Qcal:\,t\supseteq s}\hat\pi_t$, and
\[
\hat\omega_{i,r}=\sum_{s:\,r\subseteq s\subseteq[p]}(-1)^{|s|-|r|}\frac{\Ind(R_i\supseteq s)}{\hat\lambda_s}.
\]
The RAY estimator solves
\begin{equation}
\label{eqn: empirical_ray}
\hat U_N(\theta;\alpha):=\frac1N\sum_{i=1}^N\left\{
\frac{\Ind(R_i=[p])}{\hat\pi_{[p]}}\psi^F(Z_i,\theta)
+\sum_{r\in\Qcal,\,r\neq[p]}\alpha_r\hat\omega_{i,r}\Func_r(Z_{i,r};\theta)
\right\}=\mathbf 0_d.
\end{equation}
The empirical weights are exactly calibrated: $N^{-1}\sum_i\hat\omega_{i,r}=0$. Consequently, replacing $\Func_r$ by $\Func_r-c_r$ for any constant vector $c_r$ leaves~\eqref{eqn: empirical_ray} unchanged. We use this invariance to center the imputation functions when estimating covariance matrices and tuning parameters. Replacing $\hat\omega_{i,r}$ by $\Ind(R_i=r)/\hat\pi_r-\Ind(R_i=[p])/\hat\pi_{[p]}$ gives the IPI estimator of \citet{duan2025imputation, zhao2025imputation}; the result below applies to either calibrated weight choice.

\begin{theorem}
    \label{thm: asymptotic_behavior}
    Let $\bar\Func_r=\Func_r(Z_r;\theta^\star)-\Exp[\Func_r(Z_r;\theta^\star)]$, $A = \Exp[\partial \psi^F(Z,\theta^\star)/\partial \theta^\top]$, $\gamma_r = \Exp[\omega_r \Ind(R=[p])/\pi_{[p]}]$, and $\nu_{r,r'} = \Exp[\omega_r\omega_{r'}]$. Suppose the imputation functions are independent and given. Under Assumption~\ref{assumption: missing_positivity} and the regularity conditions in Appendix~\ref{proof: asymp_behavior}, for every fixed tuning vector $\alpha$, the solution of~\eqref{eqn: empirical_ray} satisfies
    \begin{enumerate}[label=(\alph*)]
        \item $\hat\theta(\alpha) \pto \theta^\star$;
        \item $\sqrt{N}\{\hat\theta(\alpha) -\theta^\star\} \dto \Normal(\mathbf{0}_d, \Sigma_{\alpha})$, with
        \begin{equation}
            \label{eqn: avar}
            \begin{split}
            \Sigma_{\alpha} = A^{-1}\Bigg\{\frac{\Var(\psi^F)}{\pi_{[p]}}
            &+ \sum_{r\neq [p]}\alpha_r\gamma_r[\Cov(\psi^F, \bar\Func_{r}) + \Cov(\bar\Func_{r}, \psi^F)] \\
            &+\sum_{r,r'\neq [p]}\alpha_r\alpha_{r'}\nu_{r, r'}\Cov(\bar\Func_{r}, \bar\Func_{r'}) \Bigg\}A^{-\top}.
            \end{split}
        \end{equation}
    \end{enumerate}
    The conclusions do not require $\Func_r$ to equal the corresponding conditional expectation. If a tuning estimator computed on an independent fold satisfies $\hat\alpha\pto\alpha^\star$, then the estimator on the evaluation fold has the same first-order expansion as $\hat\theta(\alpha^\star)$.
\end{theorem}

The first term of~\eqref{eqn: avar} is the asymptotic variance of the complete-case estimator. The remaining terms describe how the imputation functions change variance across patterns. Let $\ell$ be a nonnegative, transpose-invariant linear scalarization of a covariance matrix, such as $\ell(\Sigma)=\operatorname{tr}(\Sigma)$ or $\ell(\Sigma)=c^\top\Sigma c$. Then $\ell(\Sigma_\alpha)$ is quadratic in $\alpha$.

\begin{corollary}
    \label{coro: min_alpha}
    Suppose
    \[
    G = \big(\nu_{r,r'}\ell\{A^{-1}\Cov(\bar\Func_{r}, \bar\Func_{r'})A^{-\top}\}\big)_{r,r'\neq [p]}
    \]
    is positive definite, and define
    $L = (\gamma_r\ell\{A^{-1}\Cov(\psi^F, \bar\Func_{r})A^{-\top}\})_{r\neq [p]}$. Then the unique minimizer is $\alpha^\star = -G^{-1}L$, and
    \[
    \ell(\Sigma_{\alpha^\star}) = \frac{\ell\{A^{-1}\Var(\psi^F)A^{-\top}\}}{\pi_{[p]}} - L^\top G^{-1}L.
    \]
    Thus $L^\top G^{-1}L\ge0$ is the improvement over complete-case analysis under the selected scalar criterion.
\end{corollary}

We estimate $\alpha^\star$ and $\theta^\star$ on separate folds as in Algorithm~\ref{alg: mmpp}. The imputation functions remain fixed; cross-fitting is used only for the finite-dimensional tuning vector. Because $\alpha=0$ is feasible, population-optimal tuning cannot increase the selected scalar variance criterion relative to complete-case analysis.

\begin{algorithm}[H]
\caption{Cross-fitted estimation of the RAY and adaptive RAY estimators}
\label{alg: mmpp}
\begin{algorithmic}
\State \textbf{Step 1.} Randomly split $\Data$ into two approximately equal folds $\Data^{(1)}$ and $\Data^{(2)}$.
\State \textbf{Step 2.} On $\Data^{(k)}$, obtain a consistent preliminary estimate of $\theta^\star$, evaluate and center the fixed imputation functions, estimate $A$, $G$, and $L$, and compute $\hat{\alpha}^{(k)} = -(\hat G^{(k)})^{-1}\hat L^{(k)}$, for $k=1,2$.
\State \textbf{Step 3.} On the opposite fold $\Data^{(k')}$, form fold-specific pattern proportions and RAY weights, solve~\eqref{eqn: empirical_ray} with $\hat\alpha^{(k)}$ plugged in, and obtain $\hat{\theta}^{(k')}$ and a per-observation sandwich covariance estimate $\hat{\Sigma}^{(k')}$, for $k\neq k'$.
\State \textbf{Step 4.} If the fold sizes are $N_1$ and $N_2$, report $\hat\theta_{\mathrm{RAY}} = (\hat{\theta}^{(1)} + \hat{\theta}^{(2)})/2$ with estimated variance $\{\hat{\Sigma}^{(1)}/N_1 + \hat{\Sigma}^{(2)}/N_2\}/4$.
\end{algorithmic}
\end{algorithm}

If $N_1/N\to1/2$ and both fold-specific tuning estimates converge to $\alpha^\star$, Theorem~\ref{thm: asymptotic_behavior} gives $\sqrt N(\hat\theta_{\mathrm{RAY}}-\theta^\star)\dto\Normal(\mathbf 0_d,\Sigma_{\alpha^\star})$, and the variance formula in Step 4 is consistent for $\Sigma_{\alpha^\star}/N$.

Theorem~\ref{thm: asymptotic_behavior} establishes imputation-model-robust validity under MCAR. The result treats the imputation functions as fixed and given, but does not require them to equal the corresponding conditional expectations. Validity follows because the RAY weights are mean-zero and independent of functions of $Z$ under MCAR; sample calibration preserves this property at the estimator level. The quality of the imputation functions affects $\Sigma_\alpha$ and therefore the attainable efficiency gain, not the target of the estimating equation.

\begin{remark}[Connection to prediction-powered inference]
\label{remark: ppi}
In the special case of outcome-mean estimation with a single predictor and the two-pattern semi-supervised design, \eqref{eqn: mmpp_aef} reduces to the \texttt{PPI++} estimator of \citet{angelopoulos2023ppi++}, with $\alpha_r$ playing the role of its power-tuning parameter. The RAY estimator therefore extends the prediction-powered construction to nonmonotone missing data with several modalities and observed patterns while retaining imputation-model-robust validity under MCAR.
\end{remark}

\subsection{The adaptive RAY estimator}
\label{subsec: adaptive-ray-estimator}

Theoretically, neither RAY nor IPI dominates the other, as analyzed in Appendix~\ref{app_sec: compare_two_ibm}. We therefore embed both in a larger class, in which a given imputation function can be evaluated on every pattern that observes the input modalities. Consider
\begin{equation}
    \label{eqn: adaptive_ef}
    \psi_{\alpha}(R,Z_R,\theta) = \frac{\Ind(R=[p])}{\pi_{[p]}}\psi^F(Z, \theta) + \sum_{r\in \Qcal,\, r\neq [p]}\Big[\sum_{s\in\Qcal:\,s\supseteq r}\frac{\Ind(R=s)}{\pi_s}\alpha_{r, s}\Big]\Func_{r}(Z_r;\theta),
\end{equation}
with tuning parameters $\alpha = \{\alpha_{r,s} : r,s \in \Qcal, r \neq [p], s \supseteq r\}$ subject to
\begin{equation}
    \label{eqn: alpha_constraint}
    \sum_{s\in\Qcal:\,s\supseteq r}\alpha_{r, s} = 0 \quad \text{for all } r\in \Qcal,\, r\neq [p].
\end{equation}
Each $\Func_r$ is now evaluated on $\Data_s$ for every $s\in\Qcal$ with $s\supseteq r$. Under~\eqref{eqn: alpha_constraint}, the augmentation has population mean zero under MCAR, and both IPI and RAY are special cases (Appendix~\ref{app_sec: valid_ee} and Appendix~\ref{app_sec: two_estimator_char}). The sample estimator replaces $\pi_s$ by the corresponding sample proportions, as in~\eqref{eqn: empirical_ray}.

\begin{theorem}
    \label{thm: adaptive_asymptotic_behavior}
    Let $A = \Exp[\partial \psi^F/\partial \theta^\top]$, let $\bar\Func_r$ be as in Theorem~\ref{thm: asymptotic_behavior}, and define $H_t=\sum_{r\in\Qcal:\,r\subseteq t,\,r\neq[p]}\alpha_{r,t}\bar\Func_r$. Under Assumption~\ref{assumption: missing_positivity} and the regularity conditions of Theorem~\ref{thm: asymptotic_behavior}, for any fixed $\alpha$ satisfying~\eqref{eqn: alpha_constraint}, the sample-calibrated estimator is consistent and $\sqrt{N}(\hat\theta_{\mathrm{aRAY}} -\theta^\star) \dto \Normal(\mathbf{0}_d, \Sigma_{\alpha})$, with
    \begin{equation*}
        \Sigma_{\alpha} = A^{-1}\left\{
        \frac{\Var(\psi^F)}{\pi_{[p]}}
        +\frac{\Cov(\psi^F,H_{[p]})+\Cov(H_{[p]},\psi^F)}{\pi_{[p]}}
        +\sum_{t\in\Qcal}\frac{\Var(H_t)}{\pi_t}
        \right\}A^{-\top}.
    \end{equation*}
\end{theorem}

For the scalarization $\ell$ used in Corollary~\ref{coro: min_alpha}, minimizing $\ell(\Sigma_\alpha)$ subject to~\eqref{eqn: alpha_constraint} is a constrained quadratic program. Its population minimizer is optimal within the class~\eqref{eqn: adaptive_ef} and cannot have a larger value of $\ell(\Sigma)$ than either RAY or IPI, both of which are feasible points. We estimate the tuning coefficients with Algorithm~\ref{alg: mmpp}, replacing the unconstrained calculation in Step 2 by the constrained program. If the fold-specific minimizers are consistent, estimating these finite-dimensional coefficients does not alter the first-order law in Theorem~\ref{thm: adaptive_asymptotic_behavior}.

The class~\eqref{eqn: adaptive_ef} also clarifies the difference between RAY and IPI. The IPI correction~\eqref{eqn: warmup_ipi} pairs each $\Data_r$ with the complete cases, whereas RAY and aRAY evaluate $\Func_r$ on every subsample $\Data_s$ with $s\in\Qcal$ and $s\supseteq r$. This broader use of the observed patterns also resonates with the multiple block imputation method in \citep{xue2021integrating}.

\section{Efficiency gap of the RAY estimator}
\label{sec: exactness}

The RAY estimating function~\eqref{eqn: aef} discards the remainders of Section~\ref{subsec: almost-eigen-decomposition} and the cross-pattern compositions of~\eqref{eqn: L_surrogate_split}, so it need not attain the fixed-$\psi^F$ efficiency lower bound of Theorem~\ref{thm: oef}. To isolate the loss caused by these two approximations, this section evaluates~\eqref{eqn: aef} with the true conditional expectations $\mathcal{A}_r[\psi^F]$ and the RAY coefficients in~\eqref{eqn: aef}, before safeguard tuning. Write $\psi_{\mathrm{RAY}}(R,Z_R,\theta^\star)$ for this estimating function, and use $\psi_{\mathrm{RAY}}$ as shorthand below. Both $\psi_{\mathrm{opt}}$ in Theorem \ref{thm: oef} and $\psi_{\mathrm{RAY}}$ have Jacobian $A = \Exp[\partial \psi^F/\partial\theta^\top]$ (Appendix~\ref{app_sec: gap_proofs}), so the \emph{efficiency gap}
\begin{equation}\label{eqn: Eloss}
\mathcal E := A^{-1}\big[\Var(\psi_{\mathrm{RAY}}) - \Var(\psi_{\mathrm{opt}})\big]A^{-\top} \succeq 0
\end{equation}
is the excess asymptotic variance of $\hat\theta_{\mathrm{RAY}}$ over the efficiency lower bound.

Both estimating functions act pattern by pattern. In particular, $\psi_{\mathrm{opt}} = \sum_{r\in \Qcal}\Ind(R=r)q_r$ with $q_r := \mathcal{A}_r[\Mop^{-1}\psi^F]$, a well-defined population function of $Z_r$, whereas $\psi_{\mathrm{RAY}} = \sum_{r\in \Qcal}\Ind(R=r) \rho_r$ with $\rho_r := \sum_{s\subseteq r}\lambda_s^{-1}\mathcal{P}_s\psi^F$. Their difference is supported pattern by pattern, and the loss is determined by the \emph{per-pattern residual}
\begin{equation}\label{eqn: er}
e_r := q_r - \rho_r = \mathcal{A}_r[\Mop^{-1}\psi^F] - \sum_{s\subseteq r}\lambda_s^{-1}\mathcal{P}_s\psi^F, \qquad r\in\Qcal,
\end{equation}
which is the gap on the pattern $R = r$ between the optimal and RAY estimating functions. Since the patterns are disjoint, these residuals do not interact, and the loss is their $\boldsymbol\pi$-weighted average.

\begin{proposition}[Efficiency gap]\label{prop: lossid}
Under Assumption~\ref{assumption: missing_positivity}, $\;A\,\mathcal E\,A^\top = \sum_{r\in\Qcal}\pi_r\,\Exp[e_re_r^\top].$
\end{proposition}

Questions about exactness and the magnitude of the efficiency gap therefore reduce to the residuals $e_r$. Section~\ref{subsec: exact} gives conditions under which all residuals vanish, and Section~\ref{subsec: bound} bounds the loss otherwise.

\subsection{When the RAY estimator is exact}
\label{subsec: exact}

By Proposition~\ref{prop: lossid}, the loss vanishes exactly when every residual does, yielding a necessary and sufficient condition for the RAY estimator to attain the efficiency lower bound.

\begin{theorem}[Necessary and sufficient exactness]\label{thm: ns}
The RAY estimator considered in this section attains the fixed-$\psi^F$ efficiency lower bound, that is $\mathcal E = 0$, if and only if $e_r = 0$ almost surely for every $r \in \Qcal$; equivalently,
\begin{equation}\label{eqn: nscond}
\sum_{s\subseteq r}\lambda_s^{-1}\mathcal{P}_s\psi^F = \mathcal{A}_r[\Mop^{-1}\psi^F]\quad\text{a.s., for all } r\in\Qcal.
\end{equation}
\end{theorem}

The characterization is exact, but it depends on the law and the target through $\Mop^{-1}\psi^F$. It also separates, at the level of the residual, the two approximations behind RAY. Write
\begin{equation}\label{eqn: gb}
e_r = \mathcal{A}_r(g) + b_r, \qquad g := (\Mop^{-1}-\Mray)\psi^F, \qquad b_r := \sum_{s\subseteq[p]:\, s\not\subseteq r}\lambda_s^{-1}\mathcal{A}_r[\mathcal{P}_s\psi^F],
\end{equation}
where the sum defining $b_r$ runs over all subsets $s\subseteq[p]$ that are \emph{not} contained in $r$. The first term $\mathcal{A}_r(g)$ measures the surrogate approximation $(\Mop^{-1}-\Mray)$; the second term $b_r$ collects the cross-pattern compositions $\mathcal{P}_s\psi^F$ with $s\not\subseteq r$ that pattern $r$ cannot resolve. The complete pattern carries no such compositions, so $b_{[p]} = 0$ and $e_{[p]} = g$. Requiring only $e_{[p]} = 0$, that is $\Mray\psi^F = \Mop^{-1}\psi^F$, controls the complete-case residual alone; it does not on its own imply estimator exactness, which additionally requires the incomplete-pattern terms $b_r$ to vanish.

We next give three sufficient conditions under which every residual vanishes and RAY attains the fixed-$\psi^F$ efficiency lower bound. These conditions are complementary rather than nested in strength.

First, we define the intersection-closed observed pattern set that is essential for those conditions:
\begin{definition}
$\Qcal$ is \emph{intersection-closed} if $r \cap r' \in \Qcal$ for all $r, r' \in \Qcal$ with $r \cap r' \neq \emptyset$; that is, every nonempty overlap of two observed patterns is itself observed.
\end{definition}

\begin{theorem}[Sufficient conditions for exactness]\label{thm: sufficient} The RAY estimator attains the efficiency lower bound, $\mathcal E = 0$, under any one of the following conditions on the observed-pattern set and the law:
\begin{enumerate}[label=\textup{(\roman*)}, itemsep=3pt, topsep=3pt]
\item \emph{(Intersection-closed, independent modalities)} $\Qcal$ is intersection-closed and the modalities $Z_1,\dots,Z_p$ are independent.
\item \emph{(Monotone)} $\Qcal$ is a chain $r_1 \subsetneq \cdots \subsetneq r_K = [p]$.
\item \emph{(Anchored)} $\Qcal = \{r_0 \cup v : v \subseteq [p]\setminus r_0\}$ for a core $r_0$ observed in every pattern, and the non-core modalities $\{Z_j : j \notin r_0\}$ are mutually conditionally independent given $Z_{r_0}$.
\end{enumerate}
\end{theorem}

The three conditions relax one another along different axes rather than forming a hierarchy. A monotone chain is intersection-closed, so (ii) strengthens (i) in that case by dropping the independence requirement: for a monotone observed-pattern set, the RAY weight is equivalent to the fixed-function oracle weight of \citet{tsiatis2006semiparametric} for any law (Appendix~\ref{app_sec: gap_proofs}). An anchored family is also intersection-closed, and (iii) likewise relaxes full independence to conditional independence given the core $Z_{r_0}$. When these sufficient conditions do not hold, the residual characterization in Theorem~\ref{thm: ns} still applies, and the following subsection bounds any resulting loss.

\subsection{Bounding the efficiency gap}
\label{subsec: bound}

Outside the sufficient cases of Section~\ref{subsec: exact}, RAY may incur a positive efficiency gap. The bound below separates a \emph{correlation} component, driven by dependence between the private modalities of overlapping patterns, from a \emph{structural} component, whose coefficients record whether required pattern intersections are absent. The magnitude of the structural component also depends on the full-data law and $\psi^F$.

The efficiency gap follows from the residual split~\eqref{eqn: gb}, $e_r = \mathcal{A}_r(g) + b_r$. Its two terms correspond to the two approximations of Section~\ref{sec: estimation-mcar} for pattern $r$: the first, with $g = (\Mop^{-1}-\Mray)\psi^F$, is the error from replacing the oracle inverse by the RAY surrogate; the second, $b_r$, consists of the cross-pattern compositions that pattern $r$ cannot resolve. Summing across patterns gives $\psi_{\mathrm{opt}} - \psi_{\mathrm{RAY}} = \Lop(g) + \sum_{r\in\Qcal}\Ind(R=r)\,b_r$, so Proposition~\ref{prop: lossid} and the triangle inequality yield
\begin{equation}\label{eqn: loss_split}
\sqrt{\operatorname{tr}(A\,\mathcal E\,A^\top)}\ \le\ \|\Lop(g)\|_{L^2}\ +\ \Big(\textstyle\sum_{r\in\Qcal}\pi_r\,\Exp\|b_r\|^2\Big)^{1/2}.
\end{equation}
Here $\|X\|_{L^2}^2 := \Exp\|X\|^2$.

The first term is the square root of the \emph{operator discrepancy}:
\begin{equation}\label{eqn: gapdef}
\Delta(\psi^F) := \big\|\Mop^{1/2}g\big\|_d^2 = \sum_{j=1}^d\langle g_j,\Mop g_j\rangle = \operatorname{tr}\Var(\Lop(g)) \;\le\; \pi_{[p]}^{-1}\big\|\Mop g\big\|_d^2,
\end{equation}
which measures the surrogate error in the geometry induced by $\Mop$. The spectral bound $\Mop \succeq \pi_{[p]}\mathcal I$ and the identity $\Mop(\Mop^{-1}-\Mray)\psi^F = -\sum_s\lambda_s^{-1}\Rem_s\psi^F$ show that $\Delta$ is controlled by the discarded remainders (Lemma~\ref{lem:rem}).

Substituting the RAY components~\eqref{eqn: ray_decomp} into the remainder~\eqref{eqn: Mop_eigen} shows that each $\Rem_s$ is a signed sum of \emph{pairwise compositions} of conditioning operators,
\begin{equation}\label{eqn: rem_expand}
\Rem_s = \sum_{r\in\Qcal:\, s\not\subseteq r}\pi_r\sum_{t\in\Qcal:\, t\subseteq s}(-1)^{|s|-|t|}\,\mathcal{A}_r\mathcal{A}_t,
\end{equation}
where $\mathcal{A}_r\mathcal{A}_t$ first conditions on $Z_t$ and then on $Z_r$. We split each composition into $\mathcal{A}_r\mathcal{A}_t-\mathcal{A}_{r\cap t}$, which measures dependence between the patterns' private parts, and the reference projection $\mathcal{A}_{r\cap t}$. This gives the following operators; the term-by-term derivation is in Appendix~\ref{app_sec: gap_proofs}.

\begin{definition}[Correlation and structural operators]\label{def:UV}
We define the \emph{correlation operator} $\Vcorr := \sum_s\lambda_s^{-1}\Vcorr_s$ and the \emph{structural operator} $\Ustr := \sum_s\lambda_s^{-1}\Ustr_s$, where
\[
\Vcorr_s := \sum_{r\in\Qcal:\, s\not\subseteq r}\pi_r\sum_{t\in\Qcal:\, t\subseteq s}(-1)^{|s|-|t|}\big(\mathcal{A}_r\mathcal{A}_t - \mathcal{A}_{r\cap t}\big), \qquad \Ustr_s := \sum_{r\in\Qcal:\, s\not\subseteq r}\pi_r\sum_{t\in\Qcal:\, t\subseteq s}(-1)^{|s|-|t|}\mathcal{A}_{r\cap t},
\]
so that $\Rem_s = \Vcorr_s + \Ustr_s$ and $\sum_s\lambda_s^{-1}\Rem_s = \Vcorr + \Ustr$.
\end{definition}

For $r,t\in\Qcal$, let $\rstar(Z_{r\setminus t};Z_{t\setminus r}\mid Z_{r\cap t})$ denote the largest value of $|\Corr(f(Z_r),g(Z_t))|$ over square-integrable $f$ and $g$ that are mean-zero given $Z_{r\cap t}$; set it to zero for nested patterns, and let $\rstar_{\max}$ be its maximum over $r,t\in\Qcal$. Lemma~\ref{lem:maxcorr} shows that this global conditional maximal correlation \citep{hirschfeld1935connection, gebelein1941statistische, renyi1959measures} equals $\|\mathcal{A}_r\mathcal{A}_t-\mathcal{A}_{r\cap t}\|$. Thus $\Vcorr$ is controlled by $\rstar_{\max}$ and vanishes when every pair of private blocks is conditionally independent given its overlap.

The structural operator $\Ustr$ is a combination of single conditioning maps with coefficients determined by $(\boldsymbol\pi,\Qcal)$. It vanishes whenever $\Qcal$ is intersection-closed (Lemma~\ref{lem: Uvanish}). Thus, when the correlation component is zero, intersection-closedness removes the remaining discrepancy.

The incomplete-pattern term $b_r$ admits the same decomposition, $b_r=b_r^{\mathrm{cor}}+b_r^{\mathrm{str}}$. Its correlation part is bounded by $C_r\rstar_{\max}\|\psi^F\|$, and its structural part vanishes for intersection-closed $\Qcal$. Combining these bounds with~\eqref{eqn: loss_split} gives the following result.

\begin{theorem}[Two-factor bound on the loss]\label{thm: Ebound}
Under Assumption~\ref{assumption: missing_positivity}, for any $\Qcal$ and any $\psi^F$,
\begin{equation}\label{eqn: Etwofactor}
\sqrt{\operatorname{tr}\big(A\,\mathcal E\,A^\top\big)}\ \le\ \mathsf C(\boldsymbol\pi,\Qcal)\,\rstar_{\max}\,\|\psi^F\| + \mathsf S(\psi^F),
\end{equation}
where the \emph{correlation} and \emph{structural} factors are
\[
\mathsf C(\boldsymbol\pi,\Qcal) = \pi_{[p]}^{-1/2}C(\boldsymbol\pi,\Qcal) + \Big(\sum_{r\in\Qcal}\pi_r\,C_r^2\Big)^{1/2}, \qquad
\mathsf S(\psi^F) = \pi_{[p]}^{-1/2}\|\Ustr\psi^F\| + \Big(\sum_{r\in\Qcal}\pi_r\,\|b_r^{\mathrm{str}}\|_{L^2}^2\Big)^{1/2},
\]
with $C(\boldsymbol\pi,\Qcal) = \sum_s\lambda_s^{-1}\big(\sum_{r:\,s\not\subseteq r}\pi_r\big)\,|\{t\in\Qcal:t\subseteq s\}|$ and $C_r = \sum_{s\not\subseteq r}\lambda_s^{-1}|\{t\in\Qcal:t\subseteq s\}|$. The correlation factor $\mathsf C(\boldsymbol\pi, \Qcal) < \infty$ depends only on $(\boldsymbol\pi, \Qcal)$; the structural factor $\mathsf S(\psi^F)$ depends only on $(\boldsymbol\pi, \Qcal)$, $\psi^F$, and the noncomposed conditional-expectation operators, and vanishes when $\Qcal$ is intersection-closed. In particular the right-hand side is zero when $\rstar_{\max} = 0$ and $\Qcal$ is intersection-closed.
\end{theorem}

As a triangle-inequality bound, ~\eqref{eqn: Etwofactor} can be conservative because it ignores cancellations among signed operator terms. Estimating either factor in a high-dimensional application would require additional modeling of conditional expectations or maximal correlation. Therefore, the bound \eqref{eqn: Etwofactor} is to identify the two population mechanisms behind the loss, rather than acting as a directly computable finite-sample diagnostic.

We close the discussion by making the two factors concrete on the two-modality design, where the bound takes an explicit form in the modality correlation.

\begin{example}[Two-modality Gaussian design]
\label{ex: two-modality-gaussian}
Let $\Qcal = \{\{1\}, \{2\}, \{1,2\}\}$ with $\pi_1, \pi_2, \pi_{12} > 0$ summing to one, let $(Z_1, Z_2)$ be standard bivariate normal with $\Corr(Z_1, Z_2) = \rho$, and take $\psi^F = Z_2 - \theta$. We evaluate the two factors of Theorem~\ref{thm: Ebound} in closed form.

\emph{Structural factor.} The design is intersection-closed, since the nonempty overlaps $\{1\}\cap\{1,2\} = \{1\}$ and $\{2\}\cap\{1,2\} = \{2\}$ are observed (and $\{1\}\cap\{2\} = \emptyset$). Hence $\Ustr = 0$ (Lemma~\ref{lem: Uvanish}) and each $b_r^{\mathrm{str}} = 0$, so $\mathsf S(\psi^F) = 0$ for every $\rho$.

\emph{Correlation scalar.} The only non-nested pair is $(\{1\}, \{2\})$, with empty overlap and private blocks $Z_1, Z_2$; their Gaussian maximal correlation is $\rstar(Z_1; Z_2) = |\rho|$. The nested pairs $(\{1\},\{1,2\})$ and $(\{2\},\{1,2\})$ give $\rstar = 0$ by convention, so $\rstar_{\max} = |\rho|$.

\emph{Correlation factor.} With $\lambda_1 = \pi_1 + \pi_{12}$, $\lambda_2 = \pi_2 + \pi_{12}$ and $\pi_{[p]} = \pi_{12}$, the constants of Theorem~\ref{thm: Ebound} evaluate to
\[
C(\boldsymbol\pi,\Qcal) = \frac{\pi_2}{\lambda_1} + \frac{\pi_1}{\lambda_2} + \frac{3(\pi_1+\pi_2)}{\pi_{12}}, \qquad C_{\{1\}} = \frac{1}{\lambda_2} + \frac{3}{\pi_{12}}, \qquad C_{\{2\}} = \frac{1}{\lambda_1} + \frac{3}{\pi_{12}}, \qquad C_{\{1,2\}} = 0,
\]
so $\mathsf C(\boldsymbol\pi,\Qcal) = \pi_{12}^{-1/2}\,C(\boldsymbol\pi,\Qcal) + \big(\pi_1 C_{\{1\}}^2 + \pi_2 C_{\{2\}}^2\big)^{1/2}$, a finite constant depending only on $\boldsymbol\pi$.

\emph{The bound.} Since $\mathsf S(\psi^F) = 0$, $\rstar_{\max} = |\rho|$, and $\|\psi^F\|^2 = \Var(Z_2) = 1$, Theorem~\ref{thm: Ebound} reduces to
\[
\operatorname{tr}\big(A\,\mathcal E\,A^\top\big) \;\le\; \mathsf C(\boldsymbol\pi,\Qcal)^2\,\rho^2 .
\]
As $\rho\to0$, the upper bound is $O(\rho^2)$, and at $\rho=0$ the bound is zero, recovering Theorem~\ref{thm: sufficient}. A direct computation (Appendix~\ref{app_sec: gaussian_loss}) gives $\operatorname{tr}(A\mathcal E A^\top) = \pi_1\pi_2^2\rho^2/(\lambda_1\lambda_2^2\pi_{12}) + O(\rho^4)$, so the two-factor bound captures the local quadratic scaling, with a larger constant because it uses triangle inequalities.
\end{example}

\section{Toward an extension under MAR}
\label{sec: estimation-mar}

The estimator developed in Sections~\ref{sec: estimation-mcar}--\ref{sec: exactness} relies on MCAR. It is natural to ask whether the same RAY device extends to missing at random (MAR), where the pattern probabilities depend on the observed values. Our investigation shows that some properties of RAY survive while others do not. We obtain two valid population constructions. The direct construction preserves a strong conditional mean-zero identity and has ordinary oracle root-$N$ behavior, but its weights generally require unobserved modalities and hence do not produce an observed-data estimator. Projecting those weights onto observable modalities produces a second mean-zero identity and an observable oracle equation, but introduces a law-dependent conditional-expectation nuisance. The resulting plug-in map can have a nonzero partial derivative with respect to this nuisance when the observed-data law and complete-case propensity are held fixed. When the nuisance functions admit root-$N$ consistent parametric estimators, this contribution can be incorporated through stacked inference. We therefore present the MAR development as an exploratory population extension that identifies a sufficient parametric route and the questions that remain for more general nuisance estimators.

\subsection{The almost-eigen identity under MAR}
\label{subsec: mar-almost-eigen}

Under Assumption~\ref{assumption: mar}, the observed-data operator is
$\Mop=\sum_{r\in\Qcal}\pi_r(Z_r)\mathcal A_r$. For a RAY component
$\mathcal P_s(f)$, every pattern $r\supseteq s$ leaves the component unchanged,
whereas the other patterns contribute a remainder. Thus
\begin{equation}\label{eqn: mar_eigen_raw}
\Mop[\mathcal P_s(f)]
=
\Lambda_s(Z)\mathcal P_s(f)
+
\sum_{r\in\Qcal:\,r\not\supseteq s}
\pi_r(Z_r)\mathcal A_r[\mathcal P_s(f)],
\qquad
\Lambda_s(Z):=
\sum_{t\in\Qcal:\,t\supseteq s}\pi_t(Z_t)
=
\mathbb P(R\supseteq s\mid Z).
\end{equation}
We denote the second term on the right-hand side of~\eqref{eqn: mar_eigen_raw} by $\Rem_s(f)$.
Equation~\eqref{eqn: mar_eigen_raw} is the exact MAR counterpart of
\eqref{eqn: Mop_eigen}. Its interpretation is nevertheless different. Under
MCAR, $\lambda_s=\mathbb P(R\supseteq s)$ is a scalar and commutes with every
conditional-expectation operator. Under MAR, $\Lambda_s(Z)$ is a nondegenerate random variable. In general,
\[
\Mop\{\Lambda_s(Z)^{-1}\mathcal P_s(f)\}
\neq
\Lambda_s(Z)^{-1}\Mop\{\mathcal P_s(f)\}.
\]
Consequently, the exact identity alone does not justify the formal substitution
$\lambda_s^{-1}\mapsto\Lambda_s(Z)^{-1}$ as an approximation to
$\Mop^{-1}$. A separate argument is needed even for population validity.

\subsection{A direct population extension}
\label{subsec: mar-direct}

The direct analogue of the MCAR weight is, for $r\in\Qcal$ with $r\neq[p]$,
\begin{equation}\label{eqn: omega_mar_direct}
\omega_r^{\mathrm{dir}}
:=
\sum_{s:\,r\subseteq s\subseteq[p]}
(-1)^{|s|-|r|}
\frac{\Ind(R\supseteq s)}{\Lambda_s(Z)},
\end{equation}
Strict positivity in Assumption \ref{assumption: mar} implies $\Lambda_s(Z)\geq c$ almost surely.

\begin{proposition}[Population validity of direct MAR weights]
\label{prop:mar-direct-mean-zero}
Under Assumption~\ref{assumption: mar},
\[
\Exp[\omega_r^{\mathrm{dir}}\mid Z]=0
\]
for every $r\neq[p]$. Consequently,
$\Exp[\omega_r^{\mathrm{dir}}F_r(Z_r;\theta)]=0$ for any square-integrable
$\sigma(Z_r)$-measurable function $F_r$.
\end{proposition}

The proposition follows from
$\Exp\{\Ind(R\supseteq s)/\Lambda_s(Z)\mid Z\}=1$ and the alternating-sum
identity. It gives the direct population estimating function
\begin{equation}\label{eqn: mar_direct_ee}
\psi_\alpha^{\mathrm{dir}}(Z,R;\theta)
=
\frac{\Ind(R=[p])}{\pi_{[p]}(Z)}\psi^F(Z,\theta)
+
\sum_{r\in\Qcal,\,r\neq[p]}
\alpha_r\omega_r^{\mathrm{dir}}F_r(Z_r;\theta).
\end{equation}
For every $\theta$,
\begin{equation}\label{eqn: mar_direct_population}
\Exp[\psi_\alpha^{\mathrm{dir}}(Z,R;\theta)]
=
\Exp[\psi^F(Z,\theta)].
\end{equation}
Thus the direct population equation has the correct target and the full-data
Jacobian $A$, regardless of the fixed tuning coefficients or the quality of the
functions $F_r$. If all terms in~\eqref{eqn: mar_direct_ee} were available for
every unit, standard Z-estimation gives
\[
\sqrt N(\hat\theta_{\mathrm{dir}}^{\mathrm{or}}-\theta^\star)
\dto
\Normal\!\left(
0,\,
A^{-1}\Var\{\psi_\alpha^{\mathrm{dir}}(Z,R;\theta^\star)\}A^{-\top}
\right).
\]
However, this is a pointwise oracle statement, not a theorem for a feasible observed-data estimator.

Indeed, knowing the propensity scores $\pi_t(z_t)$'s does not make
\eqref{eqn: omega_mar_direct} computable. On a unit with $R=r'\supseteq s$,
the denominator $\Lambda_s(Z)$ contains terms $\pi_t(Z_t)$ for
$t\supseteq s$ that need not satisfy $t\subseteq r'$. Their arguments are then
unobserved. Hence $\omega_r^{\mathrm{dir}}$ is generally not measurable with
respect to the observed data $(R,Z_R)$, so direct RAY is not an observed-data estimator. It becomes feasible only under additional
observability restrictions, for example when every required $\Lambda_s(Z)$ is
$\sigma(Z_s)$-measurable. Under that restriction the direct construction
coincides with the projected construction below.

\subsection{A projection extension}
\label{subsec: mar-projected}

The natural observable projection of the full-data tail propensity is
\begin{equation}\label{eqn: lambda_proj}
\lambda_s^{\mathrm{proj}}(Z_s)
:=
\Exp[\Lambda_s(Z)\mid Z_s]
=
\mathbb P(R\supseteq s\mid Z_s).
\end{equation}
Writing
$\Lambda_s(Z)=\lambda_s^{\mathrm{proj}}(Z_s)+\delta_s(Z)$, where
$\Exp[\delta_s(Z)\mid Z_s]=0$, gives
\begin{equation}\label{eqn: mar_eigen_proj}
\Mop[\mathcal P_s(f)]
=
\lambda_s^{\mathrm{proj}}(Z_s)\mathcal P_s(f)
+
\delta_s(Z)\mathcal P_s(f)
+
\Rem_s(f).
\end{equation}
The additional term $\delta_s\mathcal P_s(f)$ measures the price of
replacing the full-data tail propensity by an observable function. It vanishes
when $\Lambda_s(Z)$ is already $\sigma(Z_s)$-measurable, including under MCAR.

Define the projected weight
\begin{equation}\label{eqn: omega_proj}
\omega_r^{\mathrm{proj}}
:=
\sum_{s:\,r\subseteq s\subseteq[p]}
(-1)^{|s|-|r|}
\frac{\Ind(R\supseteq s)}
{\lambda_s^{\mathrm{proj}}(Z_s)}.
\end{equation}

\begin{proposition}[Population validity of projected MAR weights]
\label{prop:mean-zero}
Under Assumption~\ref{assumption: mar},
\[
\Exp[\omega_r^{\mathrm{proj}}\mid Z_r]=0
\]
for every $r\neq[p]$. Consequently,
$\Exp[\omega_r^{\mathrm{proj}}F_r(Z_r;\theta)]=0$ for every square-integrable
$\sigma(Z_r)$-measurable function $F_r$.
\end{proposition}

If the true $\pi_{[p]}$ and $\lambda_s^{\mathrm{proj}}$ were supplied, the
corresponding oracle equation would be
\begin{equation}\label{eqn: proj_oracle_ee}
\psi_\alpha^{\mathrm{proj}}(R,Z_R,\theta)
=
\frac{\Ind(R=[p])}{\pi_{[p]}(Z)}\psi^F(Z,\theta)
+
\sum_{r\in\Qcal,\,r\neq[p]}
\alpha_r\omega_r^{\mathrm{proj}}F_r(Z_r;\theta).
\end{equation}
Unlike the direct equation, every evaluation of~\eqref{eqn: proj_oracle_ee} is
observable: a term involving $Z_s$ is active only when $R\supseteq s$.
Proposition~\ref{prop:mean-zero} implies
$\Exp[\psi_\alpha^{\mathrm{proj}}(R,Z_R,\theta)]
=\Exp[\psi^F(Z,\theta)]$. Treating the nuisance functions as fixed, the
empirical root therefore has the same standard pointwise root-$N$ expansion as
the direct oracle equation, with $\psi_\alpha^{\mathrm{proj}}$ replacing
$\psi_\alpha^{\mathrm{dir}}$.

To estimate the projected tail propensity $\mathbb P(R\supseteq s|Z_s)$, note that the indicator
$\Ind(R\supseteq s)$ is observed for every unit, but $Z_s$ is observed only
when $R\supseteq s$, where the indicator is identically one on those units.
Therefore a regression of $\Ind(R\supseteq s)$ on $Z_s$ using units with
$R\supseteq s$ is degenerate and does not estimate
\eqref{eqn: lambda_proj}. Under a correctly specified structural propensity
model $\pi_r(Z_r;\gamma_0)$, the projection can instead be identified from
complete cases. Let
\[
W=\frac{\Ind(R=[p])}{\pi_{[p]}(Z;\gamma_0)}.
\]
Since $\Exp(W\mid Z)=1$,
\begin{equation}\label{eqn: lambda_proj_ratio}
\lambda_s^{\mathrm{proj}}(z_s)
=
\frac{
\Exp[W\Lambda_s(Z;\gamma_0)\mid Z_s=z_s]
}{
\Exp[W\mid Z_s=z_s]
}
=
\frac{
\Exp[\Lambda_s(Z;\gamma_0)/\pi_{[p]}(Z;\gamma_0)
\mid Z_s=z_s,R=[p]]
}{
\Exp[1/\pi_{[p]}(Z;\gamma_0)
\mid Z_s=z_s,R=[p]]
}.
\end{equation}
The second representation involves only complete cases, for which all arguments of
$\Lambda_s$ are observed. Thus, once the propensity mechanism is identified or
correctly modeled, \eqref{eqn: lambda_proj_ratio} provides an observed-data
route to $\lambda_s^{\mathrm{proj}}$. If the two conditional mean functions in
\eqref{eqn: lambda_proj_ratio} admit correctly specified finite-dimensional
parametric models, and their parameters together with $\gamma_0$ have
root-$N$ consistent and asymptotically linear estimators, then the resulting
estimator of $\lambda_s^{\mathrm{proj}}$ is also root-$N$ consistent. Under
standard joint Z-estimation conditions, stacking the corresponding nuisance
estimating equations with~\eqref{eqn: proj_oracle_ee} then yields a root-$N$
consistent and asymptotically normal estimator of $\theta^\star$. Its
asymptotic variance includes the first-order contribution from nuisance
estimation and can be estimated by the corresponding sandwich formula.

The following calculation describes this first-order contribution. Let
$\theta(\lambda)$ be the population root of the projected equation given
the projected tails $\lambda$. Then
$\theta(\lambda_0)=\theta^\star$, and for a nearby working collection $\lambda$,
\[
\theta(\lambda)-\theta^\star
\approx
-A^{-1}
\{\Psi(\theta^\star,\lambda)-\Psi(\theta^\star,\lambda_0)\}.
\]
Appendix \ref{app_sec: first-order} shows that the moment difference on the right is generally
linear in $\lambda-\lambda_0$. Thus nuisance estimation can shift the
population root at first order. The empirical estimator has the
decomposition
\[
\widehat\theta(\hat{\lambda})-\theta^\star
=
\{\widehat\theta(\hat{\lambda})-\theta(\hat\lambda)\}
+
\{\theta(\hat\lambda)-\theta^\star\},
\]
where the first term is ordinary $N^{-1/2}$ sampling error and the second is
the plug-in contribution from $\hat{\lambda} - \lambda_0$. When
$\hat\lambda-\lambda_0=O_p(N^{-1/2})$, both terms enter a joint root-$N$
expansion, as in the parametric route above. For more general nuisance
estimators, cross-fitting controls the empirical-process contribution but does
not remove this first-order term. Establishing suitable rate conditions,
special cancellation, or an observed-data orthogonalization for such
estimators is left for future work.

\section{Simulation study}
\label{sec: sim}

In this section, we present four simulation experiments to illustrate the finite-sample performance of the proposed estimators. The first three experiments evaluate performance under MCAR: outcome-mean estimation with varying sample size, outcome-mean estimation with varying imputation quality, and low-dimensional linear regression with varying sample size. The fourth experiment applies the MCAR estimators under controlled MAR and MNAR departures, thereby assessing the numerical sensitivity of the proposed estimators. The benchmark methods include the complete-case (CC) estimator, IPI \citep{duan2025imputation, zhao2025imputation}, and \texttt{PPI++} \citep{angelopoulos2023ppi++} when applicable. Each configuration uses $B=1000$ Monte Carlo replications and reports root-mean-square error (RMSE) and empirical coverage of nominal $95\%$ Wald confidence intervals. We also report the average interval width for the scalar target and the average trace of the estimated covariance matrix for the regression target.

Throughout, the observed-pattern collection is
$\Qcal=\{\{X_1,X_2,Y\},\{X_1,Y\},\{X_1,X_2\},\{X_1\}\}$, as in Figure~\ref{figure:prototype}. In the first three experiments, we fix the four pattern counts and randomly assign the resulting pattern labels independently of the generated full data. This produces a MCAR design. The outcome-mean CC benchmark averages $Y$ over both patterns that observe it, as in~\eqref{eqn: warmup_cc}; the regression CC estimator uses only the complete pattern. Unless imputation quality is varied deliberately, the imputation functions are trained on an independent, fully observed sample of size $2000$ and then held fixed relative to every estimation sample. The finite-dimensional tuning parameters for \texttt{PPI++}, IPI, RAY, and aRAY are estimated by the two-fold procedure in Algorithm~\ref{alg: mmpp}.

For numerical stability in these finite-sample experiments, coefficient tuning is regularized on the training fold only. If that fold has size $m$ and the empirical quadratic criterion has matrix $\hat G$ of dimension $q_\alpha$, we replace $\hat G$ by $\hat G+\tau_m I$, where $\tau_m=m^{-1/2}\operatorname{tr}(\hat G)/q_\alpha$, and constrain the Euclidean norm of the tuning vector to be at most $2$. The same regularization is used in the constrained aRAY program. The ridge vanishes with $m$, and the norm constraint is asymptotically inactive whenever the population optimum lies in its interior; neither step uses information from the evaluation fold.

\subsection{Outcome mean under MCAR: varying sample size}
\label{subsec: sim-mcar-mean}

\begin{figure}[htbp]
\centering
\includegraphics[width=0.85\textwidth]{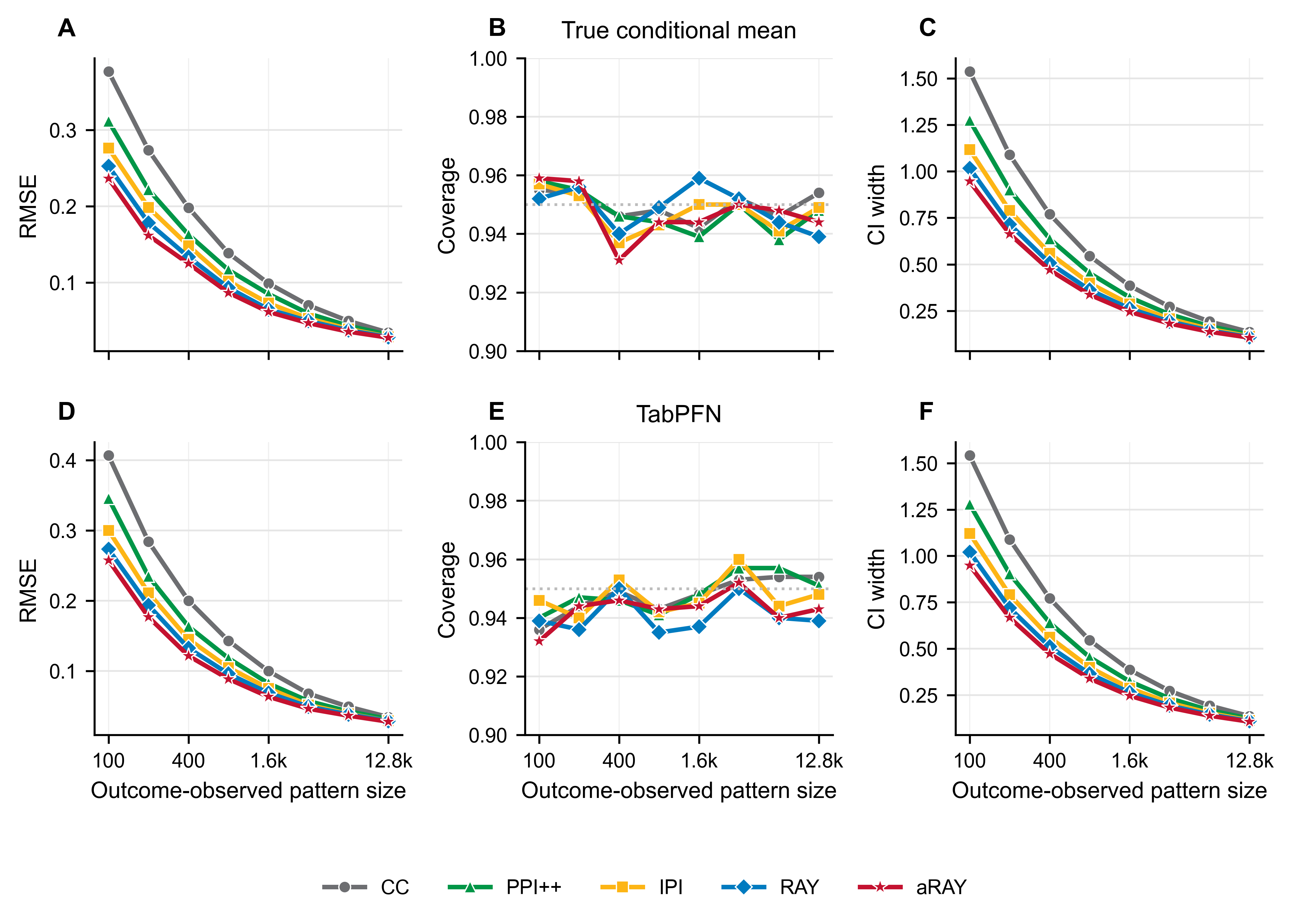}
\caption{Outcome mean under MCAR as the size $n$ of each outcome-observed pattern varies. The two outcome-missing patterns each contain $20000$ observations. The panels report RMSE, empirical coverage, and confidence-interval width for the five estimators using oracle conditional-mean and TabPFN imputations.}
\label{fig: performance_comparison_outcome_mean}
\end{figure}

The full data satisfy $X_1 \sim \Normal(\mathbf{1}_{10}, I_{10})$ and $X_2 \sim \Normal(\mathbf{1}_{20}, I_{20})$ independently, with
\[
    Y=X_1^\top\mathbf{1}_{10}+X_2^\top\mathbf{1}_{20}+\epsilon,
    \qquad \epsilon\sim\Normal(0,1).
\]
The target is $\theta^\star=\Exp[Y]=30$. For each
$n\in\{100,200,400,800,1600,3200,6400,12800\}$, we set
$n_{\{X_1,X_2,Y\}}=n_{\{X_1,Y\}}=n$ and
$n_{\{X_1,X_2\}}=n_{\{X_1\}}=20000$. Thus, $n$ denotes the size of each outcome-observed pattern, while the two outcome-missing pattern sizes remain fixed. We consider analytic true conditional means and expectation-mode predictions from TabPFN \citep{hollmann2022tabpfn}. Both prediction sources give the same qualitative ordering, as shown in Figure~\ref{fig: performance_comparison_outcome_mean}. Across the full range of $n$, IPI, RAY, and aRAY reduce RMSE and interval width relative to \texttt{PPI++} and CC while maintaining coverage near $95\%$. RAY improves on IPI, and aRAY has the smallest RMSE and shortest intervals throughout.

\subsection{Outcome mean under MCAR: varying imputation quality}
\label{subsec: sim-mcar-quality}

\begin{figure}[htbp]
\centering
\includegraphics[width=0.85\textwidth]{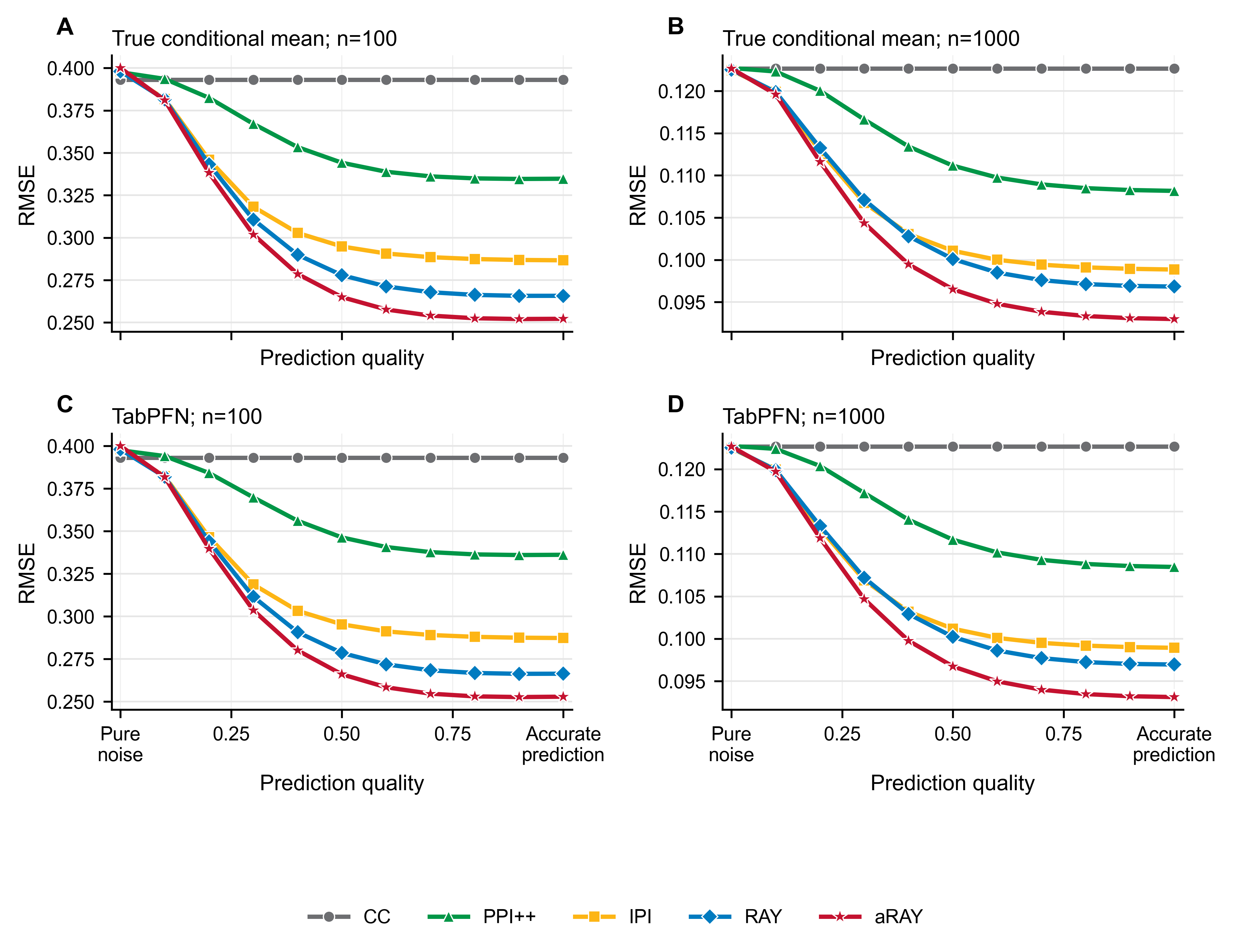}
\caption{Outcome-mean RMSE under MCAR as prediction quality $1-q$ varies, for $n=100$ and $n=1000$ observations in each outcome-observed pattern. Each outcome-missing pattern contains $2000$ observations. The top row uses the analytic conditional mean as the base predictor, and the bottom row uses TabPFN.}
\label{fig: predictor_quality}
\end{figure}

We next hold the pattern sizes fixed and vary the quality of the imputation functions. For each outcome-missing pattern $r$, let $m_r^{\mathrm{base}}$ denote either the true conditional mean or the prediction from a fixed TabPFN model. We define
\[
    m_{r,q}(Z_r)=(1-q)m_r^{\mathrm{base}}(Z_r)+q\varepsilon_r,
    \qquad \varepsilon_r\sim\Normal(0,2),
\]
and set $\Func_{r,q}(Z_r;\theta)=m_{r,q}(Z_r)-\theta$. The noise variables are independent across observations and patterns. We take $q\in\{0,0.1,0.2,\ldots,1\}$, so prediction quality $1-q$ ranges from pure noise to the selected base predictor. We set $n_{\{X_1,X_2,Y\}}=n_{\{X_1,Y\}}=n$ for $n\in\{100,1000\}$ and $n_{\{X_1,X_2\}}=n_{\{X_1\}}=2000$. Figure~\ref{fig: predictor_quality} reports results for both base predictors. As prediction quality improves, all augmented estimators reduce RMSE, with aRAY attaining the smallest RMSE and RAY the next smallest across the nondegenerate settings. Under pure noise, the tuned estimators return approximately to the CC benchmark, illustrating the safeguard supplied by variance-based tuning. The true conditional mean and TabPFN panels are nearly indistinguishable in this linear data-generating process.

\subsection{Linear regression under MCAR: increasing sample size}
\label{subsec: sim-regression}

\begin{figure}[htbp]
    \centering
    \includegraphics[width=1.0\linewidth]{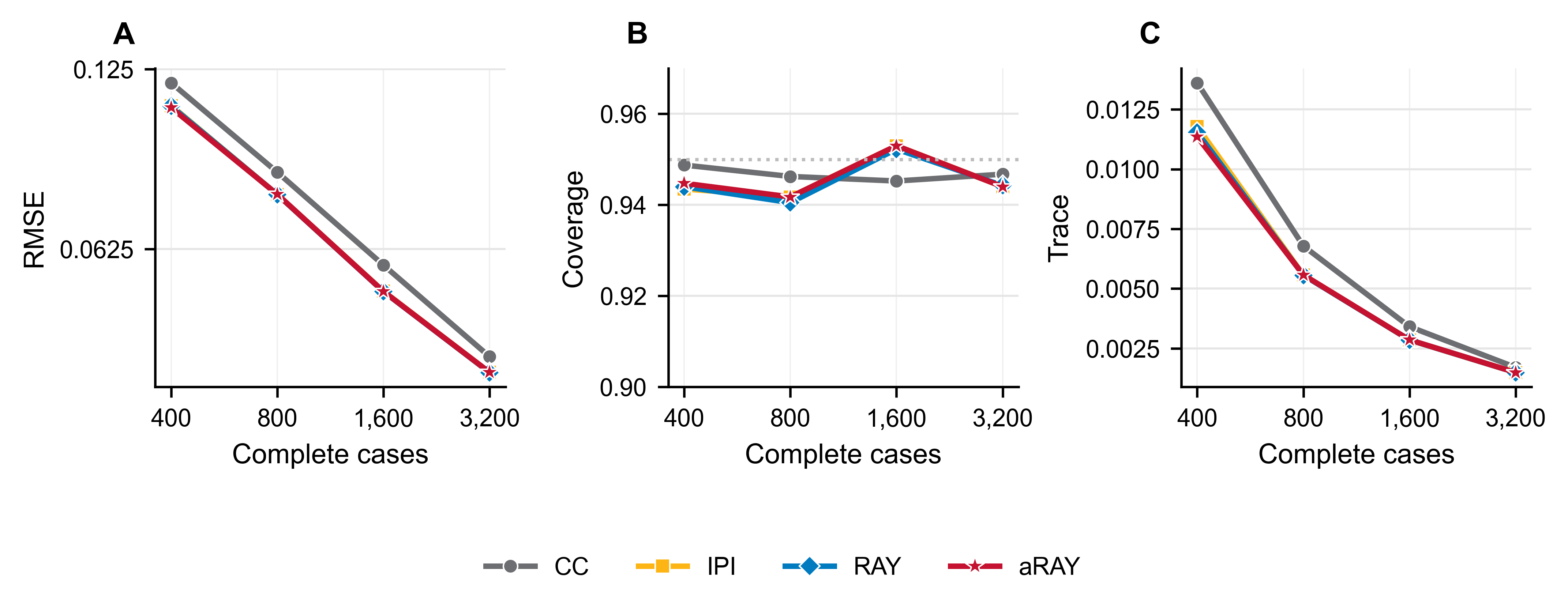}
    \caption{Linear regression coefficients under MCAR as the complete-pattern size varies. Each incomplete pattern contains $3000$ observations. The panels report RMSE, mean marginal coverage, and the average trace of the estimated covariance matrix for the four estimators.}
    \label{fig: low-dim-lr}
\end{figure}

We next consider linear regression without an intercept. Let $X_1,X_2\in\Real^2$ and $X=(X_1^\top,X_2^\top)^\top\sim\Normal(\mathbf{0}_4,\Sigma)$, where $\Sigma$ has unit diagonal and all off-diagonal entries equal to $0.4$. We generate
\[
    Y=X_1^\top\mathbf{1}_2+X_2^\top\mathbf{1}_2+\epsilon,
    \qquad \epsilon\sim\Normal(0,1),
\]
so that the target coefficient is $\theta^\star=\mathbf{1}_4$. The complete-pattern size varies over $\{400,800,1600,3200\}$, and each of the three incomplete patterns contains $3000$ observations. The pattern-specific functions are constructed in imputation mode: TabPFN regressors fitted on the independent fully observed sample impute the missing modalities, after which the full-data regression estimating function is evaluated on the completed observation. The tuning criterion is the estimated variance trace, and the full empirical tuning matrix $\hat G$ is used before applying the common ridge and norm constraint described above. Figure~\ref{fig: low-dim-lr} reports RMSE, coverage, and trace. IPI, RAY, and aRAY reduce RMSE by approximately $6\%$--$10\%$ and variance trace by approximately $12\%$--$18\%$ relative to CC, while retaining coverage near $95\%$. The three augmented estimators are nearly indistinguishable in this low-dimensional design.

\subsection{Sensitivity to MAR and MNAR departures}
\label{subsec: sim-missingness-sensitivity}

The final experiment examines the robustness of the MCAR estimators. Let $X_1$, $X_2$, and $\epsilon$ be independent standard-normal variables and generate
\[
    Y=X_1+X_2+0.5(X_1^2-1)+\epsilon,
    \qquad \theta^\star=\Exp[Y]=0.
\]
Let $L$ indicate that $Y$ is observed. For the MAR path, outcome observation depends on the always-observed $X_1$ through $S_{\mathrm{MAR}}=(X_1^2-1)/\sqrt{2}$. For the MNAR path, it depends on $S_{\mathrm{MNAR}}=0.3X_2+\sqrt{1-0.3^2}\,\epsilon$, which includes the unobserved outcome residual. In either case,
\[
    \operatorname{logit}\Pr(L=1\mid Z)=a_\delta+\delta S,
\]
where $a_\delta$ is calibrated so that $\Pr(L=1)=0.25$. Independently, $X_2$ is observed with probability $0.5$. The two indicators generate the four patterns in $\Qcal$ with marginal probabilities $1/8$, $1/8$, $3/8$, and $3/8$, respectively. We use $N=8000$ and $\delta\in\{0,0.25,0.5,0.75,1,1.25\}$. Analytic true conditional means under the full-data law provide the imputation functions.

\begin{figure}[htbp]
    \centering
    \includegraphics[width=1.0\linewidth]{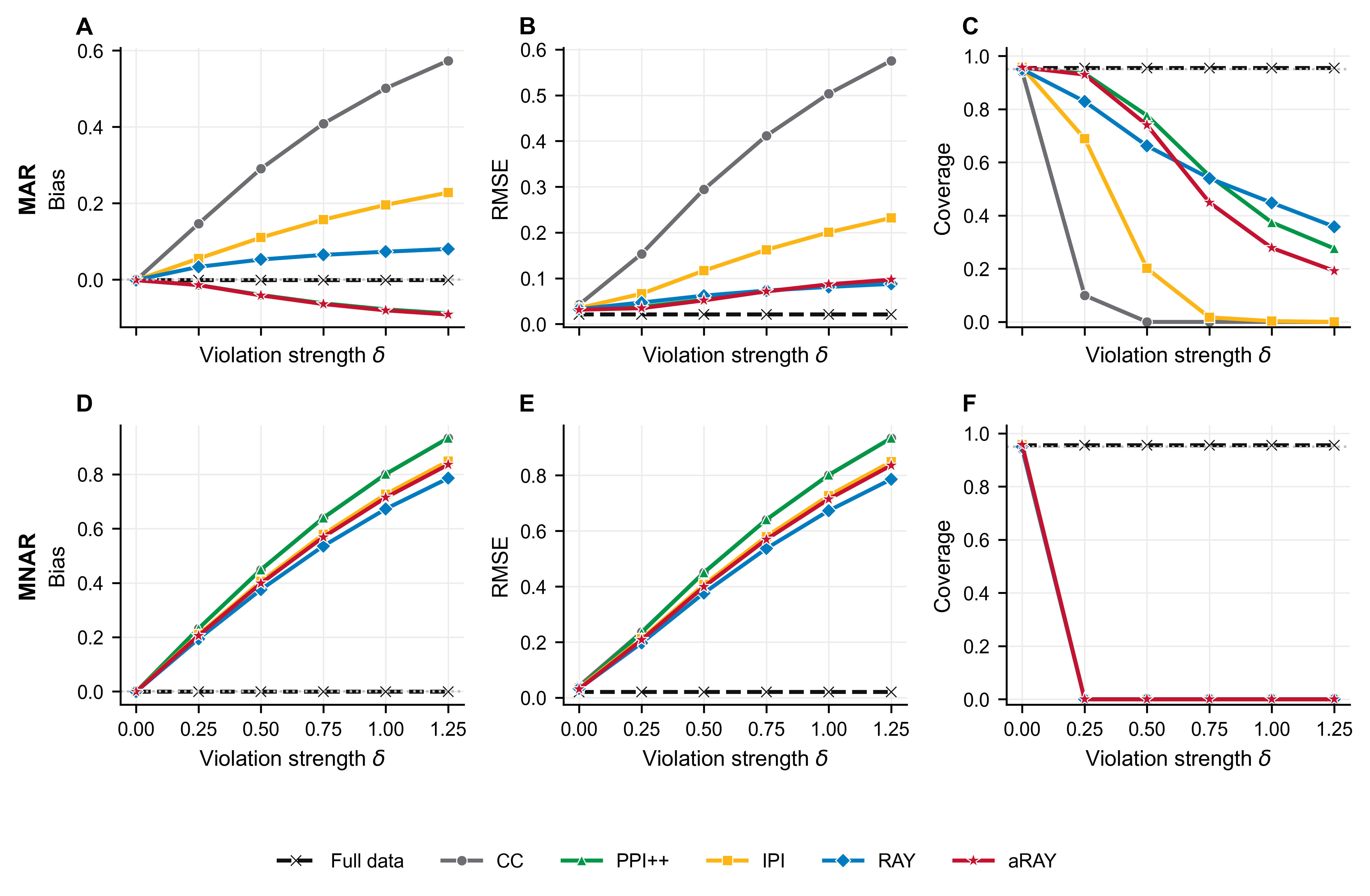}
    \caption{Sensitivity of the outcome-mean estimators to departures from MCAR. The top row follows the MAR path and the bottom row follows the MNAR path. The columns report bias, RMSE, and empirical coverage as the violation strength $\delta$ increases. The estimators use the MCAR weights unchanged; the full-data sample mean is an oracle benchmark.}
    \label{fig: missingness_sensitivity}
\end{figure}

We apply the MCAR versions of CC, \texttt{PPI++}, IPI, RAY, and aRAY without propensity correction; the full-data sample mean is included only as an oracle benchmark. At $\delta=0$, both paths reduce to MCAR: the five observed-data estimators have negligible bias and nominal coverage, and the augmented estimators reduce RMSE relative to CC. Under MAR, bias and undercoverage increase with $\delta$ for every method, although the augmentation estimators, especially RAY, adaptive RAY, and PPI++, can remain substantially less biased than CC and IPI. Under MNAR, all observed-data estimators exhibit substantial bias and essentially zero coverage once $\delta>0$.

\section{Application to surface protein abundance data}
\label{sec: real_data}

We apply the RAY-family estimators to surface protein abundance estimation in single-cell multi-omics. Recent assays measure different combinations of modalities: DOGMA-seq jointly measures chromatin accessibility (ATAC), gene expression (RNA), and cell-surface protein levels (ADT) \citep{mimitou2021scalable}; CITE-seq measures RNA and ADT; ASAP-seq measures ATAC and ADT. Estimating mean protein abundance across cells is of interest for understanding cell function, but ADT is often unmeasured while RNA or ATAC is available. Because protein is the downstream product of transcription and translation, RNA and ATAC are correlated with protein abundance, so using all available data should improve the estimate. We use these complementary assay configurations to train modality-specific predictors. Figure~\ref{fig: real_data} summarizes the modalities available in each sequencing assay; it does not depict the randomized observation patterns used in the subsequent estimation experiment. Since these datasets are used for following estimation and inference, it is expected that trained neural nets are inherently biased. This would mimic the real applications where pretrained imputation algorithms are usually trained from different distributions.

\begin{figure}[htbp]
    \centering
    \input{real_data_illustration}
    \caption{Modality availability in the three sequencing assays used for predictor pretraining. DOGMA-seq records RNA, ATAC, and ADT; CITE-seq records RNA and ADT; and ASAP-seq records ATAC and ADT. These assay-source configurations are distinct from the randomized observation patterns used for estimation.}
    \label{fig: real_data}
\end{figure}

\paragraph{Predictor pretraining.} We use PBMC datasets and target CD4 cells. After quality control and batch correction, we retain $880$ genes (RNA), $3288$ peaks (ATAC), and $208$ proteins (ADT). We pretrain three feedforward neural networks with ReLU activations: one predicts ADT from RNA using CITE-seq, one predicts ADT from ATAC using ASAP-seq, and one predicts ADT from both modalities using $20\%$ of DOGMA-seq cells.

\paragraph{Estimation benchmark.} For estimation, we use the remaining $80\%$ of the DOGMA-seq cells, for which all three modalities are available before masking. We independently assign each cell to one of four analysis patterns: $\Data_1 = \{\text{RNA, ATAC, ADT}\}$, $\Data_2 = \{\text{RNA, ATAC}\}$, $\Data_3 = \{\text{RNA}\}$, and $\Data_4 = \{\text{ATAC}\}$, with probabilities $10\%, 30\%, 30\%$, and $30\%$, respectively. These randomized analysis patterns, rather than the assay-source configurations shown in Figure~\ref{fig: real_data}, define the missingness variable $R$. Because pattern assignment is independent of the complete multimodal measurements, the resulting benchmark satisfies MCAR and therefore evaluates the estimator developed in Section~\ref{sec: estimation-mcar}, not the exploratory MAR constructions of Section~\ref{sec: estimation-mar}. Since protein abundance is in fact observed for all DOGMA-seq cells, we treat the full-sample protein means as ground truth. \texttt{PPI}-type estimators do not naturally apply here and are omitted.

\begin{figure}[htbp]
    \centering
    \includegraphics[width=0.8\linewidth]{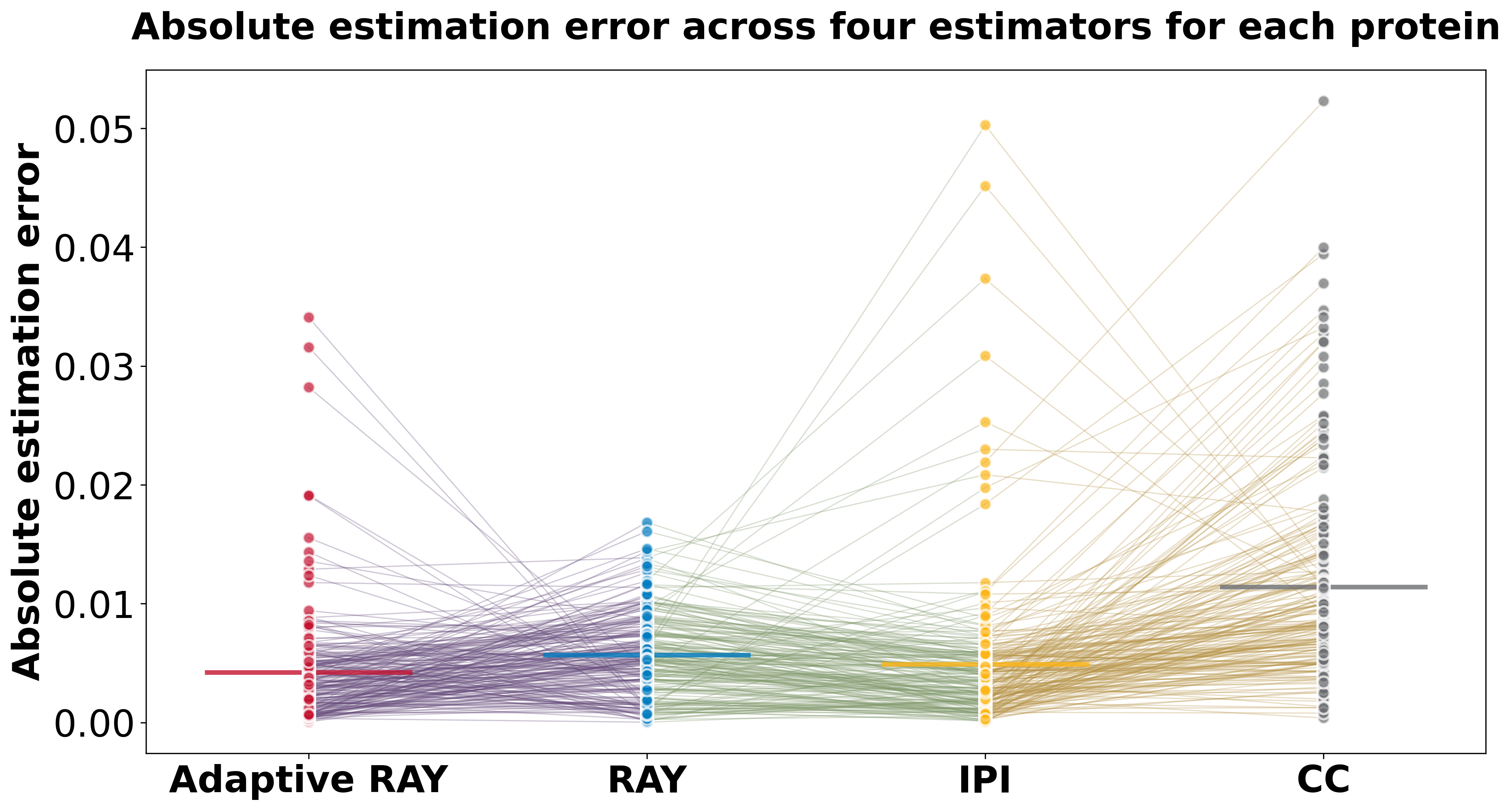}
    \caption{Per-protein absolute error for the four estimators; horizontal bars are averages across proteins.}
    \label{fig: dogma_absolute_error}
\end{figure}

\begin{figure}[htbp]
    \centering
    \includegraphics[width=0.8\linewidth]{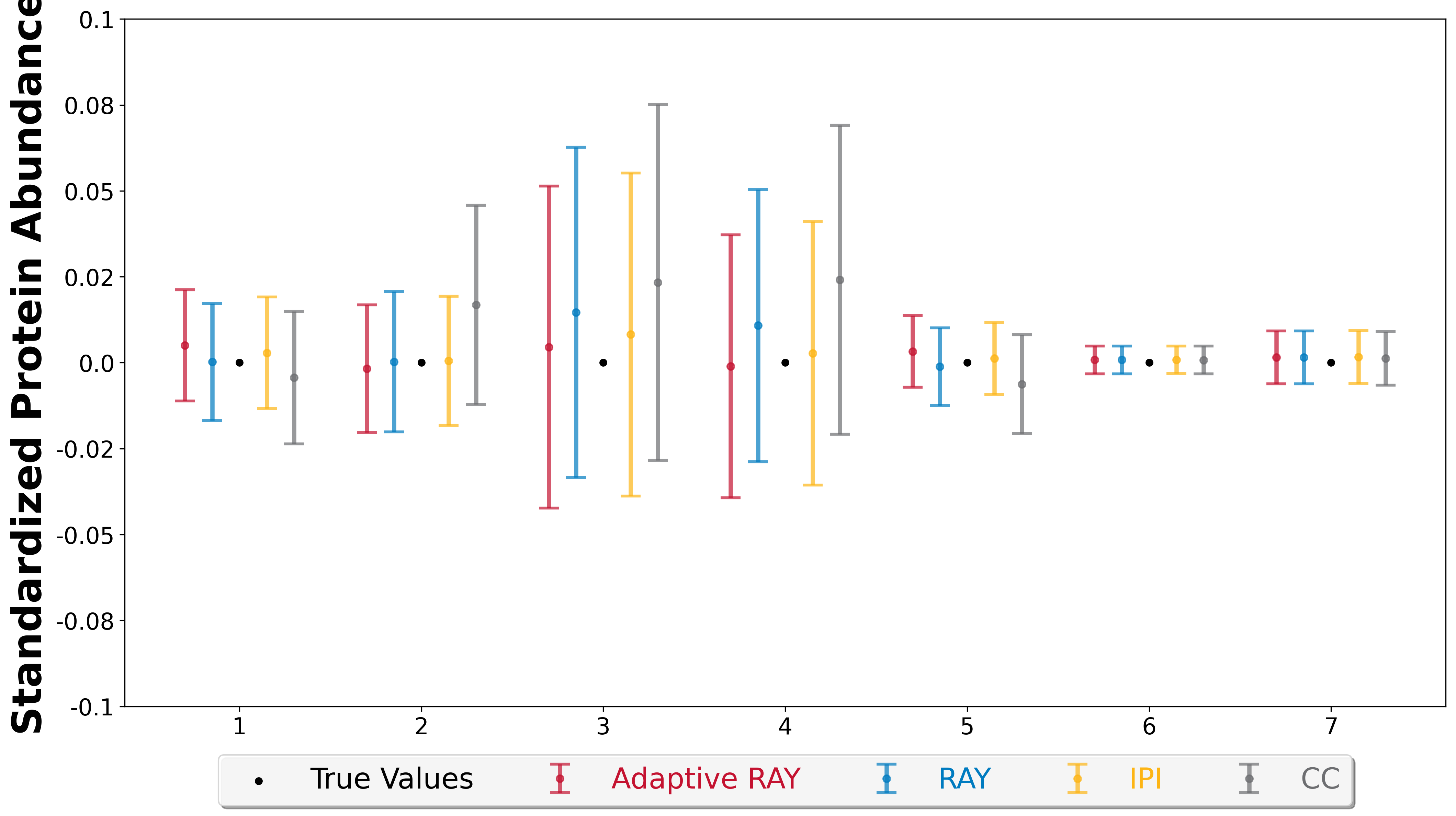}
    \caption{Confidence intervals of the evaluated estimators, for the top $5\%$ most abundant proteins.}
    \label{fig: dogma_ci_efficiency}
\end{figure}

Figure~\ref{fig: dogma_absolute_error} reports the per-protein absolute error against the ground truth. The three augmentation estimators, RAY, IPI, and aRAY, achieve a lower average error than the complete-case estimator. Among them, aRAY and IPI have the lowest error for most proteins, while RAY is more stable across proteins at a slightly higher average error. Figure~\ref{fig: dogma_ci_efficiency} reports confidence intervals for the highest-abundance proteins. All three augmented estimators are more efficient than the complete-case estimator for every such protein, and aRAY is uniformly the most efficient, achieving up to 35\% shorter confidence interval.

\section{Discussion}
\label{sec: discussion}

We studied estimation with nonmonotone observed patterns through the RAY decomposition, which converts the inverse-operator characterization of the optimal estimating equation into a closed-form, pattern-wise approximation under MCAR. The resulting RAY estimator remains valid for arbitrary fixed, possibly misspecified imputation functions, while variance-based tuning retains complete-case analysis as a feasible choice under the selected scalar variance criterion. Adaptive RAY further embeds RAY and IPI in a broader calibrated class and is population-optimal within that class under the same criterion. We characterized exactness through a necessary and sufficient pattern-wise residual condition, identified several verifiable sufficient cases, and bounded the remaining efficiency gap by correlation and structural factors. The simulations support the predicted efficiency gains and nominal coverage under MCAR while showcasing deterioration under MAR and MNAR departures. The randomized single-cell benchmark demonstrates the gains from combining incomplete multimodal samples when estimating mean surface-protein abundance.

Two limitations remain. First, our structural result is one-sided: intersection-closedness implies that the structural factor vanishes, but we do not establish a converse or otherwise fully characterize the observed-pattern collections for which the structural contribution is zero. The current theory therefore does not classify possible cancellations for non-intersection-closed designs. Second, the practical estimator and its closed-form advantage remain fundamentally MCAR results. Under general MAR, the direct weights satisfy valid population identities but are generally not measurable with respect to the observed data, whereas projection restores measurability by introducing projected tail propensities whose estimation contributes at first order. Correctly specified finite-dimensional models with root-$N$ consistent nuisance estimators provide one route to regular inference through stacked estimation. Our partial sensitivity calculation does not characterize the full MAR nuisance tangent space or rule out cancellations along admissible joint submodels, and regular inference with more general nuisance estimators remains unresolved without additional structural restrictions.

These limitations suggest several directions for future work. For MAR, one sufficient route is to establish scientifically interpretable conditions under which the nonmonotone pattern propensity scores can be estimated at a parametric rate. Some explorations are in \citep{sun2018inverse, zhao2020statistical, zhao2023maximum}. Such conditions could support estimators of both $\pi_r(Z_r)$ and $\lambda_s^{\mathrm{proj}}(Z_s)$. More generally, it may be enough to estimate the particular nuisance-error functional entering the target expansion at the required rate. Combining these estimators with observed-data orthogonalization or augmented-IPW corrections may yield feasible root-$N$ procedures while retaining some of RAY's patternwise structure. A sharper combinatorial characterization of structural cancellation would further clarify when the closed-form surrogate can be exact beyond the sufficient cases established here. Finally, although RAY is defined for arbitrary nonmonotone observed-pattern sets, studies with many modalities can generate large, sparse, and highly overlapping pattern collections. Scalable truncation, pattern aggregation, or selection of informative RAY components may extend the method to these more complex missing-pattern regimes.

\section*{Acknowledgment}
This project was supported by National Institute of Mental Health grant R01MH123184 and NSF grants DMS-2310764 and DMS-2515687.

\bibliographystyle{apalike}
\bibliography{reference}

\newpage
\begin{center}
    \Large Supplementary Materials
\end{center}
\appendix
\pagenumbering{arabic}
\renewcommand\thefigure{\thesection.\arabic{figure}}
\renewcommand\thetable{\thesection.\arabic{table}}
\setcounter{figure}{0}
\setcounter{table}{0}

\section{Proofs for the MCAR development}
\label{app_sec: technical}

For any logical statement $B$, $[B]$ denotes its Iverson indicator, equal to one when $B$ is true and zero otherwise.

\subsection{Proof of Lemma~\ref{lemma: ray}}
\label{app_sec: ray_property}
\begin{proof}
We verify separately the zero-order component and the sum of all components. For $s=\emptyset$, the only possible term in the definition of $\mathcal P_s(f)$ is $\mathcal A_\emptyset(f)=\Exp_P[f]$, which is zero because $f\in L_0^2(P)$. Thus $\mathcal P_\emptyset(f)=0$, as stated.

For the full decomposition, substitute the definition of $\mathcal P_s$ and exchange the two finite sums:
\[
\sum_{s\subseteq [p]}\mathcal{P}_s(f) = \sum_{s\subseteq [p]}\sum_{r\subseteq s,\, r\in \Qcal}(-1)^{|s|-|r|}\mathcal{A}_r(f)
= \sum_{r\in \Qcal}\Big(\sum_{s:\,r\subseteq s\subseteq[p]}(-1)^{|s|-|r|}\Big)\mathcal{A}_r(f).
\]
It remains to evaluate the coefficient of each conditional-expectation term $\mathcal A_r(f)$. For $r \neq [p]$, write $s = r \cup s'$ with $s' \subseteq [p]\setminus r$. The coefficient is then
$\sum_{s'\subseteq [p]\setminus r}(-1)^{|s'|} = (1-1)^{|[p]\setminus r|} = 0$.
For $r = [p]$, the Boolean interval contains only $s=[p]$, so the coefficient is $1$. Since conditioning on the full data leaves $f$ unchanged, $\mathcal{A}_{[p]}(f) = f$. All terms except the full-pattern term therefore cancel, and $\sum_s \mathcal{P}_s(f) = f$.
\end{proof}

\subsection{Proof of Propositions~\ref{prop: valid_ee} and~\ref{prop: valid_aee}}
\label{app_sec: valid_ee}

\begin{proposition}
    \label{prop: valid_aee}
    For any $\alpha$ satisfying~\eqref{eqn: alpha_constraint}, the function in~\eqref{eqn: adaptive_ef} satisfies $\Exp[\psi_{\alpha}(R,Z_R,\theta^\star)] = \mathbf{0}_d$.
\end{proposition}

\begin{proof}
We first prove Proposition~\ref{prop: valid_ee}. For the complete-case term, MCAR separates the pattern indicator from the full-data estimating function, so
$\Exp[\Ind(R=[p])\pi_{[p]}^{-1}\psi^F] = \Exp[\Ind(R=[p])]\pi_{[p]}^{-1}\Exp[\psi^F] = 0$.
For a RAY augmentation term, MCAR likewise gives $\Exp[\omega_r\Func_r]=\Exp[\omega_r]\Exp[\Func_r]$. We therefore only need to show that the RAY weight has mean zero. Writing $\Ind(R\supseteq s) = \sum_{t\in\Qcal:\,t\supseteq s}\Ind(R=t)$ and then using the definition of $\lambda_s$ gives
\[
\Exp[\omega_r] = \sum_{s:\,r\subseteq s\subseteq[p]}\frac{(-1)^{|s|-|r|}}{\lambda_s}\sum_{t\in\Qcal:\,t\supseteq s}\pi_t
= \sum_{s:\,r\subseteq s\subseteq[p]}(-1)^{|s|-|r|} = 0,
\]
where the last equality is the same Boolean cancellation used in Appendix~\ref{app_sec: ray_property}; it applies because the augmentation index satisfies $r\neq[p]$. Thus every fixed augmentation has mean zero, regardless of the choice of $\Func_r$ or its tuning coefficient. Together with the complete-case calculation, this proves Proposition~\ref{prop: valid_ee} and the validity of~\eqref{eqn: mmpp_aef}.

It remains to prove Proposition~\ref{prop: valid_aee}. Fix an incomplete pattern $r$. For its contribution to~\eqref{eqn: adaptive_ef}, MCAR gives
\[
\Exp\left[\sum_{s\in\Qcal:\,s\supseteq r}\frac{\Ind(R=s)}{\pi_s}\alpha_{r,s}\Func_r\right]
=\left(\sum_{s\in\Qcal:\,s\supseteq r}\alpha_{r,s}\right)\Exp[\Func_r]=0
\]
by~\eqref{eqn: alpha_constraint}. Hence the augmentation associated with each $r$ is mean-zero. Summing these identities over the incomplete patterns, and adding the already mean-zero complete-case term, proves Proposition~\ref{prop: valid_aee}.
\end{proof}

\subsection{Proof of Theorem~\ref{thm: asymptotic_behavior}}
\label{proof: asymp_behavior}
For clarity, the regularity conditions invoked in Theorem~\ref{thm: asymptotic_behavior} are the standard local Z-estimation conditions used below. On a neighborhood of $\theta^\star$, $\psi^F(Z,\theta)$ and each fixed $\Func_r(Z_r;\theta)$ are almost surely continuously differentiable in $\theta$ and, together with their derivatives, admit integrable envelopes sufficient for the uniform laws of large numbers below; their values at $\theta^\star$ have finite second moments. The population map $\theta\mapsto\Exp[\psi^F(Z,\theta)]$ has $\theta^\star$ as its unique, well-separated zero, and its Jacobian $A$ at $\theta^\star$ exists and is nonsingular. We consider a sequence of empirical roots in this neighborhood. These conditions, the finiteness of $\Qcal$, and Assumption~\ref{assumption: missing_positivity} also justify the multivariate central limit theorem and the Taylor expansions used in the proof.

\begin{proof}
The proof has three parts: we first remove the effect of estimating the pattern proportions, then apply the standard Z-estimation expansion, and finally calculate the variance of the resulting score.

\emph{Calibrated score expansion.} Let $\mathbb P_N$ denote the empirical average and write $\bar\Func_r=\Func_r-\Exp[\Func_r]$, with all functions evaluated at $\theta^\star$. The exact sample calibration identity $\mathbb P_N\hat\omega_r=0$ removes the constant part of every imputation function:
\[
\mathbb P_N[\hat\omega_r\Func_r]=\mathbb P_N[\hat\omega_r\bar\Func_r].
\]
Because $\Qcal$ is finite and $\pi_{[p]}$ is bounded away from zero, $\omega_r(R;\boldsymbol\pi)$ is continuously differentiable in the pattern proportions in a neighborhood of the truth. A Taylor expansion, $\hat{\boldsymbol\pi}-\boldsymbol\pi=O_p(N^{-1/2})$, and MCAR yield
\[
\mathbb P_N[(\hat\omega_r-\omega_r)\bar\Func_r]=o_p(N^{-1/2}),
\]
because each derivative of $\omega_r(R;\boldsymbol\pi)$ is a function of $R$ alone and is therefore independent of the centered $\bar\Func_r$. Its product with $\bar\Func_r$ has mean zero, so the corresponding empirical average is $O_p(N^{-1/2})$; multiplication by $\hat{\boldsymbol\pi}-\boldsymbol\pi$ makes the first-order plug-in term $O_p(N^{-1})$. The Taylor remainder is of the same or smaller order. Similarly,
\[
\mathbb P_N\left[\left\{\hat\pi_{[p]}^{-1}-\pi_{[p]}^{-1}\right\}\Ind(R=[p])\psi^F\right]=o_p(N^{-1/2}),
\]
since $\Exp[\Ind(R=[p])\psi^F]=0$ and the inverse pattern proportion is smooth near $\pi_{[p]}>0$. Consequently, estimating the pattern proportions does not contribute to the first-order score, and
\begin{equation}
\label{eqn: calibrated_expansion}
\sqrt N\,\hat U_N(\theta^\star;\alpha)
=\frac1{\sqrt N}\sum_{i=1}^N\xi_{\alpha,i}+o_p(1),\qquad
\xi_\alpha=\frac{\Ind(R=[p])}{\pi_{[p]}}\psi^F+\sum_{r\neq[p]}\alpha_r\omega_r\bar\Func_r.
\end{equation}

\emph{Z-estimation expansion.} Under the stated uniform convergence conditions, $\hat U_N(\theta;\alpha)$ converges uniformly to $\Exp[\psi^F(Z,\theta)]$, because every augmentation has expectation zero for every $\theta$ by the argument in Appendix~\ref{app_sec: valid_ee}. The unique, well-separated root condition therefore gives $\hat\theta(\alpha)\pto\theta^\star$. The empirical Jacobian converges to $A$: the complete-case contribution has expectation $A$, while the expected derivative of each augmentation is zero by MCAR and the mean-zero weight identity. A Taylor expansion of the empirical equation around $\theta^\star$, followed by~\eqref{eqn: calibrated_expansion}, gives
\[
\sqrt N\{\hat\theta(\alpha)-\theta^\star\}
=-A^{-1}\frac1{\sqrt N}\sum_{i=1}^N\xi_{\alpha,i}+o_p(1).
\]
The multivariate central limit theorem now yields asymptotic normality with covariance $A^{-1}\Var(\xi_\alpha)A^{-\top}$.

\emph{Variance calculation.} It remains to compute $\Var(\xi_\alpha)$. Write $M = \Ind(R=[p])\pi_{[p]}^{-1}\psi^F$ and $B_r = \omega_r\bar\Func_r$. Each term is mean-zero, and MCAR separates every $R$-measurable factor from every $Z$-measurable factor. The three types of covariance terms are:
\begin{itemize}
\item \emph{Main--main:} $\Exp[MM^\top] = \pi_{[p]}^{-2}\Exp[\Ind(R=[p])]\,\Exp[\psi^F\psi^{F\top}] = \pi_{[p]}^{-1}\Var(\psi^F)$.
\item \emph{Main--augmentation:} $\Exp[MB_r^\top]=\gamma_r\Cov(\psi^F,\bar\Func_r)$, and the transpose term is $\gamma_r\Cov(\bar\Func_r,\psi^F)$.
\item \emph{Augmentation--augmentation:} $\Exp[B_rB_{r'}^\top]=\nu_{r,r'}\Cov(\bar\Func_r,\bar\Func_{r'})$.
\end{itemize}
Combining these terms and conjugating by $A^{-1}$ and $A^{-\top}$ gives~\eqref{eqn: avar}.

Finally, consider tuning estimated on an independent fold. Conditional on that training fold, $\hat\alpha-\alpha^\star=o_p(1)$, while the calibrated empirical augmentation average on the evaluation fold is $O_p(N^{-1/2})$. Replacing $\alpha^\star$ by $\hat\alpha$ therefore changes the score by $(\hat\alpha-\alpha^\star)^\top\mathbb P_N(\hat\omega_r\Func_r)=o_p(N^{-1/2})$. Thus tuning estimation does not change the first-order expansion.
\end{proof}

\subsection{Proof of Corollary~\ref{coro: min_alpha}}
\label{proof: corollary_min_alpha}
\begin{proof}
By linearity and transpose invariance of $\ell$, the objective is the quadratic
$f(\alpha) = C_0 + \alpha^\top G\alpha + 2L^\top\alpha$,
with $C_0 = \ell\{A^{-1}\Var(\psi^F)A^{-\top}\}/\pi_{[p]}$ and $G,L$ as stated. Positive definiteness of $G$ is assumed in the corollary; a sufficient condition is that both $(\nu_{r,r'})$ and $(\ell\{A^{-1}\Cov(\bar\Func_r,\bar\Func_{r'})A^{-\top}\})$ are positive definite, by the Schur product theorem.

The gradient and Hessian are $\nabla f(\alpha)=2G\alpha+2L$ and $\nabla^2f(\alpha)=2G$, respectively. Since $G$ is positive definite, the stationary point is the unique global minimizer. Solving the first-order condition gives $\alpha^\star=-G^{-1}L$, and substitution gives the stated minimum. Moreover, $L^\top G^{-1}L\geq0$ because $G^{-1}$ is positive definite, which establishes the claimed nonnegative improvement over the complete-case criterion.
\end{proof}

\subsection{Proof of Theorem~\ref{thm: adaptive_asymptotic_behavior}}
\begin{proof}
The argument parallels the proof of Theorem~\ref{thm: asymptotic_behavior}; the only new point is the aggregation of several imputation functions within each realized pattern.

\emph{Centering and first-order score.} With sample pattern proportions, the empirical coefficient multiplying any constant shift of $\Func_r$ is $\sum_{s\in\Qcal:\,s\supseteq r}\alpha_{r,s}=0$. Thus the adaptive estimating equation is unchanged when every $\Func_r$ is replaced by $\bar\Func_r$. After this centering, the same Taylor and calibration argument as in~\eqref{eqn: calibrated_expansion} removes the first-order effect of estimating the pattern proportions. The resulting first-order score is
\[
\xi_\alpha^{\mathrm{aRAY}}
=\frac{\Ind(R=[p])}{\pi_{[p]}}\psi^F
+\sum_{t\in\Qcal}\frac{\Ind(R=t)}{\pi_t}H_t,
\qquad
H_t=\sum_{r\in\Qcal:\,r\subseteq t,\,r\neq[p]}\alpha_{r,t}\bar\Func_r.
\]

\emph{Consistency and linearization.} The augmentation has mean zero by~\eqref{eqn: alpha_constraint}, and its expected derivative is zero by the same calculation applied to the derivatives of the fixed imputation functions. Therefore, the population equation remains $\Exp[\psi^F(Z,\theta)]=0$, and the empirical Jacobian converges to $A$. The consistency and Z-estimation arguments in the proof of Theorem~\ref{thm: asymptotic_behavior} then apply without change.

\emph{Variance calculation.} The pattern indicators are disjoint, so augmentation terms associated with different realized patterns have zero product. In addition, $\Exp[H_t]=0$ for every $t$ because each $\bar\Func_r$ is centered. The complete-pattern augmentation $H_{[p]}$ is the only augmentation that has a nonzero cross-covariance with the complete-case term. Expanding $\Var(\xi_\alpha^{\mathrm{aRAY}})$ therefore gives
\[
\frac{\Var(\psi^F)}{\pi_{[p]}}
+\frac{\Cov(\psi^F,H_{[p]})+\Cov(H_{[p]},\psi^F)}{\pi_{[p]}}
+\sum_{t\in\Qcal}\frac{\Var(H_t)}{\pi_t}.
\]
Conjugating this score variance by $A^{-1}$ and $A^{-\top}$ proves the stated asymptotic variance and completes the proof.
\end{proof}

\subsection{Characterization of IPI and RAY within~\eqref{eqn: adaptive_ef}}
\label{app_sec: two_estimator_char}
We show that the $\omega_r$ of RAY and IPI correspond to $\alpha$ satisfying~\eqref{eqn: alpha_constraint}. For RAY, expanding $\Ind(R\supseteq t) = \sum_{s\in\Qcal:\,s\supseteq t}\Ind(R=s)$,
\[
\omega_r = \sum_{t:\,r\subseteq t\subseteq[p]}(-1)^{|t|-|r|}\frac{\Ind(R\supseteq t)}{\lambda_t}
= \sum_{s\in\Qcal:\,s\supseteq r}\frac{\Ind(R=s)}{\pi_s}\,\pi_s\Big(\sum_{t:\,r\subseteq t\subseteq s}\frac{(-1)^{|t|-|r|}}{\lambda_t}\Big),
\]
so $\alpha_{r,s} = \pi_s\sum_{t: r\subseteq t\subseteq s}(-1)^{|t|-|r|}/\lambda_t$, and
\[
\sum_{s\in\Qcal:\,s\supseteq r}\alpha_{r,s}
= \sum_{t:\,r\subseteq t\subseteq[p]}\frac{(-1)^{|t|-|r|}}{\lambda_t}
\sum_{s\in\Qcal:\,s\supseteq t}\pi_s
= \sum_{t:\,r\subseteq t\subseteq[p]}(-1)^{|t|-|r|} = 0.
\]
For IPI, $\omega_r = \Ind(R=r)/\pi_r - \Ind(R=[p])/\pi_{[p]}$ corresponds to $\alpha_{r,r} = 1$, $\alpha_{r,[p]} = -1$, which satisfies the constraint.

\subsection{Comparison of efficiency gains}
\label{app_sec: compare_two_ibm}
By Theorem~\ref{thm: asymptotic_behavior} and Corollary~\ref{coro: min_alpha}, comparing RAY and IPI amounts to comparing $L^\top G^{-1}L$ under their respective $\omega_r$. When $\Cov(\bar\Func_r, \bar\Func_{r'}) = 0$ for $r \neq r'$, $G$ is diagonal and the gain is $\sum_{r\neq[p]}\gamma_r^2\,\ell_r/\nu_{rr}$, where $\ell_r = \ell^2\{A^{-1}\Cov(\bar\Func_r, \psi^F)A^{-\top}\}/\ell\{A^{-1}\Var(\bar\Func_r)A^{-\top}\}$ is fixed given the imputation functions. The two estimators differ only through $\gamma_r^2/\nu_{rr}$, which for IPI equals $\pi_r/(\pi_{[p]}^2 + \pi_{[p]}\pi_r)$ and for RAY is the corresponding ratio of alternating sums of $\lambda^{-1}$. There is no uniform ordering over the pattern-probability simplex, motivating optimization over the enlarged adaptive class.

\section{Proofs for the efficiency-loss theory}
\label{app_sec: gap_proofs}

Throughout, $\psi_{\mathrm{RAY}}$ is the RAY estimating function studied in Section~\ref{sec: exactness}, $\psi_{\mathrm{opt}} = \Lop\{\Mop^{-1}\psi^F\}$, and $e_r$ is the per-pattern residual~\eqref{eqn: er}.

\subsection{Common Jacobian}
We prove the result from the population means of the two estimating
functions. This avoids applying the RAY decomposition to
$\dot\psi^F=\partial\psi^F(\cdot,\theta^\star)/\partial\theta^\top$,
which need not be mean-zero.

For RAY, the same mean-zero augmentation calculation used in
Proposition~\ref{prop: valid_ee} gives, for every $\theta$ in a
neighborhood of $\theta^\star$,
\[
\Exp\{\psi_{\mathrm{RAY}}(R,Z_R,\theta)\}
=\Exp\{\psi^F(Z;\theta)\}.
\]
Indeed, under MCAR the complete-case term has expectation
$\Exp\{\psi^F(Z;\theta)\}$, while every augmentation has expectation
zero. Hence, under the differentiation-under-the-expectation condition
used throughout,
\[
\left.
\frac{\partial}{\partial\theta^\top}
\Exp\{\psi_{\mathrm{RAY}}(R,Z_R,\theta)\}
\right|_{\theta=\theta^\star}
=
\Exp\!\left[
\frac{\partial\psi^F(Z,\theta^\star)}{\partial\theta^\top}
\right]
=A.
\]

For the fixed-$\psi^F$ oracle equation, the tower property and the
definition $\Mop=\sum_r\pi_r\mathcal A_r$ give, again for every nearby
$\theta$,
\[
\begin{split}
\Exp\{\psi_{\mathrm{opt}}(R,Z_R,\theta)\}
&=
\Exp\!\left[
\Lop\{\Mop^{-1}\psi^F(\cdot,\theta)\}
\right]\\
&=
\Exp\!\left[
\Mop\{\Mop^{-1}\psi^F(\cdot,\theta)\}
\right]\\
&=
\Exp\{\psi^F(Z;\theta)\}.
\end{split}
\]
Differentiating this identity at $\theta^\star$ yields the same
Jacobian $A$. This direct argument remains valid when
$\emptyset\in\Qcal$ and does not require the identity
$\sum_s\mathcal P_s=\mathcal I$ outside $L_0^2(P)$.

\subsection{Proof of Proposition~\ref{prop: lossid}}
\begin{lemma}[Aggregate residual]\label{app_lem: aggregate}
$\sum_{r\in\Qcal}\pi_r\,e_r = 0$.
\end{lemma}
\begin{proof}
Recall that $e_r$ is the oracle pattern function minus its RAY counterpart. Averaging the oracle term over the pattern probabilities gives
$\sum_r\pi_r\mathcal{A}_r(\Mop^{-1}\psi^F) = \Mop(\Mop^{-1}\psi^F) = \psi^F$.
For the RAY term, exchange the sums over $r$ and $s$. Since the total probability of the patterns containing $s$ is $\lambda_s$,
$\sum_r\pi_r\sum_{s\subseteq r}\lambda_s^{-1}\mathcal{P}_s\psi^F = \sum_s\mathcal{P}_s\psi^F = \psi^F$,
where the final equality is Lemma~\ref{lemma: ray}. The two pattern averages are identical, so subtracting them gives $\sum_r\pi_r e_r = 0$.
\end{proof}
\begin{proof}[Proof of Proposition~\ref{prop: lossid}]
By the common-Jacobian identity, it is enough to compare the variances of the two estimating functions before conjugation by $A^{-1}$ and $A^{-\top}$. Write $\psi_{\mathrm{opt}} = \sum_r\Ind(R=r)q_r$ with $q_r = \mathcal{A}_r(\Mop^{-1}\psi^F)$, and write $\psi_{\mathrm{RAY}} = \sum_r\Ind(R=r)\rho_r$ with $\rho_r = \sum_{s\subseteq r}\lambda_s^{-1}\mathcal{P}_s\psi^F$. Their difference is patternwise:
$\psi_{\mathrm{opt}} - \psi_{\mathrm{RAY}} = \sum_r\Ind(R=r)e_r$, where $e_r = q_r-\rho_r$.

Each $e_r$ is $\sigma(Z_r)$-measurable. It is also mean-zero: the tower property gives $\Exp[q_r] = \Exp[\Mop^{-1}\psi^F] = 0$, and every $\mathcal{P}_s\psi^F$ has mean zero because it is a linear combination of conditional expectations of the mean-zero function $\psi^F$. Thus the variance of the patternwise difference is obtained by summing its within-pattern second moments.

We next show that this difference is uncorrelated with the optimal estimating function. Since $\psi_{\mathrm{RAY}} - \psi_{\mathrm{opt}} = -\sum_r\Ind(R=r)e_r$, disjointness of the pattern indicators and MCAR give
$\Cov(\psi_{\mathrm{opt}}, \psi_{\mathrm{RAY}} - \psi_{\mathrm{opt}}) = -\sum_r\pi_r\Exp[q_re_r^\top]$.
Because $e_r$ is $\sigma(Z_r)$-measurable and $q_r = \Exp[\Mop^{-1}\psi^F\mid Z_r]$, the tower property gives $\Exp[q_re_r^\top] = \Exp[\Mop^{-1}\psi^F\,e_r^\top]$. Hence
$\sum_r\pi_r\Exp[q_re_r^\top] = \Exp\big[\Mop^{-1}\psi^F(\sum_r\pi_r e_r)^\top\big] = 0$
by Lemma~\ref{app_lem: aggregate}.

The zero cross-covariance yields
$\Var(\psi_{\mathrm{RAY}}) - \Var(\psi_{\mathrm{opt}}) = \Var(\psi_{\mathrm{RAY}} - \psi_{\mathrm{opt}})$.
Finally, disjointness of the pattern indicators and $\Exp[e_r]=0$ give
$\Var(\psi_{\mathrm{RAY}} - \psi_{\mathrm{opt}}) = \sum_r\pi_r\Exp[e_re_r^\top]$.
Conjugating this identity by $A^{-1}$ and $A^{-\top}$ proves the proposition.
\end{proof}

\subsection{Proof of Theorem~\ref{thm: ns}}
\begin{proof}
We first establish the necessary and sufficient condition. By Proposition~\ref{prop: lossid} and invertibility of $A$,
$\mathcal E = 0 \iff \sum_r\pi_r\Exp[e_re_r^\top] = 0$.
Each matrix $\Exp[e_re_r^\top]$ is positive semidefinite, and every observed pattern has $\pi_r>0$. The weighted sum can therefore be zero only if $\Exp[e_re_r^\top]=0$ for every $r$, which is equivalent to $e_r=0$ almost surely. The converse is immediate. Substituting the definition~\eqref{eqn: er} of $e_r$ gives~\eqref{eqn: nscond}.

We next verify the residual split in~\eqref{eqn: gb}. The component $\mathcal{P}_s\psi^F$ is $\sigma(Z_s)$-measurable. Hence, when $s\subseteq r$, conditioning it on $Z_r$ leaves it unchanged:
$\mathcal{A}_r[\mathcal{P}_s\psi^F] = \mathcal{P}_s\psi^F$.
Separate the terms in $\mathcal{A}_r(\Mray\psi^F)$ according to whether $s\subseteq r$. The contained terms form $\rho_r$, while the remaining terms form
$b_r = \sum_{s\not\subseteq r}\lambda_s^{-1}\mathcal{A}_r[\mathcal{P}_s\psi^F]$.
It follows that $\mathcal{A}_r(\Mray\psi^F)=\rho_r+b_r$ and therefore
$e_r=q_r-\rho_r=\mathcal{A}_r[(\Mop^{-1}-\Mray)\psi^F]+b_r=\mathcal{A}_r(g)+b_r$.

For the complete pattern $r=[p]$, every subset $s$ is contained in $r$. The sum defining $b_{[p]}$ is therefore empty, so $b_{[p]}=0$. Since $\mathcal A_{[p]}$ is the identity, the complete-pattern residual is $e_{[p]}=g$, as claimed.
\end{proof}

\subsection{The remainder bound}
\begin{lemma}[Remainder bound]\label{lem:rem}
$\Mop(\Mop^{-1}-\Mray)\psi^F = -\sum_s\lambda_s^{-1}\Rem_s(\psi^F)$, and hence $\Delta(\psi^F) \le \pi_{[p]}^{-1}\|\sum_s\lambda_s^{-1}\Rem_s(\psi^F)\|^2$, with $\Delta(\psi^F) = 0$ iff $\sum_s\lambda_s^{-1}\Rem_s(\psi^F) = 0$.
\end{lemma}
\begin{proof}
By Lemma~\ref{lemma: ray}, $\sum_s \mathcal{P}_s = \mathcal I$. Using the almost-eigen identity~\eqref{eqn: Mop_eigen},
\[
\Mop\Mray\psi^F = \sum_s\lambda_s^{-1}\Mop \mathcal{P}_s\psi^F = \sum_s\lambda_s^{-1}(\lambda_s \mathcal{P}_s + \Rem_s)\psi^F
= \psi^F + \sum_s\lambda_s^{-1}\Rem_s\psi^F,
\]
so $\Mop(\Mop^{-1}-\Mray)\psi^F = -\sum_s\lambda_s^{-1}\Rem_s\psi^F$. Therefore
\[
\Mop^{1/2}(\Mop^{-1}-\Mray)\psi^F
=\Mop^{-1/2}\Mop(\Mop^{-1}-\Mray)\psi^F
=-\Mop^{-1/2}\sum_s\lambda_s^{-1}\Rem_s\psi^F.
\]
Applying $\Mop \succeq \pi_{[p]}\mathcal I$ componentwise to any $h=(h_1,\ldots,h_d)^\top$ gives
\[
\|\Mop^{-1/2}h\|_d^2
=\sum_{j=1}^d\langle h_j,\Mop^{-1}h_j\rangle
\le \pi_{[p]}^{-1}\sum_{j=1}^d\|h_j\|_2^2
=\pi_{[p]}^{-1}\|h\|_d^2.
\]
This proves the stated bound. Equality $\Delta = 0$ holds iff $\sum_s\lambda_s^{-1}\Rem_s\psi^F = 0$ because $\Mop^{-1/2}$ is injective.
\end{proof}

\subsection{An operator identity for intersection-closed designs}
The following identity, used both in the proof of Theorem~\ref{thm: sufficient}(i) and in the defect analysis of Appendix~\ref{app_sec: defect}, holds under intersection-closedness and independence.
\begin{lemma}[Operator identity]\label{lem: opident}
Assume $[p]\in\Qcal$, $\Qcal$ is intersection-closed, and the modalities $Z_1,\dots,Z_p$ are independent. Then $\Mray = \Mop^{-1}$.
\end{lemma}
\begin{proof}
\emph{Step 1: construct the orthogonal ANOVA decomposition.} We first give a self-contained description of the ANOVA spaces used in the proof. For the empty set, let $V_\emptyset=\operatorname{span}\{1\}$ be the space of constant functions. For every nonempty $u\subseteq[p]$, define
\[
V_u
:=
\left\{
g(Z_u)\in L^2(P):
\Exp_P\!\left[g(Z_u)\mid Z_{u\setminus\{j\}}\right]=0
\ \text{a.s. for every }j\in u
\right\}.
\]
Thus a function in $V_u$ depends only on the modalities indexed by $u$ and is \emph{centered in each coordinate}: after all coordinates in $u$ except $j$ are fixed, averaging over $Z_j$ gives zero. For a singleton $u=\{j\}$, this condition reduces to $\Exp_P[g(Z_j)]=0$. For $|u|>1$, $V_u$ contains the pure interaction involving exactly the modalities in $u$, after all lower-order effects have been removed. Each $V_u$ is a closed linear subspace of $L^2(P)$ because the space of $\sigma(Z_u)$-measurable functions is closed and the centering restrictions are kernels of continuous conditional-expectation operators.

Independence implies that conditional-expectation operators indexed by coordinate sets commute according to
\[
\mathcal A_a\mathcal A_b=\mathcal A_{a\cap b},
\qquad a,b\subseteq[p].
\]
Indeed, under the product law of $(Z_1,\ldots,Z_p)$, applying $\mathcal A_b$ integrates out the coordinates outside $b$, and subsequently applying $\mathcal A_a$ integrates out the coordinates outside $a$. The only coordinates retained after both operations are those in $a\cap b$. This identity can also be verified first for product functions and then extended to $L^2(P)$ by density.

For every $u\subseteq[p]$, define the operator
\begin{equation}\label{eqn: anova_projection}
\Pi_u
:=
\sum_{v\subseteq u}(-1)^{|u|-|v|}\mathcal A_v.
\end{equation}
In particular, $\Pi_\emptyset=\mathcal A_\emptyset$ is the projection onto the constants. When $p=2$, the four operators are
\[
\Pi_\emptyset=\mathcal A_\emptyset,\qquad
\Pi_{\{1\}}=\mathcal A_{\{1\}}-\mathcal A_\emptyset,\qquad
\Pi_{\{2\}}=\mathcal A_{\{2\}}-\mathcal A_\emptyset,\qquad
\Pi_{\{1,2\}}=\mathcal I-\mathcal A_{\{1\}}-\mathcal A_{\{2\}}+\mathcal A_\emptyset.
\]
The first two nonconstant terms are the main effects of $Z_1$ and $Z_2$, and the last term is their centered interaction.

We next verify that $\Pi_u(f)$ belongs to $V_u$. The function $\Pi_u(f)$ is $\sigma(Z_u)$-measurable because every term $\mathcal A_v(f)$ in~\eqref{eqn: anova_projection} is measurable with respect to $Z_v$, and $v\subseteq u$. Fix $j\in u$ and apply $\mathcal A_{u\setminus\{j\}}$. Using the commuting identity above and pairing each $v=w$ with $v=w\cup\{j\}$ gives
\[
\begin{split}
\mathcal A_{u\setminus\{j\}}\Pi_u(f)
&=
\sum_{w\subseteq u\setminus\{j\}}
\left\{
(-1)^{|u|-|w|}
+(-1)^{|u|-|w|-1}
\right\}
\mathcal A_w(f)\\
&=0.
\end{split}
\]
The two signs in each pair are opposite. Hence $\Pi_u(f)$ is centered in coordinate $j$; since this holds for every $j\in u$, $\Pi_u(f)\in V_u$.

The operators also sum to the identity. Exchanging the two finite sums in~\eqref{eqn: anova_projection},
\[
\begin{split}
\sum_{u\subseteq[p]}\Pi_u(f)
&=
\sum_{v\subseteq[p]}
\left\{
\sum_{u:\,v\subseteq u\subseteq[p]}
(-1)^{|u|-|v|}
\right\}
\mathcal A_v(f)\\
&=\mathcal A_{[p]}(f)
=f.
\end{split}
\]
The inner coefficient is $(1-1)^{|[p]\setminus v|}$, so it is zero unless $v=[p]$, in which case it is one.

Finally, the spaces are mutually orthogonal. Let $g_u\in V_u$ and $g_v\in V_v$ with $u\neq v$. After interchanging $u$ and $v$ if necessary, choose $j\in u\setminus v$. The function $g_v$ does not depend on $Z_j$ and is therefore measurable with respect to $Z_{[p]\setminus\{j\}}$. Moreover, independence of the modalities implies
\[
\Exp_P[g_u\mid Z_{[p]\setminus\{j\}}]
=
\Exp_P[g_u\mid Z_{u\setminus\{j\}}]
=0,
\]
where the last equality is the defining centering condition for $V_u$. Therefore
\[
\Exp_P[g_ug_v]
=
\Exp_P\!\left[
g_v\,
\Exp_P[g_u\mid Z_{[p]\setminus\{j\}}]
\right]
=0.
\]
We have shown that the ranges of the $\Pi_u$ lie in mutually orthogonal spaces and that their sum is the identity. To see that the range is all of $V_u$, take any $g_u\in V_u$ and decompose it as $g_u=\sum_v\Pi_v(g_u)$. The summands lie in the mutually orthogonal spaces $V_v$, whereas $g_u$ itself lies in $V_u$; uniqueness of an orthogonal decomposition therefore gives $\Pi_u(g_u)=g_u$ and $\Pi_v(g_u)=0$ for $v\neq u$. Thus $\Pi_u$ is the orthogonal projection onto $V_u$, and
\[
L^2(P)=\bigoplus_{u\subseteq[p]}V_u.
\]
For $f\in L_0^2(P)$, the constant component is $\Pi_\emptyset(f)=\Exp_P[f]=0$, so the relevant decomposition is
\[
L_0^2(P)=\bigoplus_{\emptyset\neq u\subseteq[p]}V_u.
\]

\emph{Step 2: calculate the action of $\mathcal A_r$, $\Mop$, and $\Mray$ on $V_u$.} Fix a nonempty $u$ and a function $f_u\in V_u$. We first show
\begin{equation}\label{eqn: Ar_on_Vu}
\mathcal A_r(f_u)
=
\begin{cases}
f_u, & u\subseteq r,\\
0, & u\not\subseteq r.
\end{cases}
\end{equation}
If $u\subseteq r$, then $f_u$ is a function of $Z_u$ and is therefore $\sigma(Z_r)$-measurable, giving $\mathcal A_r(f_u)=f_u$.

Now suppose $u\not\subseteq r$ and choose $j\in u\setminus r$. Because $f_u$ depends only on $Z_u$ and the modalities are independent, conditioning additionally on the variables $Z_{r\setminus u}$ does not change its conditional expectation. Consequently,
\[
\mathcal A_r(f_u)
=
\Exp_P[f_u\mid Z_{u\cap r},Z_{r\setminus u}]
=
\Exp_P[f_u\mid Z_{u\cap r}].
\]
The first equality only partitions the variables in $Z_r$ into those whose indices are in $u$ and those whose indices are outside $u$. For the second equality, $Z_{r\setminus u}$ is independent of the entire vector $Z_u$ and hence supplies no additional information about the $Z_u$-measurable function $f_u$ after $Z_{u\cap r}$ is given.

Since $j\notin r$, we have $u\cap r\subseteq u\setminus\{j\}$. The tower property and the defining centering condition for $V_u$ give
\[
\begin{split}
\Exp_P[f_u\mid Z_{u\cap r}]
&=
\Exp_P\!\left[
\Exp_P[f_u\mid Z_{u\setminus\{j\}}]
\ \middle|\ Z_{u\cap r}
\right]\\
&=0.
\end{split}
\]
This proves the second case of~\eqref{eqn: Ar_on_Vu}. In words, if pattern $r$ omits even one coordinate involved in the pure interaction $f_u$, integrating over that omitted independent coordinate removes the entire interaction.

For example, with two independent modalities, a function $f_{\{1,2\}}\in V_{\{1,2\}}$ satisfies $\Exp_P[f_{\{1,2\}}\mid Z_1]=0$ and $\Exp_P[f_{\{1,2\}}\mid Z_2]=0$. Hence $\mathcal A_{\{1\}}(f_{\{1,2\}})=\mathcal A_{\{2\}}(f_{\{1,2\}})=0$, while $\mathcal A_{\{1,2\}}(f_{\{1,2\}})=f_{\{1,2\}}$. This is the two-modality instance of~\eqref{eqn: Ar_on_Vu}.

Applying~\eqref{eqn: Ar_on_Vu} to the ANOVA decomposition of an arbitrary $f\in L_0^2(P)$ gives
\[
\mathcal A_r(f)
=
\sum_{\emptyset\neq u\subseteq[p]}\mathcal A_r\Pi_u(f)
=
\sum_{\emptyset\neq u\subseteq r}\Pi_u(f).
\]
Thus $\mathcal A_r$ is diagonal in the ANOVA decomposition: it acts as the identity on $V_u$ when $u\subseteq r$ and as zero otherwise.

We can now calculate the action of $\Mop$. For $f_u\in V_u$,
\[
\begin{split}
\Mop(f_u)
&=
\sum_{r\in\Qcal}\pi_r\mathcal A_r(f_u)\\
&=
\left(\sum_{r\in\Qcal:\,r\supseteq u}\pi_r\right)f_u
=
\lambda_u f_u.
\end{split}
\]
Hence $\Mop|_{V_u}=\lambda_u\mathcal I$. Since every $\lambda_u\geq\pi_{[p]}>0$, the inverse is well-defined and satisfies
\[
\Mop^{-1}|_{V_u}=\lambda_u^{-1}\mathcal I.
\]

It remains to calculate the RAY surrogate on the same component. From the definition of $\mathcal P_s$ and~\eqref{eqn: Ar_on_Vu},
\[
\begin{split}
\mathcal P_s(f_u)
&=
\sum_{\substack{t\in\Qcal\\t\subseteq s}}
(-1)^{|s|-|t|}\mathcal A_t(f_u)\\
&=
\left\{
\sum_{\substack{t\in\Qcal\\u\subseteq t\subseteq s}}
(-1)^{|s|-|t|}
\right\}f_u
=:
Z_u(s)f_u.
\end{split}
\]
Therefore
\[
\Mray(f_u)
=
\sum_{s\subseteq[p]}\lambda_s^{-1}\mathcal P_s(f_u)
=
\left\{
\sum_{s\subseteq[p]}\lambda_s^{-1}Z_u(s)
\right\}f_u
=:
\beta_u f_u.
\]
Both $\Mop^{-1}$ and $\Mray$ are thus diagonal on the spaces $V_u$. They are equal if and only if their scalar multipliers agree, that is, if and only if $\beta_u=\lambda_u^{-1}$ for every nonempty $u$. The remainder of the proof establishes this identity by grouping the terms in $\beta_u$ into closure classes.

\emph{Step 3: introduce the smallest observed pattern containing $u$.} Let $m = \lceil u\rceil = \bigcap\{r\in\Qcal: r\supseteq u\}$. The family being intersected is nonempty because $[p]\in\Qcal$. Every member contains $u$, so every finite intersection is nonempty; repeated use of intersection-closedness therefore shows that $m\in\Qcal$. By construction, $m$ contains $u$ and is contained in every observed pattern that contains $u$, so it is the smallest such observed pattern.

This definition gives two useful facts. First, for any $t\in\Qcal$, $u\subseteq t \iff m\subseteq t$. The forward implication follows from the minimality of $m$, and the reverse implication follows from $u\subseteq m$. In particular, $u$ and $m$ are contained in exactly the same observed patterns, so $\lambda_u=\lambda_m$.

Second, consider a subset $s$ for which $Z_u(s)\neq0$. The sum defining $Z_u(s)$ must then contain at least one observed pattern $t$ with $u\subseteq t\subseteq s$. The first fact gives $m\subseteq t\subseteq s$, and hence $s\supseteq m$. For such an $s$, its closure $\lceil s\rceil=\bigcap\{r\in\Qcal:r\supseteq s\}$ belongs to $\Qcal$, because all patterns in the intersection contain the nonempty set $m$. Moreover, an observed pattern contains $s$ if and only if it contains $\lceil s\rceil$, so $\lambda_s=\lambda_{\lceil s\rceil}$. Subsets with $Z_u(s)=0$ contribute nothing to $\beta_u$. We may therefore group all remaining terms by their closure $c=\lceil s\rceil$, obtaining
\[
\beta_u = \sum_{c\in\Qcal:\, c\supseteq m}\lambda_c^{-1}\Big(\sum_{s:\,\lceil s\rceil = c}Z_u(s)\Big).
\]

\emph{Step 4: reduce each closure class to Lemma~\ref{lem:closurecancel}.} Relabel $x = s\setminus m \subseteq W := [p]\setminus m$ and $\Ycal = \{t\setminus m : t\in\Qcal,\, t\supseteq m\}$. The empty set belongs to $\Ycal$ because $m\in\Qcal$, and $W$ belongs to $\Ycal$ because $[p]\in\Qcal$. Intersections in $\Ycal$ correspond to intersections of the associated observed patterns, so $\Ycal$ is intersection-closed.

Every observed pattern $t$ entering $Z_u(s)$ contains $m$ by Step 3 and can therefore be written uniquely as $t=m\cup y$ with $y\in\Ycal$. Likewise, every contributing $s$ has the form $s=m\cup x$. The relation $t\subseteq s$ is then equivalent to $y\subseteq x$, and $|s|-|t|=|x|-|y|$. Consequently, $Z_u(s) = \sum_{y\in\Ycal:\, y\subseteq x}(-1)^{|x|-|y|} =: Z(x)$. The closure operation is preserved by the same relabeling: $\lceil s\rceil = c \iff \mathrm{cl}(x) = c\setminus m$.

Lemma~\ref{lem:closurecancel} now gives $\sum_{x:\,\mathrm{cl}(x) = c\setminus m}Z(x) = [\,c\setminus m = \emptyset\,] = [\,c = m\,]$. Thus every closure class in the displayed expression for $\beta_u$ cancels except the class $c=m$. It follows that $\beta_u = \lambda_m^{-1} = \lambda_u^{-1}$, where the last equality uses the fact that $u$ and $m$ lie in the same observed patterns. Since this equality holds on every ANOVA component $V_u$, $\Mray$ and $\Mop^{-1}$ have the same action on all of $L_0^2(P)$. Hence $\Mray=\Mop^{-1}$.
\end{proof}

\begin{lemma}[Closure-class cancellation]\label{lem:closurecancel}
Let $\Ycal\subseteq 2^W$ be intersection-closed with $\emptyset\in\Ycal$, closure $\mathrm{cl}(x) = \bigcap\{z\in\Ycal: z\supseteq x\}$, and $Z(x) = \sum_{y\in\Ycal:\, y\subseteq x}(-1)^{|x|-|y|}$. Then for every $c\in\Ycal$, $\sum_{x:\,\mathrm{cl}(x) = c}Z(x) = [\,c = \emptyset\,]$.
\end{lemma}

\begin{proof}
\emph{Step 1: characterize the subsets whose closure is $c$.} Fix $c\in\Ycal$. Any $x$ satisfying $\mathrm{cl}(x)=c$ must obey $x\subseteq c$, because $x$ is contained in every set appearing in the intersection that defines its closure. We may therefore rewrite the quantity of interest as
\[
\sum_{x:\,\mathrm{cl}(x)=c}Z(x)
=
\sum_{x\subseteq c}[\mathrm{cl}(x)=c]\,Z(x).
\]

Let $z_1,\dots,z_k$ be the maximal elements of $\{z\in\Ycal: z\subsetneq c\}$. Here maximal means that $z_i\subsetneq c$ and there is no $z\in\Ycal$ strictly between $z_i$ and $c$. Because $\Ycal$ is finite, every proper member of $\Ycal$ contained in $c$ lies below at least one of these maximal elements: starting from that member, one can move upward in the finite inclusion poset until no larger proper member remains. If $c\neq\emptyset$, the collection of proper members is nonempty because it contains $\emptyset$, so $k\geq1$. If $c=\emptyset$, there are no proper subsets of $c$ in $\Ycal$, and therefore $k=0$.

Suppose $x\subseteq c$. Since $c\in\Ycal$ is one of the sets containing $x$, the intersection defining $\mathrm{cl}(x)$ is contained in $c$. We claim that $\mathrm{cl}(x) = c$ iff $x\not\subseteq z_i$ for every $i$. If $x\subseteq z_i$, then $z_i$ participates among the sets intersected in $\mathrm{cl}(x)$, so $\mathrm{cl}(x)\subseteq z_i\subsetneq c$. Conversely, if $\mathrm{cl}(x)\subsetneq c$, then $\mathrm{cl}(x)\in\Ycal$ is a proper member below $c$ and hence is contained in some maximal proper member $z_i$. Because $x\subseteq\mathrm{cl}(x)$, this implies $x\subseteq z_i$. The claim follows.

\emph{Step 2: apply inclusion--exclusion over the maximal proper members.} The preceding characterization gives
$[\mathrm{cl}(x) = c] = \prod_i(1 - [x\subseteq z_i]) = \sum_{S\subseteq[k]}(-1)^{|S|}[x\subseteq z_S]$ with $z_S = \bigcap_{i\in S}z_i\in\Ycal$ ($z_\emptyset = c$). Summing,
\[
\sum_{x:\,\mathrm{cl}(x) = c}Z(x) = \sum_{S\subseteq[k]}(-1)^{|S|}F(z_S), \qquad F(A) = \sum_{x\subseteq A}Z(x).
\]
Here $z_S\in\Ycal$ because $\Ycal$ is intersection-closed, with the empty intersection interpreted as $c$. For a fixed $S$, the indicator $[x\subseteq z_S]$ restricts the sum over $x$ to the subsets of $z_S$, which explains the appearance of $F(z_S)$.

\emph{Step 3: evaluate $F(A)$ by a second Boolean cancellation.} For $A\in\Ycal$, swapping sums, $F(A) = \sum_{y\in\Ycal:\, y\subseteq A}\sum_{x:\, y\subseteq x\subseteq A}(-1)^{|x|-|y|} = \sum_{y\subseteq A}[\,y = A\,] = 1$. To see the middle equality, fix $y\subseteq A$ and write $x=y\cup v$ with $v\subseteq A\setminus y$. The inner sum is then $\sum_{v\subseteq A\setminus y}(-1)^{|v|}=(1-1)^{|A\setminus y|}$, which equals one when $y=A$ and zero otherwise. Because $A\in\Ycal$, the surviving term $y=A$ is present, so $F(A)=1$.

Substituting $F(z_S)=1$ into the displayed inclusion--exclusion formula gives $\sum_{S\subseteq[k]}(-1)^{|S|} = [k = 0]$. If $c\neq\emptyset$, then $k\geq1$ by Step 1, and this Boolean sum is zero. If $c=\emptyset$, then $k=0$, so the sum equals one. These two cases are exactly $[\,c=\emptyset\,]$, which proves the lemma.
\end{proof}

\subsection{Proof of Theorem~\ref{thm: sufficient}(i): estimator exactness}
\begin{proof}
\emph{Step 1: reduce exactness to a componentwise comparison.} By Proposition~\ref{prop: lossid}, it is enough to show that the oracle and RAY pattern functions agree for every observed pattern $r$. Because the modalities are independent, Lemma~\ref{lem: opident} provides the orthogonal decomposition
\[
L_0^2(P)=\bigoplus_{\emptyset\neq u\subseteq[p]}V_u.
\]
Both the oracle and RAY pattern functions are linear in $\psi^F$, so we may compare their action separately on each scalar ANOVA component. The same argument is then applied coordinatewise when $\psi^F$ is vector-valued.

Fix a nonempty $u$ and let $f_u\in V_u$. From~\eqref{eqn: Ar_on_Vu},
\[
\mathcal A_r(f_u)=[u\subseteq r]f_u.
\]
Lemma~\ref{lem: opident} also gives $\Mop^{-1}(f_u)=\lambda_u^{-1}f_u$. Therefore the oracle pattern function on pattern $r$ is
\[
\begin{split}
q_r(f_u)
&:=
\mathcal A_r\{\Mop^{-1}(f_u)\}\\
&=
[u\subseteq r]\,\lambda_u^{-1}f_u.
\end{split}
\]
Equivalently, the full oracle estimating function contributes
\[
\sum_{r\in\Qcal}\Ind(R=r)q_r(f_u)
=
\Ind(R\supseteq u)\lambda_u^{-1}f_u.
\]

\emph{Step 2: write the RAY coefficient on the same component.} The proof of Lemma~\ref{lem: opident} showed that
\[
\mathcal P_s(f_u)=Z_u(s)f_u,
\qquad
Z_u(s)
=
\sum_{\substack{t\in\Qcal\\u\subseteq t\subseteq s}}
(-1)^{|s|-|t|}.
\]
In particular, $Z_u(s)=0$ unless $u\subseteq s$. Hence the RAY pattern function
\[
\rho_r(f_u)
:=
\sum_{s\subseteq r}\lambda_s^{-1}\mathcal P_s(f_u)
\]
can be written as
\[
\rho_r(f_u)
=
[u\subseteq r]\,c_u(r)f_u,
\qquad
c_u(r)
:=
\sum_{s:\,u\subseteq s\subseteq r}\lambda_s^{-1}Z_u(s).
\]
Comparing this expression with $q_r(f_u)$ shows that it remains to prove
\[
c_u(r)=\lambda_u^{-1}
\qquad
\text{for every }r\in\Qcal\text{ satisfying }r\supseteq u.
\]

\emph{Step 3: evaluate $c_u(r)$ by closure-class cancellation.} Fix an observed pattern $r\supseteq u$ and let
\[
m=\lceil u\rceil
:=
\bigcap\{q\in\Qcal:q\supseteq u\}.
\]
As established in Lemma~\ref{lem: opident}, $m\in\Qcal$, $m\subseteq r$, and $u$ and $m$ occur in exactly the same observed patterns. Consequently, $\lambda_m=\lambda_u$. The same lemma also showed that $Z_u(s)\neq0$ requires $s\supseteq m$.

For a contributing $s$, let $c=\lceil s\rceil$ be its smallest observed superset. Because $r$ is itself an observed superset of $s$, the closure satisfies $c\subseteq r$. Conversely, if $c\subseteq r$, then $s\subseteq c\subseteq r$. Thus
\[
s\subseteq r
\quad\Longleftrightarrow\quad
\lceil s\rceil\subseteq r
\]
within every contributing closure class. This equivalence is important: the restriction $s\subseteq r$ keeps either every member of a closure class or none of them.

Relabel $x=s\setminus m$, $W=[p]\setminus m$, and
\[
\Ycal
=
\{q\setminus m:q\in\Qcal,\ q\supseteq m\}.
\]
Then $Z_u(s)$ becomes $Z(x)$ in Lemma~\ref{lem:closurecancel}, and the class $\lceil s\rceil=c$ becomes $\mathrm{cl}(x)=c\setminus m$. Grouping the definition of $c_u(r)$ by $c$ therefore gives
\[
c_u(r) = \sum_{c\in\Qcal:\, m\subseteq c\subseteq r}\lambda_c^{-1}\!\!\sum_{s:\,\lceil s\rceil = c}\!\!Z_u(s)
= \sum_{c\in\Qcal:\, m\subseteq c\subseteq r}\lambda_c^{-1}\,[\,c = m\,] = \lambda_m^{-1} = \lambda_u^{-1},
\]
where Lemma~\ref{lem:closurecancel} evaluates the inner closure-class sum as $[c=m]$. The class $c=m$ is included because $m\subseteq r$.

\emph{Step 4: conclude patternwise equality.} We have shown that, for every nonempty $u$ and every observed pattern $r$,
\[
\rho_r(f_u)
=
[u\subseteq r]\lambda_u^{-1}f_u
=
q_r(f_u).
\]
By linearity and the orthogonal ANOVA decomposition, this equality extends from a single component $f_u$ to every scalar $f\in L_0^2(P)$, and then coordinatewise to $\psi^F$. Thus $e_r=q_r-\rho_r=0$ for every $r\in\Qcal$. Proposition~\ref{prop: lossid} yields $\mathcal E=0$, proving part (i).
\end{proof}

\subsection{Proof of Theorem~\ref{thm: sufficient}(ii) (monotone telescoping)}
\begin{proof}
\emph{Step 1: construct orthogonal increments along the chain.}
Let
\[
\Qcal=\{r_1\subsetneq r_2\subsetneq\cdots\subsetneq r_K=[p]\}.
\]
The proof does not require any independence assumption. Indeed, if $i\leq j$, then $\sigma(Z_{r_i})\subseteq\sigma(Z_{r_j})$, and the tower property gives
\[
\mathcal A_{r_i}\mathcal A_{r_j}f
=\Exp\{\Exp(f\mid Z_{r_j})\mid Z_{r_i}\}
=\Exp(f\mid Z_{r_i})
=\mathcal A_{r_i}f.
\]
Conditioning a $Z_{r_i}$-measurable function on $Z_{r_j}$ also leaves it unchanged, so
\[
\mathcal A_{r_i}\mathcal A_{r_j}
=\mathcal A_{r_j}\mathcal A_{r_i}
=\mathcal A_{r_{\min(i,j)}}.
\]

Set $\mathcal A_{r_0}:=0$ on $L_0^2(P)$ and define
\[
Q_k:=\mathcal A_{r_k}-\mathcal A_{r_{k-1}},
\qquad k=1,\ldots,K.
\]
These are the ``new-information'' projections introduced when one moves from $r_{k-1}$ to $r_k$. For example,
\[
\begin{split}
Q_k^2
&=(\mathcal A_{r_k}-\mathcal A_{r_{k-1}})^2\\
&=\mathcal A_{r_k}-\mathcal A_{r_{k-1}}
-\mathcal A_{r_{k-1}}+\mathcal A_{r_{k-1}}
=Q_k,
\end{split}
\]
and $Q_k$ is self-adjoint because conditional expectation is an orthogonal projection. If $i<j$, the same nesting identity gives
\[
Q_iQ_j
=(\mathcal A_{r_i}-\mathcal A_{r_{i-1}})
(\mathcal A_{r_j}-\mathcal A_{r_{j-1}})
=\mathcal A_{r_i}-\mathcal A_{r_i}
-\mathcal A_{r_{i-1}}+\mathcal A_{r_{i-1}}
=0.
\]
Thus the $Q_k$ are mutually orthogonal projections. Moreover, telescoping gives
\[
\sum_{k=1}^KQ_k=\mathcal A_{r_K}=\mathcal A_{[p]}=\mathcal I,
\qquad
\mathcal A_{r_k}=\sum_{j=1}^kQ_j.
\]
In particular, every $f\in L_0^2(P)$ has the orthogonal decomposition
$f=\sum_{j=1}^KQ_jf$.

\emph{Step 2: diagonalize the oracle operator.}
For $h\in\operatorname{ran}(Q_j)$, the preceding identities imply
\[
\mathcal A_{r_k}h=[\,j\leq k\,]h.
\]
Indeed, $\mathcal A_{r_k}=\sum_{\ell\leq k}Q_\ell$, and only the term
$Q_jh=h$ can survive. Define, as in the main text,
\[
\lambda_{r_j}
=\Pr(R\supseteq r_j)
=\sum_{i=j}^K\pi_{r_i}.
\]
Using $\mathcal A_{r_i}=\sum_{j\leq i}Q_j$ and reversing the two finite
sums,
\[
\Mop = \sum_i \pi_{r_i}\mathcal{A}_{r_i} = \sum_i\pi_{r_i}\sum_{j\le i}Q_j = \sum_j\Big(\sum_{i\ge j}\pi_{r_i}\Big)Q_j = \sum_j\lambda_{r_j}Q_j,
\]
and hence
\[
\Mop^{-1}=\sum_{j=1}^K\lambda_{r_j}^{-1}Q_j.
\]
All $\lambda_{r_j}$ are positive because $r_K=[p]$ and
$\lambda_{r_j}\geq\pi_{[p]}>0$.

\emph{Step 3: diagonalize the RAY surrogate on the same increments.}
Recall
\[
\Mray=\sum_{s\subseteq[p]}\lambda_s^{-1}\mathcal P_s,
\qquad
\mathcal P_s
=\sum_{\substack{k:\,r_k\subseteq s}}
(-1)^{|s|-|r_k|}\mathcal A_{r_k}.
\]
If $h\in\operatorname{ran}(Q_j)$, then Step 2 yields
\[
\mathcal P_s h
=\sum_{\substack{k:\,r_k\subseteq s\\k\geq j}}
(-1)^{|s|-|r_k|}h
=Z_{r_j}(s)h,
\]
where $Z_{r_j}(s)$ is exactly the coefficient $Z_u(s)$ introduced in the
proof of Lemma~\ref{lem: opident}, with $u=r_j$. The chain $\Qcal$ is
intersection-closed, and the smallest observed pattern containing $r_j$
is $r_j$ itself. The closure-class calculation in that lemma, with
Lemma~\ref{lem:closurecancel} supplying the cancellation within each
class, therefore gives
\[
\sum_{s\subseteq[p]}\lambda_s^{-1}Z_{r_j}(s)
=\lambda_{r_j}^{-1}.
\]
Consequently, $\Mray h=\lambda_{r_j}^{-1}h$ for every
$h\in\operatorname{ran}(Q_j)$. Since the ranges of the $Q_j$ form an
orthogonal decomposition of $L_0^2(P)$,
\[
\Mray
=\sum_{j=1}^K\lambda_{r_j}^{-1}Q_j
=\Mop^{-1}.
\]
Lemma~\ref{lem:rem} now implies
$\sum_s\lambda_s^{-1}\Rem_s=0$, and hence
$\Delta(\psi^F)=0$ for every full-data law.

\emph{Step 4: display the corresponding telescoping estimator weight.}
The same cancellation can be seen directly in the RAY weight
\[
\omega_{r_k}
=\sum_{s:\,r_k\subseteq s}
(-1)^{|s|-|r_k|}
\frac{\Ind(R\supseteq s)}{\lambda_s}.
\]
For any $s\supseteq r_k$, define its chain closure
\[
j(s):=\min\{j:s\subseteq r_j\}.
\]
Then $\lceil s\rceil=r_{j(s)}$, so the observed-pattern support implies
\[
\Ind(R\supseteq s)=\Ind(R\supseteq r_{j(s)}),
\qquad
\lambda_s=\lambda_{r_{j(s)}}.
\]
To identify the coefficient of the term indexed by $r_j$, let
\[
B_j
:=\sum_{\substack{s\supseteq r_k\\j(s)=j}}
(-1)^{|s|-|r_k|}.
\]
Its cumulative sum has a simple Boolean-lattice form:
\[
\sum_{\ell=k}^jB_\ell
=\sum_{r_k\subseteq s\subseteq r_j}
(-1)^{|s|-|r_k|}
=(1-1)^{|r_j\setminus r_k|}
=[\,j=k\,].
\]
It follows successively that $B_k=1$, $B_{k+1}=-1$, and
$B_j=0$ for $j\geq k+2$. Therefore, for $k<K$,
\[
\omega_{r_k} = \frac{\Ind(R\supseteq r_k)}{\lambda_{r_k}} - \frac{\Ind(R\supseteq r_{k+1})}{\lambda_{r_{k+1}}} = \frac{\Ind(R = r_k) - \Ind(R\supseteq r_k)\pi_{r_k}/\lambda_{r_k}}{\lambda_{r_k} - \pi_{r_k}},
\]
where the second equality uses
$\lambda_{r_{k+1}}=\lambda_{r_k}-\pi_{r_k}$ and
$\Ind(R\supseteq r_{k+1})
=\Ind(R\supseteq r_k)-\Ind(R=r_k)$.
For $k=K$, the absent second term is interpreted as zero, and
$\omega_{r_K}=\Ind(R=r_K)/\pi_{r_K}$. These are the oracle monotone
weights of \citet{tsiatis2006semiparametric}. Thus the operator
calculation is also visible pattern by pattern: $\psi_{\mathrm{RAY}}$
coincides with the fixed-$\psi^F$ oracle estimating function, every
residual $e_r$ is zero, and Proposition~\ref{prop: lossid} gives
$\mathcal E=0$.
\end{proof}

\subsection{Proof of Theorem~\ref{thm: sufficient}(iii) (anchored families)}
\begin{proof}
\emph{Step 1: define the conditional ANOVA projections.}
Write $W=[p]\setminus r_0$. The observed-pattern family
\[
\Qcal=\{r_0\cup v:v\subseteq W\}
\]
is intersection-closed because
$(r_0\cup v)\cap(r_0\cup v')=r_0\cup(v\cap v')$.
We will prove the stronger, patternwise identity in~\eqref{eqn: nscond}.

Fix a scalar $f\in L_0^2(P)$. Conditional on $Z_{r_0}$, the variables
$\{Z_j:j\in W\}$ are mutually independent by assumption. Hence, for
$v,w\subseteq W$, the same iterated-conditioning argument used for a
product law gives
\[
\mathcal A_{r_0\cup v}\mathcal A_{r_0\cup w}
=\mathcal A_{r_0\cup(v\cap w)}.
\]
Define, for every $u\subseteq W$,
\[
\Pi_u
:=
\sum_{w\subseteq u}
(-1)^{|u|-|w|}\mathcal A_{r_0\cup w}.
\]
This is the conditional analogue of the projection $\Pi_u$ in the proof
of Lemma~\ref{lem: opident}: the core variables $Z_{r_0}$ are retained
in every conditional expectation, while M\"obius inversion is performed
only over the non-core coordinates in $W$. The preceding composition
identity shows, by the same Boolean cancellation as in
Lemma~\ref{lem: opident}, that the $\Pi_u$ are mutually orthogonal
projections and that
\[
f=\sum_{u\subseteq W}\Pi_u f,
\qquad
\Pi_\emptyset f=\mathcal A_{r_0}f,
\qquad
\mathcal A_{r_0\cup v}f=\sum_{u\subseteq v}\Pi_u f
\quad(v\subseteq W).
\]
More concretely, $\Pi_u f$ is the conditional interaction involving
exactly the non-core coordinates indexed by $u$. It is centered in each
coordinate in $u$, conditionally on $Z_{r_0}$ and the other coordinates
in $u$. The component $\Pi_\emptyset f=\Exp(f\mid Z_{r_0})$ depends only
on the core and is therefore observed under every pattern.

\emph{Step 2: identify the RAY components.}
Recall that
\[
\mathcal P_s
=\sum_{\substack{r\in\Qcal\\r\subseteq s}}
(-1)^{|s|-|r|}\mathcal A_r.
\]
If $r_0\not\subseteq s$, no pattern $r\in\Qcal$ can be contained in $s$,
so the indexing set is empty and $\mathcal P_s=0$. If
$s=r_0\cup w$ for some $w\subseteq W$, the patterns contained in $s$
are exactly $r_0\cup v$ with $v\subseteq w$. Since
$|s|-|r_0\cup v|=|w|-|v|$,
\[
\mathcal P_{r_0\cup w}
=\sum_{v\subseteq w}(-1)^{|w|-|v|}\mathcal A_{r_0\cup v}
=\Pi_w.
\]
Thus the nonzero RAY components are precisely the conditional ANOVA
projections.

\emph{Step 3: diagonalize the oracle operator.}
If $h\in\operatorname{ran}(\Pi_u)$, the conditional ANOVA decomposition
gives
\[
\mathcal A_{r_0\cup v}h=[\,u\subseteq v\,]h.
\]
Therefore the scalar by which $\Mop$ acts on this component is
\[
\sum_{v\subseteq W}\pi_{r_0\cup v}[\,u\subseteq v\,]
=\Pr(R\supseteq r_0\cup u)
=\lambda_{r_0\cup u}.
\]
Equivalently,
\[
\Mop
=\sum_{v\subseteq W}\pi_{r_0\cup v}\mathcal A_{r_0\cup v}
=\sum_{u\subseteq W}\lambda_{r_0\cup u}\Pi_u,
\qquad
\Mop^{-1}
=\sum_{u\subseteq W}\lambda_{r_0\cup u}^{-1}\Pi_u.
\]
Every pattern contains $r_0$, so $\lambda_{r_0}=1$; hence the inverse
coefficient on the core-measurable component $u=\emptyset$ is also one.
For all $u$, $\lambda_{r_0\cup u}\geq\pi_{[p]}>0$, so the displayed
inverse is well defined.

\emph{Step 4: compare the two estimating functions pattern by pattern.}
Take an observed pattern $r=r_0\cup v$. Using Step 3 and then the fact
that $\mathcal A_r$ retains exactly the components indexed by
$u\subseteq v$, the oracle pattern function is
\[
\mathcal A_r\{\Mop^{-1}(f)\}
=\sum_{u\subseteq W}\lambda_{r_0\cup u}^{-1}
\mathcal A_r\Pi_u f
=\sum_{u\subseteq v}\lambda_{r_0\cup u}^{-1}\Pi_u f.
\]
On the other hand, Steps 1--2 give the RAY pattern function
\[
\begin{split}
\sum_{s\subseteq r}\lambda_s^{-1}\mathcal P_s f
&=\sum_{u\subseteq v}
\lambda_{r_0\cup u}^{-1}\mathcal P_{r_0\cup u}f\\
&=\sum_{u\subseteq v}
\lambda_{r_0\cup u}^{-1}\Pi_u f.
\end{split}
\]
Here every subset $s\subseteq r$ that does not contain $r_0$ contributes
zero, and every remaining subset has the unique form $s=r_0\cup u$ with
$u\subseteq v$. The two pattern functions are therefore identical.
Applying this scalar identity coordinatewise to $\psi^F$ gives
$e_r=0$ for all $r\in\Qcal$. Proposition~\ref{prop: lossid} then yields
$\mathcal E=0$.
\end{proof}

\subsection{Vanishing of the structural operator, and the defect}
\label{app_sec: defect}
The operator $\Ustr = \sum_{c\in\overline{\Qcal}}d'_c\,\mathcal{A}_c$ is a fixed combination of conditional expectations over the intersection-closure $\overline{\Qcal}$, and its coefficients $d'_c$ depend only on $(\boldsymbol\pi, \Qcal)$, not on the law. Consequently $\Ustr = 0$ as an operator (for every law) as soon as it vanishes under one convenient law under which the $\{\mathcal{A}_c : c\in\overline{\Qcal}\}$ are linearly independent, since then $\sum_c d'_c\mathcal{A}_c = 0$ forces every $d'_c = 0$. We use a nondegenerate independent law for this purpose. Under such a law, the ANOVA basis of Lemma~\ref{lem: opident} diagonalizes the construction: $\Ustr$ acts on the order-$u$ component $V_u$ by the scalar $-\lambda_u\kappa_u$, where $\kappa_u = \lambda_u^{-1} - \beta_u$ and $\beta_u = \Mray|_{V_u}$, the $\{\mathcal{A}_c\}$ are linearly independent, and $\Ustr$ vanishes if and only if $\kappa_u = 0$ for every $u$.

\begin{lemma}[Vanishing of the structural operator]\label{lem: Uvanish}
If $\Qcal$ is intersection-closed, then $\Ustr = 0$.
\end{lemma}
\begin{proof}
\emph{Step 1: separate the law-free and law-dependent parts.}
By its construction in the proof of Lemma~\ref{lem:rem}, the structural
operator has the form
\[
\Ustr=\sum_{c\in\overline{\Qcal}}d'_c\mathcal A_c,
\]
where $\overline{\Qcal}$ is the intersection-closure of $\Qcal$. The
numbers $d'_c$ are obtained only from alternating signs, the pattern
probabilities $\boldsymbol\pi$, and the incidence relations among the
patterns. In particular, they do not depend on the full-data law of
$Z$. The conditional-expectation operators $\mathcal A_c$, by contrast,
do depend on that law. Our strategy is therefore to identify the
coefficients $d'_c$ under one convenient law and then retain those same
coefficients under the original law.

\emph{Step 2: evaluate the operator under an independent reference law.}
Choose any nondegenerate product law for $Z_1,\ldots,Z_p$. Under this
reference law,
\[
\mathcal A_r\mathcal A_t=\mathcal A_{r\cap t}
\qquad(r,t\subseteq[p]),
\]
as shown in the proof of Lemma~\ref{lem: opident}. Hence every operator
difference appearing in $\Vcorr_s$ is zero, so
\[
\Vcorr_s=0\quad\text{for every }s,
\qquad\text{and therefore}\qquad
\Vcorr=0.
\]
Because $\Qcal$ is intersection-closed, Lemma~\ref{lem: opident} also
gives $\Mray=\Mop^{-1}$. Lemma~\ref{lem:rem} then implies
\[
0
=\sum_s\lambda_s^{-1}\Rem_s
=\Ustr+\Vcorr.
\]
Since $\Vcorr=0$ under the reference law, $\Ustr=0$ there.

\emph{Step 3: show that this forces the law-free coefficients to
vanish.}
We give the linear-independence argument explicitly. Under the
nondegenerate product law, let $V_u$ be the ANOVA component space from
Lemma~\ref{lem: opident}. For every nonempty $u$ and every
$f_u\in V_u$,
\[
\mathcal A_cf_u=[\,u\subseteq c\,]f_u.
\]
Suppose that
$\sum_{c\in\overline{\Qcal}}a_c\mathcal A_c=0$ on $L_0^2(P)$.
Applying this relation to a nonzero $f_u\in V_u$ gives
\[
\sum_{\substack{c\in\overline{\Qcal}\\c\supseteq u}}a_c=0.
\]
Take $u=c$ successively from the largest sets in
$\overline{\Qcal}$ down to the smallest nonempty sets. For a maximal
$c$, the displayed sum contains only $a_c$, so $a_c=0$. After all
strict supersets of a given $c$ have been eliminated, the same equation
with $u=c$ gives $a_c=0$. Thus every coefficient attached to a nonempty
$c$ is zero. If $\emptyset\in\overline{\Qcal}$, its coefficient is
irrelevant on $L_0^2(P)$ because
$\mathcal A_\emptyset f=\Exp(f)=0$.

Applying this argument to the relation
$\Ustr=\sum_cd'_c\mathcal A_c=0$ under the reference law shows that all
operator-relevant coefficients $d'_c$ vanish. Because these coefficients
are law-free, they remain zero under the original, possibly dependent,
full-data law. Hence $\Ustr=0$ without an independence assumption.
\end{proof}

Non-intersection-closed designs can leave a nonzero structural discrepancy. From the proof of Lemma~\ref{lem: opident}, $\beta_u = \sum_{t\in\Qcal:\, t\supseteq u}h(t)$ with $h(s) = \sum_{s'\supseteq s}(-1)^{|s'|-|s|}\lambda_{s'}^{-1}$, while M\"obius inversion over all supersets gives $\sum_{s\supseteq u}h(s) = \lambda_u^{-1}$. Subtracting yields the \emph{defect coefficient}
\begin{equation}\label{eqn: defect}
\kappa_u = \sum_{s\supseteq u,\, s\notin\Qcal}h(s),
\end{equation}
a signed sum of $\lambda^{-1}$ over the unobserved supersets of $u$, computable directly from $(\boldsymbol\pi, \Qcal)$. For example, $\Qcal = \{\{1,2\}, \{1,3\}, \{1,2,3\}\}$ is not closed because $\{1\} = \{1,2\}\cap\{1,3\}$ is missing, and $\kappa_{\{1\}} \neq 0$, whereas $\Qcal = \{\{1\}, \{2\}, \{1,2,3\}\}$ is closed and every $\kappa_u = 0$.

\subsection{Conditional maximal correlation and proof of Theorem~\ref{thm:main}}

We first show that the difference between the pairwise composition $\mathcal{A}_r\mathcal{A}_t$ and the intersection-based reference projection $\mathcal{A}_{r\cap t}$ has norm equal to a conditional maximal correlation. This identity is the source of the correlation factor.

\begin{lemma}[Conditional maximal correlation identity]\label{lem:maxcorr}
For $r, t\in\Qcal$, write $E = \mathcal{A}_{r\cap t}$, $P = \mathcal{A}_r$, $Q = \mathcal{A}_t$, and let $\mathcal K_r = \operatorname{ran}(P-E) = \{f \in L^2_0(P) : f \text{ is } \sigma(Z_r)\text{-measurable},\ \Exp[f\mid Z_{r\cap t}] = 0\}$, similarly $\mathcal K_t$. Then
\[
\|\mathcal{A}_r\mathcal{A}_t - \mathcal{A}_{r\cap t}\| = \sup\big\{|\Exp[fg]| : f\in\mathcal K_r,\ g\in\mathcal K_t,\ \|f\| = \|g\| = 1\big\} = \rstar(Z_{r\setminus t}; Z_{t\setminus r}\mid Z_{r\cap t}),
\]
where $\rstar$ is the global $L^2(P)$ Hirschfeld--Gebelein--R\'enyi maximal correlation between $Z_{r\setminus t}$ and $Z_{t\setminus r}$ given $Z_{r\cap t}$, that is, the supremum of $|\Corr(f(Z_r), g(Z_t))|$ over square-integrable $f, g$ that are mean-zero given $Z_{r\cap t}$ (not a fiberwise or essential-supremum version). When $r\subseteq t$ or $t\subseteq r$, one of $\mathcal K_r, \mathcal K_t$ is $\{0\}$, both sides are zero, and we set $\rstar := 0$.
\end{lemma}

\begin{proof}
\emph{Step 1: identify the two residual projection spaces.}
Work in the Hilbert space $L_0^2(P)$ with inner product
$\langle f,g\rangle=\Exp(fg)$. Conditional expectation is an orthogonal
projection, so $E$, $P$, and $Q$ are self-adjoint and idempotent.
Because $r\cap t\subseteq r$, the tower property gives
\[
EPf=PEf=Ef,
\qquad\text{that is,}\qquad EP=PE=E.
\]
Similarly, $EQ=QE=E$. It follows that $P-E$ is self-adjoint and
\[
(P-E)^2
=P^2-PE-EP+E^2
=P-E.
\]
Thus $P-E$ is an orthogonal projection. Its range is exactly
$\mathcal K_r$. To see this, if $h=(P-E)f$, then $Ph=h$, so $h$ is
$Z_r$-measurable, and
\[
Eh=E(P-E)f=(EP-E^2)f=0.
\]
Conversely, if $h$ is $Z_r$-measurable and $\Exp(h\mid Z_{r\cap t})=0$,
then $Ph=h$ and $Eh=0$, so $(P-E)h=h$. The same reasoning shows that
$Q-E$ is the orthogonal projection onto $\mathcal K_t$.

\emph{Step 2: factor the operator difference.}
Using the identities from Step 1,
\[
\begin{split}
(P-E)(Q-E)
&=PQ-PE-EQ+E^2\\
&=PQ-E-E+E\\
&=PQ-E.
\end{split}
\]
Therefore
\[
\mathcal A_r\mathcal A_t-\mathcal A_{r\cap t}
=(P-E)(Q-E)
\]
is the product of the orthogonal projections onto $\mathcal K_r$ and
$\mathcal K_t$.

\emph{Step 3: relate the norm of a product of projections to inner
products between their ranges.}
We use the following elementary Hilbert-space identity. If $P_1$ and
$P_2$ are orthogonal projections onto closed subspaces
$\mathcal K_1$ and $\mathcal K_2$, respectively, then
\[
\|P_1P_2\|
=\sup_{\substack{f\in\mathcal K_1,\ g\in\mathcal K_2\\
\|f\|=\|g\|=1}}
|\langle f,g\rangle|.
\]
For the upper bound, take any unit vector $u$. If $P_2u=0$, then
$P_1P_2u=0$. Otherwise, put
$g=P_2u/\|P_2u\|\in\mathcal K_2$. Since an orthogonal projection satisfies
\[
\|P_1g\|
=\sup_{\substack{f\in\mathcal K_1\\\|f\|=1}}
|\langle f,g\rangle|,
\]
we obtain
\[
\|P_1P_2u\|
=\|P_2u\|\,\|P_1g\|
\leq
\sup_{\substack{f\in\mathcal K_1,\ g\in\mathcal K_2\\
\|f\|=\|g\|=1}}
|\langle f,g\rangle|.
\]
Taking the supremum over unit $u$ proves ``$\leq$''. Conversely, for
unit $f\in\mathcal K_1$ and $g\in\mathcal K_2$, self-adjointness and
$P_1f=f$, $P_2g=g$ give
\[
|\langle f,g\rangle|
=|\langle f,P_1P_2g\rangle|
\leq\|P_1P_2\|.
\]
Taking the supremum over $f$ and $g$ proves the reverse inequality.
Applying this identity with $P_1=P-E$ and $P_2=Q-E$ proves the first
equality in the lemma.

\emph{Step 4: identify the supremum as conditional maximal
correlation.}
A unit vector $f\in\mathcal K_r$ is precisely a square-integrable
function of $Z_r=(Z_{r\cap t},Z_{r\setminus t})$ satisfying
$\Exp(f\mid Z_{r\cap t})=0$; similarly for $g\in\mathcal K_t$. Such
functions have unconditional mean zero, and their unit $L^2$ norms give
unit variances. Hence
\[
\Exp(fg)=\Corr(f,g).
\]
The supremum in Step 3 is therefore exactly the global conditional HGR
maximal correlation
\[
\rstar(Z_{r\setminus t};Z_{t\setminus r}\mid Z_{r\cap t})
\]
defined in the statement. If $r\subseteq t$, then $P=E$ and
$\mathcal K_r=\{0\}$; the case $t\subseteq r$ is symmetric. The operator
difference and the stipulated maximal correlation are then both zero.
\end{proof}

The two-factor bound on the operator discrepancy $\Delta$ is the ingredient the loss bound reuses.
\begin{theorem}[Two-factor bound on $\Delta$]\label{thm:main}
For any $\Qcal$ and any $\psi^F$,
\begin{equation}\label{eqn: twofactor}
\Delta(\psi^F)\ \le\ \pi_{[p]}^{-1}\big(\|\Ustr\psi^F\| + C(\boldsymbol\pi,\Qcal)\,\rstar_{\max}\,\|\psi^F\|\big)^2,
\qquad
C(\boldsymbol\pi,\Qcal) = \sum_{s}\lambda_s^{-1}\Big(\sum_{r:\,s\not\subseteq r}\pi_r\Big)\,\big|\{t\in\Qcal:t\subseteq s\}\big|.
\end{equation}
\end{theorem}

\begin{proof}[Proof of Theorem~\ref{thm:main}]
\emph{Step 1: bound one correlation remainder.}
For a fixed $s\subseteq[p]$, recall that
\[
\Vcorr_s
=\sum_{\substack{r\in\Qcal\\s\not\subseteq r}}\pi_r
\sum_{\substack{t\in\Qcal\\t\subseteq s}}
(-1)^{|s|-|t|}
\big(\mathcal A_r\mathcal A_t-\mathcal A_{r\cap t}\big).
\]
The operator norm is subadditive and
$|(-1)^{|s|-|t|}|=1$. Thus
\[
\begin{split}
\|\Vcorr_s\|
&\leq
\sum_{\substack{r\in\Qcal\\s\not\subseteq r}}\pi_r
\sum_{\substack{t\in\Qcal\\t\subseteq s}}
\|\mathcal A_r\mathcal A_t-\mathcal A_{r\cap t}\|\\
&\leq
\left(\sum_{\substack{r\in\Qcal\\s\not\subseteq r}}\pi_r\right)
\big|\{t\in\Qcal:t\subseteq s\}\big|\,
\rstar_{\max}.
\end{split}
\]
The second inequality is Lemma~\ref{lem:maxcorr}: every pairwise
difference has norm equal to the corresponding conditional maximal
correlation, which is no larger than $\rstar_{\max}$. Notice why the
decomposition into pairwise differences is useful. Each displayed
operator difference is controlled by one maximal-correlation number,
whereas the signed operator $\mathcal P_s$ appearing in $\Rem_s$ need
not itself be a contraction.

\emph{Step 2: sum the bounds over $s$.}
By definition,
\[
\Vcorr=\sum_{s\subseteq[p]}\lambda_s^{-1}\Vcorr_s.
\]
All weights are finite and positive because
$\lambda_s=\Pr(R\supseteq s)\geq\pi_{[p]}>0$. The triangle inequality
and Step 1 give
\[
\begin{split}
\|\Vcorr\|
&\leq\sum_s\lambda_s^{-1}\|\Vcorr_s\|\\
&\leq
\left\{
\sum_s\lambda_s^{-1}
\left(\sum_{\substack{r\in\Qcal\\s\not\subseteq r}}\pi_r\right)
\big|\{t\in\Qcal:t\subseteq s\}\big|
\right\}\rstar_{\max}\\
&=C(\boldsymbol\pi,\Qcal)\rstar_{\max}.
\end{split}
\]
Consequently, applying this operator to the vector-valued
$\psi^F=(\psi^F_1,\ldots,\psi^F_d)^\top$ coordinatewise yields
\[
\|\Vcorr\psi^F\|
\leq\|\Vcorr\|\,\|\psi^F\|
\leq C(\boldsymbol\pi,\Qcal)\rstar_{\max}\|\psi^F\|.
\]

\emph{Step 3: insert the structural--correlation split into the defect
bound.}
Lemma~\ref{lem:rem} states that
\[
\Delta(\psi^F)
\leq
\pi_{[p]}^{-1}
\left\|
\sum_s\lambda_s^{-1}\Rem_s(\psi^F)
\right\|^2.
\]
Since
$\sum_s\lambda_s^{-1}\Rem_s=\Ustr+\Vcorr$, the ordinary triangle
inequality followed by Step 2 gives
\[
\begin{split}
\Delta(\psi^F)
&\leq
\pi_{[p]}^{-1}
\|\Ustr\psi^F+\Vcorr\psi^F\|^2\\
&\leq
\pi_{[p]}^{-1}
\big\{\|\Ustr\psi^F\|+\|\Vcorr\psi^F\|\big\}^2\\
&\leq
\pi_{[p]}^{-1}
\big\{\|\Ustr\psi^F\|
+C(\boldsymbol\pi,\Qcal)\rstar_{\max}\|\psi^F\|\big\}^2.
\end{split}
\]
This is~\eqref{eqn: twofactor}. The two quantities inside braces are
nonnegative, so this upper bound is zero exactly when both the structural
term and the displayed correlation term are zero.
\end{proof}

\subsection{Proof of Theorem~\ref{thm: Ebound}}
\begin{proof}
\emph{Step 1: express the loss as the norm of two error terms.}
For a random vector $X$, write
$\|X\|_{L^2}^2:=\Exp\|X\|^2$. Proposition~\ref{prop: lossid} gives
\[
\operatorname{tr}(A\mathcal E A^\top)
=\sum_{r\in\Qcal}\pi_r\Exp\|e_r\|^2
=\|\psi_{\mathrm{opt}}-\psi_{\mathrm{RAY}}\|_{L^2}^2.
\]
Theorem~\ref{thm: ns} decomposes each pattern residual as
$e_r=\mathcal A_r(g)+b_r$. Therefore
\[
\psi_{\mathrm{opt}}-\psi_{\mathrm{RAY}}
=\underbrace{\sum_{r\in\Qcal}\Ind(R=r)\mathcal A_r(g)}
_{\Lop(g)}
+\underbrace{\sum_{r\in\Qcal}\Ind(R=r)b_r}_{\mathrm{II}}.
\]
The $L^2$ triangle inequality immediately yields
\[
\sqrt{\operatorname{tr}(A\mathcal E A^\top)}
\leq\|\Lop(g)\|_{L^2}+\|\mathrm{II}\|_{L^2}.
\]

\emph{Step 2: identify the first norm with $\sqrt{\Delta(\psi^F)}$.}
Write $g=(g_1,\ldots,g_d)^\top$. Each $\mathcal A_rg_j$ has mean zero
because $g_j\in L_0^2(P)$. Under MCAR, $R$ is independent of $Z$, and
the indicators $\Ind(R=r)$ are mutually exclusive. Hence, for
$j,k\in[d]$,
\[
\begin{split}
\Exp\{[\Lop(g)]_j[\Lop(g)]_k\}
&=\sum_{r\in\Qcal}
\pi_r\langle\mathcal A_rg_j,\mathcal A_rg_k\rangle\\
&=\sum_{r\in\Qcal}
\pi_r\langle g_j,\mathcal A_rg_k\rangle\\
&=\langle g_j,\Mop g_k\rangle.
\end{split}
\]
The second equality uses that $\mathcal A_r$ is a self-adjoint
idempotent projection:
$\langle\mathcal A_rg_j,\mathcal A_rg_k\rangle
=\langle g_j,\mathcal A_rg_k\rangle$.
Taking the trace of the covariance matrix gives
\[
\|\Lop(g)\|_{L^2}^2
=\operatorname{tr}\Var\{\Lop(g)\}
=\sum_{j=1}^d\langle g_j,\Mop g_j\rangle
=\Delta(\psi^F),
\]
by the definition of $\Delta(\psi^F)$. Similarly, mutual exclusivity of
the pattern indicators gives
\[
\|\mathrm{II}\|_{L^2}^2
=\sum_{r\in\Qcal}\pi_r\|b_r\|_{L^2}^2.
\]
Thus Step 1 becomes
\[
\sqrt{\operatorname{tr}(A\mathcal E A^\top)}
\leq\sqrt{\Delta(\psi^F)}+\|\mathrm{II}\|_{L^2}.
\]

\emph{Step 3: split each $b_r$ into structural and correlation parts.}
From Theorem~\ref{thm: ns},
\[
\begin{split}
b_r
&=\sum_{s\not\subseteq r}\lambda_s^{-1}
\sum_{\substack{t\in\Qcal\\t\subseteq s}}
(-1)^{|s|-|t|}\mathcal A_r\mathcal A_t\psi^F\\
&=:b_r^{\mathrm{str}}+b_r^{\mathrm{cor}},
\end{split}
\]
where, explicitly,
\[
\begin{split}
b_r^{\mathrm{str}}
&:=
\sum_{s\not\subseteq r}\lambda_s^{-1}
\sum_{\substack{t\in\Qcal\\t\subseteq s}}
(-1)^{|s|-|t|}\mathcal A_{r\cap t}\psi^F,\\
b_r^{\mathrm{cor}}
&:=
\sum_{s\not\subseteq r}\lambda_s^{-1}
\sum_{\substack{t\in\Qcal\\t\subseteq s}}
(-1)^{|s|-|t|}
\big(\mathcal A_r\mathcal A_t-\mathcal A_{r\cap t}\big)\psi^F.
\end{split}
\]
Lemma~\ref{lem:maxcorr} and the triangle inequality imply
\[
\begin{split}
\|b_r^{\mathrm{cor}}\|_{L^2}
&\leq
\sum_{s\not\subseteq r}\lambda_s^{-1}
\sum_{\substack{t\in\Qcal\\t\subseteq s}}
\|\mathcal A_r\mathcal A_t-\mathcal A_{r\cap t}\|\,
\|\psi^F\|\\
&\leq
C_r\rstar_{\max}\|\psi^F\|,
\end{split}
\]
where
\[
C_r
:=\sum_{s\not\subseteq r}\lambda_s^{-1}
\big|\{t\in\Qcal:t\subseteq s\}\big|.
\]

For completeness, the coefficients in $b_r^{\mathrm{str}}$ depend only
on $(\boldsymbol\pi,\Qcal)$: after like conditional-expectation
operators are collected, one can write
$b_r^{\mathrm{str}}=\sum_u\kappa^b_{r,u}\mathcal A_u\psi^F$ with
law-free $\kappa^b_{r,u}$. If $\Qcal$ is intersection-closed, evaluate
this relation under the same nondegenerate product law used in
Lemma~\ref{lem: Uvanish}. There,
$\mathcal A_r\mathcal A_t=\mathcal A_{r\cap t}$, so
$b_r^{\mathrm{cor}}=0$; Theorem~\ref{thm: sufficient}(i), equivalently
the calculation $c_u(r)=\lambda_u^{-1}$, gives $b_r=0$. Hence
$b_r^{\mathrm{str}}=0$ under that law. The linear-independence argument
in the proof of Lemma~\ref{lem: Uvanish} forces every
$\kappa^b_{r,u}=0$, and their law-free nature then gives
$b_r^{\mathrm{str}}=0$ for every full-data law.

\emph{Step 4: aggregate over patterns and combine the two bounds.}
Minkowski's inequality for the weighted sequence
$\{\sqrt{\pi_r}b_r:r\in\Qcal\}$ yields
\[
\begin{split}
\|\mathrm{II}\|_{L^2}
&=\left\{\sum_r\pi_r
\|b_r^{\mathrm{str}}+b_r^{\mathrm{cor}}\|_{L^2}^2\right\}^{1/2}\\
&\leq
\left\{\sum_r\pi_r\|b_r^{\mathrm{str}}\|_{L^2}^2\right\}^{1/2}
+\left\{\sum_r\pi_r\|b_r^{\mathrm{cor}}\|_{L^2}^2\right\}^{1/2}\\
&\leq
\left\{\sum_r\pi_r\|b_r^{\mathrm{str}}\|_{L^2}^2\right\}^{1/2}
+\left(\sum_r\pi_rC_r^2\right)^{1/2}
\rstar_{\max}\|\psi^F\|.
\end{split}
\]
Theorem~\ref{thm:main} also gives
\[
\sqrt{\Delta(\psi^F)}
\leq
\pi_{[p]}^{-1/2}
\{\|\Ustr\psi^F\|
+C\rstar_{\max}\|\psi^F\|\}.
\]
Substitution into the last display of Step 2 gives
~\eqref{eqn: Etwofactor} with
\[
\mathsf S(\psi^F) = \pi_{[p]}^{-1/2}\|\Ustr\psi^F\| + \Big(\sum_r\pi_r\|b_r^{\mathrm{str}}\|_{L^2}^2\Big)^{1/2}, \qquad \mathsf C(\boldsymbol\pi,\Qcal) = \pi_{[p]}^{-1/2}C + \Big(\sum_r\pi_r C_r^2\Big)^{1/2}.
\]
Lemma~\ref{lem: Uvanish} and Step 3 show that both structural terms
vanish when $\Qcal$ is intersection-closed, so
$\mathsf S(\psi^F)=0$ in that case. Finally, $C$ and all $C_r$ are
finite because the sums are over finite subsets of $[p]$ and
$\lambda_s\geq\pi_{[p]}>0$. The result is a triangle-inequality bound
and need not be tight.
\end{proof}

\subsection{The efficiency gap in the two-modality Gaussian design}
\label{app_sec: gaussian_loss}
We compute $\operatorname{tr}(A\mathcal E A^\top)$ for Example~\ref{ex: two-modality-gaussian} at $\theta^\star=0$, so $\psi^F=Z_2$. Here $\Qcal = \{\{1\},\{2\},\{1,2\}\}$ and $(Z_1,Z_2)$ is standard bivariate normal with correlation $\rho$. On the space of mean-zero linear functions $a_1 Z_1 + a_2 Z_2$, identified with coefficient vectors $(a_1, a_2)^\top$, the conditioning maps act linearly; using $\Exp[Z_2\mid Z_1] = \rho Z_1$ and $\Exp[Z_1\mid Z_2] = \rho Z_2$,
\[
\mathcal{A}_{\{1\}} = \begin{pmatrix}1 & \rho\\ 0 & 0\end{pmatrix}, \qquad \mathcal{A}_{\{2\}} = \begin{pmatrix}0 & 0\\ \rho & 1\end{pmatrix}, \qquad \mathcal{A}_{\{1,2\}} = I.
\]
Hence $\Mop = \pi_1\mathcal{A}_{\{1\}} + \pi_2\mathcal{A}_{\{2\}} + \pi_{12}I = \big(\begin{smallmatrix}\lambda_1 & \pi_1\rho\\ \pi_2\rho & \lambda_2\end{smallmatrix}\big)$ with $\det\Mop = D := \lambda_1\lambda_2 - \pi_1\pi_2\rho^2$; since $\lambda_1\lambda_2 - \pi_1\pi_2 = \pi_{12}$ (using $\pi_1+\pi_2+\pi_{12}=1$), $D = \pi_{12} + \pi_1\pi_2(1-\rho^2) > 0$. Applying the inverse to $\psi^F\leftrightarrow(0,1)^\top$,
\[
\Mop^{-1}\psi^F = \tfrac{1}{D}\big(-\pi_1\rho\,Z_1 + \lambda_1 Z_2\big).
\]
The RAY surrogate $\Mray = \lambda_1^{-1}\mathcal{P}_{\{1\}} + \lambda_2^{-1}\mathcal{P}_{\{2\}} + \pi_{12}^{-1}\mathcal{P}_{\{1,2\}}$, with $\mathcal{P}_{\{1\}} = \mathcal{A}_{\{1\}}$, $\mathcal{P}_{\{2\}} = \mathcal{A}_{\{2\}}$ and $\mathcal{P}_{\{1,2\}} = I - \mathcal{A}_{\{1\}} - \mathcal{A}_{\{2\}}$, gives
\[
\Mray\psi^F = \rho\Big(\tfrac{1}{\lambda_1} - \tfrac{1}{\pi_{12}}\Big)Z_1 + \tfrac{1}{\lambda_2}Z_2.
\]
The per-pattern residuals $e_r = \mathcal{A}_r(\Mop^{-1}\psi^F) - \sum_{s\subseteq r}\lambda_s^{-1}\mathcal{P}_s\psi^F$ are, after simplification,
\[
e_{\{1\}} = \frac{\pi_2(\pi_1\rho^2 - \lambda_1)}{D\lambda_1}\,\rho\,Z_1, \qquad e_{\{2\}} = -\frac{\pi_1\pi_{12}}{D\lambda_2}\,\rho^2 Z_2, \qquad e_{\{1,2\}} = (\Mop^{-1}-\Mray)\psi^F,
\]
the last with $Z_1$-coefficient $\pi_1\pi_2\rho/(\lambda_1\lambda_2\pi_{12}) + O(\rho^3)$ and $Z_2$-coefficient $\pi_1\pi_2\rho^2/(D\lambda_2)$. By Proposition~\ref{prop: lossid}, $\operatorname{tr}(A\mathcal E A^\top) = \sum_r\pi_r\Exp[e_r^2]$, where $\Exp[(a_1 Z_1 + a_2 Z_2)^2] = a_1^2 + a_2^2 + 2\rho a_1 a_2$. To leading order,
\[
\pi_1\Exp[e_{\{1\}}^2] = \frac{\pi_1\pi_2^2}{\lambda_1^2\lambda_2^2}\rho^2 + O(\rho^4), \qquad \pi_2\Exp[e_{\{2\}}^2] = O(\rho^4), \qquad \pi_{12}\Exp[e_{\{1,2\}}^2] = \frac{\pi_1^2\pi_2^2}{\lambda_1^2\lambda_2^2\pi_{12}}\rho^2 + O(\rho^4),
\]
and summing, with $\pi_1 + \pi_{12} = \lambda_1$,
\[
\operatorname{tr}(A\mathcal E A^\top) = \frac{\pi_1\pi_2^2}{\lambda_1^2\lambda_2^2}\Big(1 + \frac{\pi_1}{\pi_{12}}\Big)\rho^2 + O(\rho^4) = \frac{\pi_1\pi_2^2}{\lambda_1\lambda_2^2\pi_{12}}\,\rho^2 + O(\rho^4),
\]
the leading term quoted in Example~\ref{ex: two-modality-gaussian}. Because $\Qcal$ is intersection-closed, the structural factor vanishes. The local loss is therefore of correlation type; its $\rho^2$ coefficient contains the factor $\pi_{12}^{-1}$ and can become large when complete cases are scarce.

\section{Proofs for the MAR development}
\label{app_sec: mar_proofs}

\subsection{Proof of Proposition~\ref{prop:mar-direct-mean-zero}}
\begin{proof}
\emph{Step 1: compute the conditional mean of each normalized tail
indicator.}
Fix an incomplete pattern $r\neq[p]$. For every $s$ in the Boolean
interval $r\subseteq s\subseteq[p]$, Assumption~\ref{assumption: mar}
ensures that the denominator is bounded away from zero. More explicitly,
because the complete pattern $[p]$ contains every $s$,
\[
\Lambda_s(Z)
=\Pr(R\supseteq s\mid Z)
\geq\Pr(R=[p]\mid Z)
=\pi_{[p]}(Z)
\geq c>0.
\]
By the definition of $\Lambda_s(Z)$,
\[
\Exp\!\left[
\frac{\Ind(R\supseteq s)}{\Lambda_s(Z)}
\,\middle|\,Z
\right]
=
\frac{\Pr(R\supseteq s\mid Z)}{\Lambda_s(Z)}
=1.
\]

\emph{Step 2: apply Boolean cancellation to the direct weight.}
The direct RAY weight is
\[
\omega_r^{\mathrm{dir}}
=\sum_{s:\,r\subseteq s\subseteq[p]}
(-1)^{|s|-|r|}
\frac{\Ind(R\supseteq s)}{\Lambda_s(Z)}.
\]
This is a finite sum, so Step 1 can be applied term by term:
\[
\Exp[\omega_r^{\mathrm{dir}}\mid Z]
=
\sum_{s:\,r\subseteq s\subseteq[p]}(-1)^{|s|-|r|}
=
(1-1)^{|[p]\setminus r|}
=0.
\]
For clarity, write each $s$ uniquely as $s=r\cup v$ with
$v\subseteq[p]\setminus r$. The middle sum then becomes
$\sum_{v\subseteq[p]\setminus r}(-1)^{|v|}
=(1-1)^{|[p]\setminus r|}$. It is zero because $r\neq[p]$, so the
complement is nonempty.

\emph{Step 3: deduce the mean-zero estimating equation.}
The function $F_r(Z_r;\theta)$ is measurable with respect to
$\sigma(Z)$ and does not involve $R$. Therefore the tower property and
Step 2 give
\[
\begin{split}
\Exp\{\omega_r^{\mathrm{dir}}F_r(Z_r;\theta)\}
&=\Exp\!\left[
F_r(Z_r;\theta)
\Exp(\omega_r^{\mathrm{dir}}\mid Z)
\right]\\
&=0.
\end{split}
\]
This proves both assertions.
\end{proof}

\subsection{Proof of Proposition~\ref{prop:mean-zero}}
\begin{proof}
\emph{Step 1: verify positivity and the conditional mean of one
summand.}
Fix an incomplete pattern $r\neq[p]$. Since
$\Lambda_s(Z)\geq c$ as in the preceding proof,
\[
\lambda_s^{\mathrm{proj}}(Z_s)
=\Exp\{\Lambda_s(Z)\mid Z_s\}
\geq c
\]
for every $s$, so all projected weights are well defined. If
$r\subseteq s$, then
$\sigma(Z_r)\subseteq\sigma(Z_s)$. We may therefore condition first on
$Z_s$ and then on $Z_r$. Using
$\lambda_s^{\mathrm{proj}}(Z_s)
=\Pr(R\supseteq s\mid Z_s)$,
\[
\begin{split}
\Exp\!\left[
\frac{\Ind(R\supseteq s)}
{\lambda_s^{\mathrm{proj}}(Z_s)}
\,\middle|\,Z_r
\right]
&=
\Exp\!\left[
\Exp\!\left\{
\frac{\Ind(R\supseteq s)}
{\lambda_s^{\mathrm{proj}}(Z_s)}
\;\middle|\;Z_s
\right\}
\,\middle|\,Z_r
\right]\\
&=
\Exp\!\left[
\frac{\Pr(R\supseteq s\mid Z_s)}
{\Pr(R\supseteq s\mid Z_s)}
\,\middle|\,Z_r
\right]
=1.
\end{split}
\]

\emph{Step 2: cancel the projected weight over the Boolean interval.}
The preceding calculation shows that every normalized indicator has
conditional mean one given $Z_r$. Hence, by linearity,
\[
\Exp[\omega_r^{\mathrm{proj}}\mid Z_r]
=
\sum_{s:\,r\subseteq s\subseteq[p]}(-1)^{|s|-|r|}
=
(1-1)^{|[p]\setminus r|}
=0.
\]
As before, the change of variables $s=r\cup v$ turns the sum into
$\sum_{v\subseteq[p]\setminus r}(-1)^{|v|}$. This is zero because the
incomplete pattern $r$ has at least one missing coordinate.

\emph{Step 3: multiply by the pattern-specific imputation function.}
Since $F_r(Z_r;\theta)$ is $\sigma(Z_r)$-measurable, the tower property
gives
\[
\begin{split}
\Exp\{\omega_r^{\mathrm{proj}}F_r(Z_r;\theta)\}
&=\Exp\!\left[
F_r(Z_r;\theta)
\Exp(\omega_r^{\mathrm{proj}}\mid Z_r)
\right]\\
&=0.
\end{split}
\]
This proves the stated consequence.
\end{proof}

\subsection{Projection identity and first-order sensitivity}
\label{app_sec: first-order}
\emph{Step 1: verify the centering of the projection fluctuation.}
Recall that
$\delta_s(Z)=\Lambda_s(Z)-\lambda_s^{\mathrm{proj}}(Z_s)$ and
$\lambda_s^{\mathrm{proj}}(Z_s)
=\Exp\{\Lambda_s(Z)\mid Z_s\}$. Consequently,
\[
\Exp[\delta_s(Z)\mid Z_s]
=
\Exp[\Lambda_s(Z)\mid Z_s]
-\lambda_s^{\mathrm{proj}}(Z_s)
=0.
\]
Moreover, $\delta_s(Z)=0$ almost surely if and only if
$\Lambda_s(Z)$ equals its conditional expectation given $Z_s$, which
is equivalent to $\Lambda_s(Z)$ being $\sigma(Z_s)$-measurable.

\emph{Step 2: derive the inverse-probability ratio representation.}
At the true propensity parameter $\gamma_0$, define
\[
W:=\frac{\Ind(R=[p])}{\pi_{[p]}(Z;\gamma_0)}.
\]
MAR gives
$\Exp\{\Ind(R=[p])\mid Z\}=\pi_{[p]}(Z;\gamma_0)$, and hence
\[
\Exp(W\mid Z)
=\frac{\Exp\{\Ind(R=[p])\mid Z\}}
{\pi_{[p]}(Z;\gamma_0)}
=1.
\]
Applying the tower property,
\[
\Exp[W\Lambda_s(Z;\gamma_0)\mid Z_s]
=
\Exp\!\left[
\Lambda_s(Z;\gamma_0)\Exp(W\mid Z)
\,\middle|\,Z_s
\right]
=
\lambda_s^{\mathrm{proj}}(Z_s),
\]
and, similarly,
\[
\Exp(W\mid Z_s)=1,
\]
which proves the first ratio in~\eqref{eqn: lambda_proj_ratio}.

To obtain the complete-case form, let $H(Z)$ be any integrable random
variable. Conditioning on the complete-case event gives
\[
\begin{split}
\Exp\{WH(Z)\mid Z_s\}
&=
\Exp\!\left[
\frac{\Ind(R=[p])H(Z)}
{\pi_{[p]}(Z;\gamma_0)}
\,\middle|\,Z_s
\right]\\
&=
\Pr(R=[p]\mid Z_s)
\Exp\!\left[
\frac{H(Z)}{\pi_{[p]}(Z;\gamma_0)}
\,\middle|\,Z_s,R=[p]
\right].
\end{split}
\]
Use this identity first with $H=\Lambda_s(\,\cdot\,;\gamma_0)$ and then
with $H=1$. In the ratio of the resulting conditional expectations, the
common factor $\Pr(R=[p]\mid Z_s)$ cancels, yielding the second
representation in~\eqref{eqn: lambda_proj_ratio}.

\emph{Step 3: exact plug-in drift and the induced population-root shift.}

Let
\[
\mathcal S^\circ
:=
\{s\subseteq[p]:\varnothing\neq s\subsetneq[p]\}
\]
denote the collection of proper, nonempty modality subsets. Write
\[
\lambda_0
=
\{\lambda_{0,s}:s\in\mathcal S^\circ\},
\qquad
\lambda_{0,s}(Z_s)
=
\lambda_s^{\mathrm{proj}}(Z_s),
\]
for the true projected tail propensities.

We study the effect of replacing $\lambda_0$ by a collection of positive
working functions
\[
\lambda
=
\{\lambda_s:s\in\mathcal S^\circ\}.
\]
Throughout this calculation, the observed-data law $P$, the complete-case
propensity $\pi_{[p]}$, the tuning coefficients $\alpha$, and the imputation
functions $\{F_r\}$ are held fixed. We also keep
\[
\lambda_\varnothing=1,
\qquad
\lambda_{[p]}=\pi_{[p]},
\]
fixed. Thus, only the proper projected tails indexed by
$\mathcal S^\circ$ are treated as working arguments.

Let
\[
\Psi(\theta,\lambda)
:=
E\!\left[
\psi_\alpha^{\mathrm{proj}}(R,Z_R,\theta,\lambda)
\right]
\]
denote the corresponding population moment. For each
$s\in\mathcal S^\circ$, define
\[
G_s(Z_s;\theta)
:=
\sum_{\substack{
r\in\mathcal Q\setminus\{[p]\}\\
r\subseteq s
}}
\alpha_r(-1)^{|s|-|r|}
F_r(Z_r;\theta).
\]
Because $r\subseteq s$, every summand in $G_s$ is
$\sigma(Z_s)$-measurable.

Let $\widetilde\lambda$ be another positive collection of working
projected tails. Reordering the two finite sums in the augmentation part of
the estimating equation gives
\begin{align}
&\Psi(\theta^\star,\widetilde\lambda)
-
\Psi(\theta^\star,\lambda_0)
\nonumber\\
&\quad=
\sum_{s\in\mathcal S^\circ}
E\left[
I(R\supseteq s)
G_s(Z_s;\theta^\star)
\left\{
\frac{1}{\widetilde\lambda_s(Z_s)}
-
\frac{1}{\lambda_{0,s}(Z_s)}
\right\}
\right].
\label{eq:plug-moment-difference}
\end{align}
Since
\[
E\{I(R\supseteq s)\mid Z_s\}
=
\lambda_{0,s}(Z_s),
\]
and $G_s(Z_s;\theta^\star)$ and both denominators are
$\sigma(Z_s)$-measurable, \eqref{eq:plug-moment-difference} becomes
\begin{align}
&\Psi(\theta^\star,\widetilde\lambda)
-
\Psi(\theta^\star,\lambda_0)
\nonumber\\
&\quad=
-
\sum_{s\in\mathcal S^\circ}
E\left[
\frac{
\widetilde\lambda_s(Z_s)-\lambda_{0,s}(Z_s)
}{
\widetilde\lambda_s(Z_s)
}
G_s(Z_s;\theta^\star)
\right].
\label{eq:exact-plug-drift}
\end{align}
Thus \eqref{eq:exact-plug-drift} is an exact expression for the population
drift caused by separately plugging in working projected tails.

To separate its first- and higher-order parts, write
\[
\delta_s
:=
\widetilde\lambda_s-\lambda_{0,s}.
\]
The identity
\[
-\frac{\delta_s}{\lambda_{0,s}+\delta_s}
=
-\frac{\delta_s}{\lambda_{0,s}}
+
\frac{\delta_s^2}
{\lambda_{0,s}(\lambda_{0,s}+\delta_s)}
\]
then gives
\begin{align}
&\Psi(\theta^\star,\widetilde\lambda)
-
\Psi(\theta^\star,\lambda_0)
\nonumber\\
&\quad=
-
\sum_{s\in\mathcal S^\circ}
E\left[
\frac{\delta_s(Z_s)}
{\lambda_{0,s}(Z_s)}
G_s(Z_s;\theta^\star)
\right]
\nonumber\\
&\qquad+
\sum_{s\in\mathcal S^\circ}
E\left[
\frac{\delta_s(Z_s)^2}
{\lambda_{0,s}(Z_s)\widetilde\lambda_s(Z_s)}
G_s(Z_s;\theta^\star)
\right].
\label{eq:linear-quadratic-plug-drift}
\end{align}
The first sum is linear in the projected-tail error, whereas the second is
quadratic. Under positivity and the corresponding moment conditions, the
second sum is therefore a higher-order remainder.

More formally, for a collection of directions
$h=\{h_s:s\in\mathcal S^\circ\}$, consider the working-function path
\[
\lambda_{\varepsilon,s}
=
\lambda_{0,s}+\varepsilon h_s,
\qquad
s\in\mathcal S^\circ,
\]
for values of $\varepsilon$ small enough that all denominators remain
positive. The partial plug-in Gateaux derivative is defined by
\[
D_\lambda^{\mathrm{plug}}
\Psi(\theta^\star,\lambda_0)[h]
:=
\left.
\frac{d}{d\varepsilon}
\Psi(\theta^\star,\lambda_\varepsilon)
\right|_{\varepsilon=0}.
\]
Equation \eqref{eq:linear-quadratic-plug-drift} yields
\begin{equation}
D_\lambda^{\mathrm{plug}}
\Psi(\theta^\star,\lambda_0)[h]
=
-
\sum_{s\in\mathcal S^\circ}
E\left[
\frac{h_s(Z_s)}
{\lambda_{0,s}(Z_s)}
G_s(Z_s;\theta^\star)
\right].
\label{eq:partial-plug-derivative}
\end{equation}
The superscript ``plug'' emphasizes that this derivative varies the
projected tails as working arguments while holding the data law and the
other nuisance quantities fixed.

The derivative in \eqref{eq:partial-plug-derivative} can be nonzero. For
example, suppose that for some $s\in\mathcal S^\circ$ and coordinate $j$,
\[
P\{G_{s,j}(Z_s;\theta^\star)\neq0\}>0.
\]
Set all other directions to zero and choose
\[
h_s(Z_s)
=
\lambda_{0,s}(Z_s)
\tanh\{G_{s,j}(Z_s;\theta^\star)\}.
\]
This is a bounded relative perturbation, so the path remains positive for
sufficiently small $\varepsilon$. The $j$th coordinate of
\eqref{eq:partial-plug-derivative} is
\[
-
E\left[
G_{s,j}(Z_s;\theta^\star)
\tanh\{G_{s,j}(Z_s;\theta^\star)\}
\right]
<0.
\]
Hence separate plug-in estimation can perturb the population moment at
first order unless a special cancellation occurs.

We next connect this moment drift to the target parameter. Let
$\theta(\lambda)$ denote the local population root satisfying
\[
\Psi\{\theta(\lambda),\lambda\}=0.
\]
At the true projected tails, Proposition~4 gives
\[
\Psi(\theta,\lambda_0)
=
E\{\psi^F(Z,\theta)\},
\]
and therefore
\[
\theta(\lambda_0)=\theta^\star,
\qquad
\left.
\frac{\partial}{\partial\theta^\top}
\Psi(\theta,\lambda_0)
\right|_{\theta=\theta^\star}
=
A.
\]
Under the usual local implicit-function conditions and the nonsingularity
of $A$, the derivative of the population-root map is
\begin{equation}
D\theta(\lambda_0)[h]
=
-A^{-1}
D_\lambda^{\mathrm{plug}}
\Psi(\theta^\star,\lambda_0)[h].
\label{eq:population-root-derivative}
\end{equation}
Thus, when \eqref{eq:partial-plug-derivative} is nonzero, perturbing the
projected tails shifts the population root at first order.

Finally, suppose that $\widehat\lambda$ is estimated on an independent
training fold and that $\widehat\theta$ is the empirical root on the
evaluation fold. Conditional on the training fold, standard
Z-estimation gives
\[
\widehat\theta-\theta(\widehat\lambda)
=
O_p(N^{-1/2})
\]
under the usual regularity conditions. Therefore,
\begin{equation}
\widehat\theta-\theta^\star
=
\underbrace{
\widehat\theta-\theta(\widehat\lambda)
}_{\text{sampling error}}
+
\underbrace{
\theta(\widehat\lambda)-\theta^\star
}_{\text{plug-in root shift}}.
\label{eq:plug-root-decomposition}
\end{equation}
The first term in \eqref{eq:plug-root-decomposition} is the ordinary
sampling fluctuation. By
\eqref{eq:population-root-derivative}, the second term is generally
first order in the relevant projected-tail estimation error. Cross-fitting
controls the empirical dependence between the fitted nuisance functions
and the evaluation sample, but it does not generally remove this
population-root shift.

This is a partial plug-in calculation rather than a derivative along a
regular MAR statistical submodel. Along an actual MAR submodel, the
observed-data law, the complete-case propensity, and the projected tails
vary jointly and are linked by the same propensity mechanism and full-data
law. Derivatives of these additional components, including that of the
leading inverse-probability-weighted term, are not represented above and
may produce further cancellations.

\end{document}

%% file: prototype.tex
\begin{tikzpicture}[font=\small]

\definecolor{CMUred}{RGB}{196, 18, 48}
\definecolor{CMUbeige}{RGB}{228, 218, 196}
\definecolor{CMUgray1}{RGB}{109, 110, 113}
\definecolor{CMUskyblue}{RGB}{0, 123, 192}
\definecolor{CMUgold}{RGB}{253, 181, 21}
\definecolor{CMUgreen}{RGB}{0, 150, 71}

\tikzset{
  obs/.style={rounded corners=2pt,
  draw=black, fill=CMUgray1, thick},
  obsZ1/.style={rounded corners=2pt, draw=CMUgreen!55!black, fill=CMUgreen, thick},
  obsZ2/.style={rounded corners=2pt, draw=CMUgold!70!black,  fill=CMUgold,  thick},
  obsZ3/.style={rounded corners=2pt, draw=CMUskyblue!70!black,fill=CMUskyblue,thick},
  miss/.style={rounded corners=2pt, draw=black!30, fill=black!4,
               pattern=north east lines, pattern color=black!20},
  panel/.style={rounded corners=5pt, fill=CMUbeige!35, draw=none},
  hdr/.style={font=\normalsize\bfseries},
  rowlab/.style={font=\normalsize\bfseries, anchor=east},
  tag/.style={font=\footnotesize\itshape, text=CMUgray1}
}

\foreach \cx in {3.0, 5.7, 8.4, 11.1}{%
  \fill[panel] (\cx-1.2,0.6) rectangle (\cx+1.2,-2.55);
}
\draw[rounded corners=5pt, draw=CMUred, line width=1pt]
      (3.0-1.24,0.64) rectangle (3.0+1.24,-2.6);

\node[hdr] at (3.0, 1.0)  {$R = [3]$};
\node[hdr] at (5.7, 1.0)  {$R = \{1,2\}$};
\node[hdr] at (8.4, 1.0)  {$R = \{1,3\}$};
\node[hdr] at (11.1,1.0)  {$R = \{1\}$};

\node[rowlab, text=black]  at (1.35, 0)    {$Z_1$};
\node[rowlab, text=black]   at (1.35,-1.0)  {$Z_2$};
\node[rowlab, text=black]at (1.35,-2.0)  {$Z_3$};

\draw[obsZ1] (2.0,-0.4) rectangle (4.0, 0.4) node[pos=.5,white]{$Z_1$};
\draw[obsZ2] (2.0,-1.4) rectangle (4.0,-0.6) node[pos=.5,black]{$Z_2$};
\draw[obsZ3] (2.0,-2.4) rectangle (4.0,-1.6) node[pos=.5,white]{$Z_3$};
\draw[obsZ1] (4.7,-0.4) rectangle (6.7, 0.4) node[pos=.5,white]{$Z_1$};
\draw[obsZ2] (4.7,-1.4) rectangle (6.7,-0.6) node[pos=.5,black]{$Z_2$};
\draw[miss]  (4.7,-2.4) rectangle (6.7,-1.6);
\draw[obsZ1] (7.4,-0.4) rectangle (9.4, 0.4) node[pos=.5,white]{$Z_1$};
\draw[miss]  (7.4,-1.4) rectangle (9.4,-0.6);
\draw[obsZ3] (7.4,-2.4) rectangle (9.4,-1.6) node[pos=.5,white]{$Z_3$};
\draw[obsZ1] (10.1,-0.4) rectangle (12.1, 0.4) node[pos=.5,white]{$Z_1$};
\draw[miss]  (10.1,-1.4) rectangle (12.1,-0.6);
\draw[miss]  (10.1,-2.4) rectangle (12.1,-1.6);


\draw[obs] (2.0,-3.55) rectangle (2.45,-3.2);
\node[anchor=west, font=\footnotesize] at (2.55,-3.375) {observed};
\draw[miss]  (4.6,-3.55) rectangle (5.05,-3.2);
\node[anchor=west, font=\footnotesize] at (5.15,-3.375) {missing};

\end{tikzpicture}

%% file: real_data_illustration.tex
\begin{tikzpicture}[font=\small, x=0.96cm, y=0.96cm]

\tikzset{
  obsRNA/.style={rounded corners=2pt, draw=CMUgreen!55!black,
                 fill=CMUgreen, thick},
  obsATAC/.style={rounded corners=2pt, draw=CMUgold!70!black,
                  fill=CMUgold, thick},
  obsADT/.style={rounded corners=2pt, draw=CMUskyblue!70!black,
                 fill=CMUskyblue, thick},
  miss/.style={rounded corners=2pt, draw=black!30, fill=black!4,
               pattern=north east lines, pattern color=black!20},
  assaypanel/.style={rounded corners=5pt, fill=CMUbeige!35, draw=none},
  assayhdr/.style={font=\normalsize\bfseries},
  patternlab/.style={font=\scriptsize, text=CMUgray1},
  rowlab/.style={font=\normalsize\bfseries, anchor=east},
  tag/.style={font=\footnotesize\itshape, text=CMUgray1}
}

\foreach \cx in {3.0,6.4,9.8}{%
  \fill[assaypanel] (\cx-1.42,0.48) rectangle (\cx+1.42,-2.52);
}

\draw[rounded corners=5pt, draw=CMUred, line width=1pt]
      (3.0-1.46,0.52) rectangle (3.0+1.46,-2.57);

\node[assayhdr] at (3.0,1.28) {DOGMA-seq};

\node[assayhdr] at (6.4,1.28) {CITE-seq};

\node[assayhdr] at (9.8,1.28) {ASAP-seq};

\node[rowlab, text=CMUgreen!55!black]   at (1.18, 0)   {RNA};
\node[rowlab, text=CMUgold!70!black]    at (1.18,-1.0) {ATAC};
\node[rowlab, text=CMUskyblue!70!black] at (1.18,-2.0) {ADT};

\draw[obsRNA] (1.8,-0.38) rectangle (4.2, 0.38)
      node[pos=.5,white,font=\bfseries]{RNA};
\draw[obsATAC] (1.8,-1.38) rectangle (4.2,-0.62)
      node[pos=.5,black,font=\bfseries]{ATAC};
\draw[obsADT] (1.8,-2.38) rectangle (4.2,-1.62)
      node[pos=.5,white,font=\bfseries]{ADT};

\draw[obsRNA] (5.2,-0.38) rectangle (7.6, 0.38)
      node[pos=.5,white,font=\bfseries]{RNA};
\draw[miss] (5.2,-1.38) rectangle (7.6,-0.62);
\draw[obsADT] (5.2,-2.38) rectangle (7.6,-1.62)
      node[pos=.5,white,font=\bfseries]{ADT};

\draw[miss] (8.6,-0.38) rectangle (11.0, 0.38);
\draw[obsATAC] (8.6,-1.38) rectangle (11.0,-0.62)
      node[pos=.5,black,font=\bfseries]{ATAC};
\draw[obsADT] (8.6,-2.38) rectangle (11.0,-1.62)
      node[pos=.5,white,font=\bfseries]{ADT};

\draw[obsRNA] (3.25,-3.57) rectangle (3.70,-3.22);
\node[anchor=west,font=\footnotesize] at (3.80,-3.395) {observed modalities};
\draw[miss] (6.70,-3.57) rectangle (7.15,-3.22);
\node[anchor=west,font=\footnotesize] at (7.25,-3.395) {missing modalities};

\end{tikzpicture}